\documentclass[a4paper,11pt,nofootinbib,floatfix]{revtex4-2}

\usepackage{amssymb,bm,epsf}
\usepackage{setspace}
\usepackage{amsmath}
\usepackage{graphicx}
\usepackage{dcolumn}
\usepackage{color}
\usepackage{accents}
\usepackage{multirow}
\usepackage{tikz}
\usepackage{comment}
\usepackage{subfigure}
\usepackage{soul}
\usepackage[rightcaption]{sidecap}
\sidecaptionvpos{figure}{t}
\usepackage{enumerate}

\usepackage[toc,page]{appendix}
\usepackage[normalem]{ulem}

\usepackage[hyperfootnotes=false]{hyperref}
\hypersetup{
    colorlinks=true,
    linkcolor=red,
    filecolor=magenta,      
    citecolor=blue
}

\usepackage{titlesec}

\titleformat{\section}
    {\normalfont\bfseries\normalsize\centering}
    {\thesection}{1em}{\MakeUppercase}
\titlespacing{\section}
    {0pt}   
    {2.0ex plus .5ex minus .2ex}  
    {0.8ex plus .1ex}             

\titleformat{\subsection}
    {\normalfont\bfseries\centering}{\makebox[1cm][l]{\thesubsection}}{0pt}{}
\titlespacing{\subsection}
    {0pt}
    {1.5ex plus .2ex minus .2ex}
    {0.8ex plus .1ex}

\newcolumntype{P}[1]{>{\centering\arraybackslash}p{#1}}
\newcolumntype{M}[1]{>{\centering\arraybackslash}m{#1}}

\makeatletter
\@addtoreset{equation}{section}

\makeatother

\begin{document}
\begin{titlepage}
\title{Asymptotically Safe Gravitational Form Factors from the Proper-Time Flow Equation}
\author{Emiliano Maria Glaviano}
\email{emiliano.glaviano@inaf.it}
\affiliation{
INAF Osservatorio Astrofisico di Catania, Via S.Sofia 78, 95123 Catania Italy}
\affiliation{INFN, Sezione di Catania, Via S.\ Sofia 64, 95123 Catania, Italy}
\affiliation{Università di Catania, Dipartimento di Fisica e Astronomia, Via S. Sofia 64, 95123, Catania, Italy}

\pacs{}

\begin{abstract}
\noindent We study the renormalization group flow of nonlocal form factors in four-dimensional quantum gravity within the proper-time formalism at quadratic order in the curvature expansion. We show that, in the presence of a running Newtonian coupling and a fixed negative cosmological constant, the flow equations can be integrated down to $k=0$, allowing the reconstruction of the full momentum dependence of the form factors. The running of Newton's coupling generates nontrivial momentum-dependent corrections to the effective field theory behavior, with the Planck scale setting the crossover between the IR and UV regimes. In the infrared, these corrections are organized as an expansion in powers of $q^2/m_p^2$, while in the ultraviolet they are organized in powers of $m_p^2/q^2$. Within this framework, we also analyze the properties of the dimensionless flow, identifying its fixed-point structure and constructing asymptotically safe solutions. We show that the integrated asymptotically safe solutions admit a finite and cutoff-independent $\Lambda\to\infty$ limit when the non-Gaussian fixed-point solution is imposed as the ultraviolet boundary condition. This removes UV logarithmic contributions and ensures independence from the renormalization scale $\mu$. The resulting renormalized asymptotically safe form factors at $k=0$ display a power-law decay $\sim1/q^2$ in the ultraviolet and reproduce the expected logarithmic structure in the infrared, with the Planck scale replacing the renormalization scale.
\end{abstract}

\maketitle

\end{titlepage}
\newpage
\setcounter{page}{2}

\section{Introduction}\label{introduction}
\noindent Perturbative quantum gravity based on the Einstein–Hilbert action is not renormalizable beyond one loop. Several approaches have been developed to address this problem. Among them, the asymptotic safety scenario \cite{Eichhorn:2017egq,Bonanno:2020bil,Niedermaier:2006ns,Reuter:2012id,Dupuis:2020fhh, article,book,Saueressig:2023irs,Reuter_Saueressig_2019} has received considerable attention. Originally proposed by Weinberg, it is based on the existence of a nontrivial ultraviolet fixed point of the renormalization group flow, which can be explicitly identified in $d=2+\epsilon$ dimensions. This observation has motivated extensive studies in four dimensions, primarily within the functional renormalization group framework based on the Wetterich equation \cite{Morris:1993qb,WETTERICH199390,Ellwanger:1993mw}. Within this approach, a nontrivial UV-attractive fixed point has been found in a wide range of truncations, starting from the Einstein–Hilbert action and its extensions including higher-order polynomial expansions up to $R^{34}$ \cite{Machado:2007ea,Codello:2008vh,Falls:2013bv,Falls:2014tra,Falls:2018ylp,Codello:2007bd,Alkofer:2018fxj,Ohta:2015fcu,Ohta:2015efa,Falls:2016msz,Ohta:2016npm,Goncalves:2017jxq,Nink:2014yya,Falls:2015qga,Gies:2015tca,Falkenberg:1996bq,Souma:2000vs,Barra:2019rhz,Bonanno:2025tfj,Ohta:2016npm,Ohta:2016jvw}. More general truncations involving invariants such as $f(R_{\mu\nu}R^{\mu\nu})$ and $f(R_{\alpha\beta\mu\nu}R^{\alpha\beta\mu\nu})$ have been also considered and provide further evidence for the existence of such a fixed point \cite{Falls:2017lst}. Evidence for a UV-attractive fixed point has also been obtained in complementary approaches, including the inclusion of the Goroff–Sagnotti two-loop counterterm \cite{GOROFF1986709} and lattice formulations of quantum gravity \cite{Ambjorn:2005qt,Laiho:2011ya,Ambjorn:2012jv,Coumbe:2014nea,Laiho:2016nlp,Ambjorn:2018qbf,Loll:2019rdj,Ambjorn:2021yvk,Bassler:2021pzt}.

The asymptotic safety scenario aims at providing a UV-complete description of gravity by connecting a nontrivial fixed point in the ultraviolet with the infrared regime at $k=0$. However, truncations of the effective action based on a finite number of local curvature invariants inevitably miss part of the physical content of the theory, since quantum corrections generically induce nonlocal interactions \cite{Donoghue:1993eb,Donoghue:1994dn,Holstein:2004dn,Akhundov:1996jd,Codello:2015mba}. This feature already appears at quadratic order in the curvature, where one-loop effects generate logarithmic contributions to the effective action of the form $\sim R + R \ln(-\Box/\mu^2) R + R_{\mu\nu} \ln(-\Box/\mu^2) R^{\mu\nu}$ \cite{tHooft:1974toh,Donoghue:1994dn}. In a UV-complete theory, these infrared logarithmic structures are not expected to remain valid at arbitrarily high energies. Rather, they should emerge as the low-energy limit of form factors encoding the nonperturbative dynamics of the UV regime. Such form factors are expected to remain regular in the ultraviolet and to avoid uncontrolled divergences at trans-Planckian scales. In this description, the explicit dependence on the renormalization scale by $\Box/\mu^2$ is replaced by a regular dependence on the dimensionless operator $\Box/m_p^2$, providing a smooth interpolation between infrared and ultraviolet scales. From this perspective, the UV-complete theory should be understood as a consistent deformation of the infrared nonlocal structure, rather than its direct extrapolation to trans-Planckian momenta.

In \cite{Barvinsky:1990up,Barvinsky1987BeyondTS,Barvinsky:1990uq,Barvinsky:1993en,Avramidi:2000bm} it was shown that loop effects in quantum field theories on curved spacetime generate nonlocal structures that can be resummed into an expansion in curvature invariants. In this framework, these contributions organize into nonlocal form factors appearing in the effective action. To address asymptotic safety at this level, one must determine their renormalization group (RG) flow, whose $k \to 0$ limit is required to reconstruct the full physical effective action.

The asymptotic safety program for nonlocal form factors has been actively developed in recent years \cite{Knorr:2019atm,Bosma:2019aiu,Knorr:2022dsx,Knorr:2021niv}. In the purely gravitational case, a first important result at quadratic order in the curvature expansion was obtained in \cite{Codello:2010mj}, where the Wetterich equation was shown to reproduce the Polyakov action in $d=2$ at $k=0$. Subsequently, it was demonstrated in \cite{Satz:2010uu} that, in the same limit, the framework correctly reproduces in $d=4$ the well-known one-loop effective action of ’t Hooft and Veltman \cite{tHooft:1974toh}. Progress toward including the running of the Newton coupling has been achieved only more recently and mostly within simplified settings, often neglecting the cosmological constant. In \cite{Bosma:2019aiu}, a running Newton coupling was incorporated in a conformally reduced model, allowing the computation of the fixed-point form factor associated with the Weyl-squared term. Further developments were obtained in \cite{Knorr:2021niv}, where the fluctuation approach \cite{article,Christiansen:2012rx,Christiansen:2014raa,Christiansen:2015rva,Denz:2016qks,Bonanno:2021squ} was used to relate the graviton propagator to the corresponding form factors. 
Nevertheless, obtaining the complete RG flow connecting the UV fixed point to the infrared limit at $k=0$ remains challenging due to the complexity of the resulting integrodifferential equations. 

These considerations motivate the use of a framework that allows for a controlled and practically tractable reconstruction of the renormalization group flow of nonlocal gravitational form factors down to $k=0$, while preserving the correct infrared and ultraviolet limits of known results. In this work, we employ the proper-time (PT) flow equation for this purpose.

Recently, the proper-time flow equation has emerged as a useful alternative to the Wetterich formalism. Although it does not define an exact functional renormalization group equation in the effective average action (EAA) framework, it provides a one-loop improved flow that preserves the main structural features of the RG and accurately captures threshold effects. Evidence supporting its reliability comes from several results. In particular, the Reuter fixed point has been recovered within the PT framework \cite{Bonanno:2004sy}, and it has been shown that the PT flow at $k=0$ reproduces the Polyakov action \cite{Glaviano:2024hie}. Moreover, it yields accurate critical exponents in scalar theories \cite{Bonanno:2000yp,Mazza:2001bp}, and recent applications to gravity-matter systems with momentum-dependent matter operators indicate that the PT flow can also describe the fixed-point structure and the associated critical manifold of nonlocal form factors \cite{Bonanno:2026ljx}. It has also been successfully applied in a variety of contexts, including the Standard Model \cite{Giacometti:2025qyy,Giacometti:2026zrs}, pure gravity \cite{Bonanno:2025dry,Bonanno:2025tfj,Giacometti:2024qva,Bonanno:2023fij,Bonanno:2023ghc,Bonanno:2012dg}, and gravity–matter systems \cite{Bonanno:2025qsc,Bonanno:2026mzs,Spina:2025wxb,Giacometti:2026zrs}. These results indicate that, despite its approximate nature, the PT flow reliably captures the universal features of the renormalization group flow and yields results comparable to those obtained with other functional renormalization group equations. This expectation can be understood from the fact that the one-loop approximation becomes exact when integrating over infinitesimal momentum shells \cite{Wegner:1972ih}. As a consequence, the PT flow provides a controlled approximation of the full functional renormalization group dynamics, expected to correctly capture the universal momentum dependence of nonlocal form factors and their infrared behavior \cite{Bonanno:2026ljx,Glaviano:2024hie}. 

Motivated by these observations, we use the PT flow equation to compute nonlocal contributions to the gravitational effective action at quadratic order in the curvature in four dimensions, including the running of Newton’s coupling. We then integrate the flow down to $k=0$ to reconstruct the corresponding infrared effective action, providing a first step toward connecting the ultraviolet regime governed by the nontrivial fixed point with the low-energy effective theory.

The proper-time flow equation is given by
\begin{equation}
\label{PTFE}
X\partial_X S_X [\phi] = \frac{1}{2}\int_{0}^{\infty}\frac{ds}{s}\rho\left(s,X^2 Z_X\right)\mathrm{STr}\left[e^{-sS_X^{(2)}}\right],
\end{equation}
where $X$ denotes the cutoff scale, $S_X^{(2)}$ is the Hessian of the action, $Z_X$ the wave function renormalization, and $\rho(s,X^2Z_X)$ a cutoff function. Depending on whether the cutoff is interpreted as an infrared scale ($X=k$) or an ultraviolet scale ($X=\Lambda$), the equation defines, respectively, a flow for the effective average action or an exact Wilsonian RG flow \cite{Bonanno:2019ukb,deAlwis:2017ysy}. In this work, we adopt the EAA interpretation and compute a controlled one-loop resummation that captures the full momentum dependence of the effective action at quadratic order in the curvature. 

At present, the general properties of asymptotically safe form factors are not yet fully understood. Although they can be computed within different frameworks \cite{Knorr:2022dsx,Kher:2025rve,Draper:2020bop,Draper:2020knh,Knorr:2022lzn,Pawlowski:2023gym,Pawlowski:2023dda,Pastor-Gutierrez:2024sbt,Knorr:2022kqp,Knorr:2021iwv,Knorr:2018kog,Denz:2016qks,Knorr:2026vax}, a comprehensive characterization of the features that distinguish asymptotically safe form factors is still emerging. In this work, we aim to contribute to this discussion, focusing in particular on the limit $k \to 0$. We adopt an operational definition of asymptotically safe form factors as nonlocal structures whose dimensionless flow approaches a nontrivial ultraviolet fixed point and which, after integration down to $k=0$, lead to a well-defined ultraviolet behavior in dimensionful variables. This definition is motivated by the fact that, within the present framework, these two requirements are not automatically equivalent. We show that the existence of a nontrivial fixed point for the dimensionless form factors does not by itself guarantee a finite ultraviolet limit of the dimensionful effective action. When expressed in dimensionful variables at $k=0$, residual divergences may reappear in the limit $\Lambda \to \infty$, even when the dimensionless flow is asymptotically safe. This indicates that an additional renormalization condition
is required to ensure finiteness. We discuss how such a condition can be implemented within the present framework and explore its implications for the construction of asymptotically safe nonlocal form factors.

The analytic continuation to Minkowski spacetime, together with the analysis of the graviton propagator, unitarity properties, and corrections to the Newtonian potential, will be presented in a companion work.

In this work, we study the running of the form factors associated with the curvature-squared terms in the gravitational effective action within the proper-time formalism, including a running Newton coupling. In Sec.~\ref{secform} we present the formalism and the corresponding flow equations. In Sec.~\ref{secintegFF} we perform an analytic integration of the flow equations in the infrared and ultraviolet regimes, extending the results down to $k=0$. In Sec.~\ref{secASformfactor} we describe the properties of the adimensional flow and construct the asymptotically safe theory. In Sec.~\ref{secnuman} we present a numerical analysis to obtain the form factors at $k=0$ for generic external momenta $q^2$. Finally, in Sec.~\ref{secconc} we present our conclusions. Three appendices contain technical details of the computations: Appendix~\ref{appderbeta} derives the flow equations for the form factors. Appendix.~\ref{appserie} discusses the derivation of the asymptotic expansions without solving the full flow equations. Appendix.~\ref{app:techres} contains technical results of this work.

\section{The formalism}\label{secform}
We consider the Euclidean effective average action
\begin{equation}
\label{TE}
\Gamma_k[g] = \int d^d x \sqrt{g} \left[\frac{1}{16\pi G_k}\left(-R + 2\bar{\lambda}_k\right) + Rf_k^{(\mathrm{R})}(-\Box)R + R_{\mu\nu} f_k^{(\mathrm{Ricc})}(-\Box)R^{\mu\nu}\right]
\end{equation}
Here $\Box$ denotes the Laplace–Beltrami operator. $f_k^{(\mathrm{R})}(-\Box)$ and $f_k^{(\mathrm{Ricc})}(-\Box)$ are nonlocal form factors encoding the momentum dependence of the quadratic curvature sector. The truncation of the EAA is restricted to diffeomorphism-invariant operators up to second order in the curvature.

Following \cite{Bonanno:2019ukb}, we choose a spectrally adjusted cutoff function for the PT flow equation,
\begin{equation}
\label{cutoff}
\rho_m(s,k^2 Z_k) = \left(2 + \epsilon\frac{k \partial_k Z_k}{Z_k}\right)\frac{(s\, m\, k^2 Z_k)^m}{\Gamma(m)}e^{-s\, m\, k^2 Z_k}\,,
\end{equation}
where $m>0$ controls the behavior of $\rho(s,k^2 Z_k)$ in the interpolating region. The parameter $\epsilon$ distinguishes between two cutoff prescriptions referred to as type B scheme ($\epsilon=1$) and type C scheme ($\epsilon=0$) in the literature \cite{Bonanno:2019ukb}. The difference between them is that scheme B retains the contribution of the anomalous dimension generated by the scale dependence of $Z_k$, whereas in scheme C this contribution is omitted. The effect of $\epsilon$ on the running of the Newton coupling and cosmological constant was analyzed in \cite{Bonanno:2025tfj}, while its effect on gravity-matter systems was studied in \cite{Bonanno:2025qsc,Bonanno:2026ljx}. In addition to these two uniform choices, we will also consider a mixed prescription, in which scheme C is used for the running couplings and scheme B for the form factors. The motivation for this mixed choice will become clear after describing the corresponding flow equations.

\subsection{The flow equations for the form factors}
The proper-time flow equations for the theory in Eq.~(\ref{TE}) are derived using the background field method combined with the nonlocal heat-kernel expansion developed in \cite{Barvinsky:1990up,Barvinsky1987BeyondTS,Barvinsky:1990uq,Barvinsky:1993en,Avramidi:2000bm}. Within this framework, the resulting expansion corresponds to the leading order of the vertex expansion of the effective action in the fluctuation field $h_{\mu\nu}$, with vertices evaluated on a fixed background metric \cite{Codello:2012kq}. The derivation relies on the approximations discussed in Sec.~\ref{subsecappro} and Appendix~\ref{appderbeta}, which ultimately lead to the flow equations presented below.

Introducing the dimensionless variables $g_k=k^{d-2}G_k$ and $\lambda_k=k^{-2}\bar\lambda_k$, the flow equations in $d=4$ with $m=3$ for the two couplings read (the general expressions are given in Appendix~\ref{appderbeta})
\begin{equation}\begin{split}
\label{eqglambdam3}
&k\partial_k\lambda_k=g_k\left(-\frac{18}{\pi}+\frac{135\left(2-\epsilon\eta_k\right)}{4\pi\left(3-2\lambda_k\right)}\right)-\lambda_k\left(2-\eta_k\right)\\
&k\partial_kg_k=g_k\left(2+\eta_k\right)
\end{split}\end{equation}
where the gravitational anomalous dimension is given by
\begin{equation}
\eta_k=\frac{2g_k\ \left(207+40\left(\lambda_k-3\right)\lambda_k\right)}{117\epsilon g_k-4\pi\left(3-2\lambda_k\right)^2}.
\end{equation}
The flow equations for the two form factors share the same structure and take the form
\begin{equation}\begin{split}
\label{eqFFm3}
&k\partial_kf_k^{\left(i\right)}\left(z\right)=A^{(i)}\left(k,z;m\right)+B^{(i)}\left(k,z;m\right)\tanh^{-1}\left(\sqrt{\frac{z}{4mk^2+z}}\right)+\\
&+\left(2-\epsilon\eta_k\right)\left(C^{(i)}\left(k,z,\bar{\lambda}_k;m\right)+D^{(i)}\left(k,z,\bar{\lambda}_k;m\right)\tanh^{-1}\left(\sqrt{\frac{z}{4mk^2-8\bar\lambda_k+z}}\right)\right)
\end{split}\end{equation}
where $i=\mathrm{R}$, $\mathrm{Ricc}$ and $z=-\Box$. Different values of the cutoff parameter $m$ modify the form of the coefficients but leave the structure of the flow equations unchanged. The contributions in the first and second lines originate from the ghost and gravitational sectors, respectively. The scale dependence of Newton's coupling enters the flow through $\eta_k$, while no explicit overall dependence on $g_k$ appears in the flow equation. This is a consequence of the background approximation and of the structure of the proper-time flow equation adopted here. A detailed explanation is given in Appendix~\ref{appderbeta}. The explicit expressions for the coefficients are quite long and are given in Appendix~\ref{app:techres} for $m=3$.

\subsection{Approximations and domain of validity}\label{subsecappro}
The derivation of the flow equations for the form factors relies on a set of approximations, which we summarize here together with their domain of validity.

A first limitation concerns the structure of the flow equation, which is determined by the heat-kernel trace in Eq.~(\ref{PTFE}). For the action in Eq.~(\ref{TE}), the exact Hessian has a nontrivial tensorial structure and does not fall into the class of Laplace-type operators $\Gamma^{(2)} \sim f(-\Box)+\mathbf{U}$ (see Appendix~\ref{appderbeta}). This prevents the direct use of standard Laplace-type heat-kernel formulas, and more general techniques, such as off-diagonal \cite{Barvinsky:1985an,Benedetti:2010nr,Decanini:2005gt,Groh:2011dw,Groh:2011vn,Decanini:2007gj,Anselmi:2007eq} or nonminimal heat-kernel methods \cite{Barvinsky:2025jbw,Sauro:2025sbt}, would in principle be required. To retain computational control, we restrict the analysis to the Einstein-Hilbert Hessian in $d=4$, which allows the use of the nonlocal heat-kernel expansion of Barvinsky and Vilkovisky \cite{Barvinsky:1990up, Barvinsky1987BeyondTS, Barvinsky:1990uq, Barvinsky:1993en, Avramidi:2000bm}. In this setup, the form factors are not included in the operator entering the trace but are instead reconstructed from the flow via a projection onto curvature invariants quadratic in the curvature, where the full momentum dependence is encoded.

As a consequence, the resulting system does not define a closed renormalization group flow for the full effective action. The computation therefore corresponds to a one-loop approximation, in which the resulting flow equations resum the Einstein-Hilbert contributions, providing a controlled approximation that captures quantum corrections at quadratic order in the curvature while retaining the full nonlocal momentum dependence. This approximation has been shown to be reliable, correctly reproducing the known nonlocal structure and universal momentum dependence of the form factors in both $d=2$ and $d=4$.  \cite{Codello:2010mj,Satz:2010uu,Codello:2015oqa,Ribeiro:2018pyo,Codello:2011js,Glaviano:2024hie,Ohta:2020bsc}.
Self-consistent treatments that retain the momentum-dependent feedback in the flow have also been developed in the context of the graviton spectral function \cite{Pawlowski:2025etp}.

A second limitation arises from the presence of a positive cosmological constant, which introduces an infrared scale affecting the spectrum of the Laplacian. This modifies the contribution of low-lying modes and can restrict the validity of the nonlocal expansion \cite{Codello:2015mba}. In the FRG flow, this manifests as a singularity at a finite value of $\lambda_k$ \cite{Reuter:2001ag}, which in the proper-time flow with the cutoff Eq.(\ref{cutoff}) is $\lambda_k = m/2$. To avoid these infrared issues, we restrict the analysis to a negative cosmological constant. This is a technical simplification ensuring the validity of the expansion but at the same time does not affect the ultraviolet structure of the form factors.

Extending the analysis to fluctuation form factors would require going beyond the background-field approximation, leading to coupled non-linear integrodifferential equations \cite{Knorr:2019atm,Bonanno:2026ljx,Bosma:2019aiu}, schematically represented in Eq.~(\ref{integroDE}) for perturbations around flat space. At present, no general nonperturbative method exists to solve the coupled integrodifferential systems for $f_k(q^2)$ along the full RG trajectory down to $k=0$, except in simplified settings where only fixed-point or perturbative solutions are considered \cite{Knorr:2019atm,Bonanno:2026ljx,Bosma:2019aiu}. For this reason, such an extension lies beyond the scope of this work. In this sense, the one-loop approximation corresponds to retaining only the first iteration of the full integrodifferential system.

\section{The flow of the form factors}\label{secintegFF}
\noindent The form factors are obtained by solving Eq.~(\ref{eqFFm3}). The general solution reads
\begin{equation}
\label{solFFdim}
f_{k}^{(i)}\left(q^2\right)=f_{\Lambda}^{(i)}\left(q^2\right)+\int_{\Lambda}^{k}\frac{dk^\prime}{k^\prime}\,\beta^{(i)}_{k'}\left(q^2,\bar\lambda_{k'}, \eta_{k^\prime}\right),
\end{equation}
where $i=\mathrm{R},\mathrm{Ricc}$ and $f_{\Lambda}^{(i)}$ encodes the boundary conditions at the UV scale $k=\Lambda$. In the standard approach, the boundary conditions implement perturbative renormalization prescriptions, but nonperturbative renormalization conditions can also be adopted, as discussed in Sec.~\ref{secASformfactor}.

When both the Newton coupling and the cosmological constant are allowed to run, the flow equations can in general be solved only numerically. However, to isolate the main features of the flow, an analytical treatment is desirable. To this end, we consider the simplified case of a nonrunning negative dimensionful cosmological constant $\bar\lambda$. This is a specific approximation adopted throughout our analysis, rather than a consequence of the full coupled flow equations. The effects of the cosmological constant on the form factors have been previously analyzed in \cite{Codello:2015mba} within the effective field theory (EFT) framework. The emergence of these results from a genuinely nonperturbative RG flow, however, is not yet fully understood. For this reason, we first revisit the EFT case in the presence of a cosmological constant, and then analyze how the running of Newton’s coupling modifies the picture.

It is important to note that, in general,
\begin{equation}
\lim_{q^2\to0}\int_{k}^{\Lambda}\frac{\beta(q^2,k')}{k'}\,dk'
\;\neq\;
\int_{k}^{\Lambda}\lim_{q^2\to0}\frac{\beta(q^2,k')}{k'}\,dk'\,.
\end{equation}
This arises from nonanalytic contributions in $q^2$ and implies that a naïve $q^2\to0$ projection does not reproduce the running of the local couplings in $\Gamma_k\sim R + c_k^{(R)} R^2 + c_k^{(Ricc)} R_{\mu\nu}R^{\mu\nu} + O(R^3)$.

\subsection{The EFT case in the presence of a nonrunning \texorpdfstring{$\bar \lambda$}{N}}
In the EFT case the flow can be computed exactly. 
At $k=0$ the EFT result is given by
\begin{equation}\begin{split}
\label{EFTresult}
&f_{k=0}^{\left(\mathrm{R,EFT}\right)}\left(q^2,u;\mu^2\right)=c_\mathrm{R}(m)+\frac{\ln{\left(\frac{q^2}{\mu^2}\right)}}{1920\pi^2}-\\
&-\frac{7\ln{\left(\frac{u}{2}\right)}}{384\pi^2}+\frac{24-107u}{288\pi^2u^2}+\frac{\sqrt{u+8}\left(7u^2+72u-16\right)\ln{\left(\frac{1+\sqrt{\frac{u}{8+u}}}{1-\sqrt{\frac{u}{8+u}}}\right)}}{384\pi^2u^\frac{5}{2}}\\
&f_{k=0}^{\left(\mathrm{Ric,EFT}\right)}\left(q^2,u;\mu^2\right)=c_\mathrm{Ric}(m)+\frac{7\ln{\left(\frac{q^2}{\mu^2}\right)}}{320\pi^2}-\\
&-\frac{7\ln{\left(\frac{u}{2}\right)}}{192\pi^2}+\frac{-24+29u}{36\pi^2u^2}+\frac{\sqrt{u+8}\left(7u^2-80u+64\right)\ln{\left(\frac{1+\sqrt{\frac{u}{8+u}}}{1-\sqrt{\frac{u}{8+u}}}\right)}}{192\pi^2u^\frac{5}{2}}
\end{split}\end{equation}
where the $c_i(m)$ are two scheme dependent constants that can be removed by the boundary conditions and where we introduced the dimensionless variable $u=\frac{q^2}{-\bar{\lambda}}$. These expressions exhibit the well-known logarithmic divergences with the correct universal coefficients \cite{tHooft:1974toh}, here $\mu^2$ denotes the renormalization scale introduced to absorb such divergences. The presence of a nonvanishing cosmological constant generates additional nonlocal contributions.

The asymptotic behavior in $u$ identifies two regimes: $u \ll 1$, corresponding to the decoupling limit, and $u \gg 1$, describing the massless regime \cite{Codello:2015mba}. In the former case, one obtains
\begin{equation}\begin{split}
\label{EFTIR}
&f_{k=0}^{\left(\mathrm{R,EFT}\right)}\left(u\rightarrow0\right)=\frac{\ln{\left(\frac{q^2}{\mu^2}\right)}}{1920\pi^2}-\frac{7\ln{\left(\frac{u}{2}\right)}}{384\pi^2}+\frac{301}{5760\pi^2}+\frac{13u}{17920\pi^2}+\frac{37u^2}{1935360\pi^2}+O\left(u^3\right)\\
&f_{k=0}^{\left(\mathrm{Ric,EFT}\right)}\left(u\rightarrow0\right)=\frac{7\ln{\left(\frac{q^2}{\mu^2}\right)}}{320\pi^2}-\frac{7\ln{\left(\frac{u}{2}\right)}}{192\pi^2}-\frac{53}{1440\pi^2}+\frac{131u}{26880\pi^2}-\frac{55u^2}{193536\pi^2}+O\left(u^3\right)
\end{split}\end{equation}
which shows an analytic dependence on momenta supplemented by logarithmic contributions. 

In the massless limit we find 
\begin{equation}\begin{split}
\label{EFTUV}
&f_{k=0}^{\left(\mathrm{R,EFT}\right)}\left(u\rightarrow\infty\right)=\frac{\ln{\left(\frac{q^2}{\mu^2}\right)}}{1920\pi^2}+\frac{\frac{25\ln{\left(\frac{u}{2}\right)}}{96\pi^2}-\frac{43}{144\pi^2}}{u}+\frac{\frac{9\ln{\left(\frac{u}{2}\right)}}{16\pi^2}+\frac{29}{32\pi^2}}{u^2}+O\left(\frac{1}{u^3}\right)\\
&f_{k=0}^{\left(\mathrm{Ric,EFT}\right)}\left(u\rightarrow\infty\right)=\ \frac{7\ln{\left(\frac{q^2}{\mu^2}\right)}}{320\pi^2}+\frac{\frac{137}{144\pi^2}-\frac{13\ln{\left(\frac{u}{2}\right)}}{48\pi^2}}{u}+\frac{-\frac{13\ln{\left(\frac{u}{2}\right)}}{8\pi^2}-\frac{35}{16\pi^2}}{u^2}+O\left(\frac{1}{u^3}\right)
\end{split}\end{equation}
This corresponds to a nonlocal expansion where the effects of the cosmological constant appear as a power series in $1/u=-\bar\lambda/q^2$, reflecting its role as an IR scale in the propagator. The form factors therefore remain regular in the limit $\bar\lambda\to0$. 

Equations~(\ref{EFTIR}) and (\ref{EFTUV}) reproduce the results of \cite{Codello:2015mba}, now obtained from the asymptotic expansion of a nonperturbative flow evaluated at $k=0$. This provides a nontrivial consistency check of the truncation, showing that the approximation used in the trace of the flow equation correctly reproduces the one-loop contributions to the effective action, despite the absence of explicit form-factor terms in the Hessian.

\subsection{The scheme dependence}\label{subsectionscheme}
If $\eta \neq 0$, the flow acquires a genuinely Wilsonian character. From the structure of the proper-time flow equation, one might expect a fixed choice of $\epsilon$ should be used throughout the computation, in particular $\epsilon=1$ (scheme B). This requirement can, however, be relaxed. Within the approximation adopted here, as shown in Eqs.~(\ref{eqglambdam3}) and (\ref{eqFFm3}) and, more generally, in Appendix~\ref{appderbeta}, the flows of the Newton coupling and
cosmological constant decouple from those of the form factors because the feedback of the latter is neglected. Consequently, $g_k$ and $\lambda_k$ enter the form factor flows only as external inputs. This allows to adopt a mixed prescription, in the following called \textit{mixed scheme}, in which the running couplings are computed with $\epsilon=0$ (scheme C), while the form factors are evaluated with $\epsilon=1$ (scheme B). In this interpretation, using different values of $\epsilon$ in different sectors corresponds to fixing the scheme after projection rather than introducing distinct regulators. For comparison, we also consider the uniform choice $\epsilon=1$ in both sectors.

It is important to distinguish the $\epsilon$ prescription dependence from the general scheme dependence associated with the parameter $m$ in Eq.~(\ref{cutoff}). Other sources of scheme dependence, such as the gauge-fixing choice, are not investigated here. Within the present approximations, changing $m$ modifies the numerical coefficients entering the flow, while leaving its qualitative momentum dependence unchanged. The role of $\epsilon$ dependence differs between sectors. For running couplings, changing $\epsilon$ only reparametrizes the flow without affecting the space of solutions. For form factors, instead, $\epsilon$ is physically relevant: $\epsilon=0$ corresponds to an EFT flow, while $\epsilon=1$ captures the full momentum dependence, including trans-Planckian regimes.

As a result, using the flow of the Newton coupling computed in scheme C as input for the form factors in scheme B does not alter the physical content of the results. Differences are limited to scheme-dependent numerical values, while the momentum dependence, in particular the universal logarithmic terms, remains unaffected. The validity of these statements can be established analytically in the asymptotic regimes using the techniques of Appendix~\ref{appserie}, and are further supported by the numerical analysis of Sec.~\ref{secnuman}.

\subsection{The running Newtonian constant in the C scheme}
The properties of the running Newton coupling are determined by its fixed-point structure. Since the dimensionful cosmological constant $\bar\lambda$ is kept fixed in the present analysis, it enters as an external parameter in the flow of the dimensionful Newton coupling $G_k$. We therefore regard the dimensionless Newton coupling as a function of the dimensionless cosmological constant,
\begin{equation}
g_k(\lambda_k)=k^{d-2}G_k\left(\bar{\lambda}=\lambda_k k^2\right),
\end{equation}
where $\lambda_k$ is treated as an independent variable. The flow
equation for $g_k(\lambda_k)$ can then be written as
\begin{equation}
\label{eqgnewlambdaconst}
k\partial_k g_k(\lambda_k)=(d-2+\eta_k(\lambda_k))g_k(\lambda_k)+2\lambda_k\frac{\partial g_k(\lambda_k)}{\partial\lambda_k}.
\end{equation}
where $\eta_k$ is given in Eq.(\ref{eqeta1}). This equation should be compared with the standard $k\partial_kg_k=(d-2+\eta_k)g_k$. The additional term in Eq.(\ref{eqgnewlambdaconst}) arises because $\lambda_k$ is not treated as an independent running coupling with its own flow equation, rather it is fixed by $\lambda_k=\bar\lambda/k^2$.

In the C scheme, where $\epsilon=0$, Eq.(\ref{eqgnewlambdaconst}) can be solved
analytically. The fixed-point equation is obtained by imposing
$k\partial_k g_k(\lambda_k)=0$, yielding
\begin{equation}
0=(d-2+\eta_*)g_*(\lambda_k)+2\lambda_k\frac{\partial g_*(\lambda_k)}{\partial\lambda_k}.
\end{equation}
The Gaussian fixed point, $g_*=0$, is present independently of the
value of $\lambda_k$. In addition, the equation admits a nontrivial
fixed-point solution which, in $d=4$, is given by
\begin{equation}\begin{split}
\label{solFPgnew}
&g_*(\lambda_k)=\\
&=\frac{6 \pi (m-1)}{13 m^{m-1}(m-2\lambda_k)^{1-m}\left[m-2\lambda_k\,{}_2F_1\left(1,1-m;2-m;1-\frac{2\lambda_k}{m}\right)\right]+2\left[5m+3\pi(m-1)\lambda_k c\right]},
\end{split}\end{equation}
where $c$ is an integration constant fixed by the boundary conditions.

The linear perturbations around this fixed-point solution determine the
critical exponents. We write
\begin{equation}
\label{critgngen}
g_k(\lambda_k)=g_*(\lambda_k)+k^{-\theta}\delta g(\lambda_k),
\end{equation}
where $\theta$ denotes the critical exponent. Substituting this
ansatz into the flow equation and linearizing in the perturbation
yields a first-order differential equation for $\delta g(\lambda_k)$.
Requiring the perturbation to remain regular at $\lambda_k=0$ quantizes the allowed critical exponents as
\begin{equation}
\theta=2j,
\qquad
j=1,2,\ldots.
\end{equation}
In particular, the leading perturbation corresponds to
$\theta=2$, showing that the nontrivial fixed point is ultraviolet attractive along the leading perturbative direction.

The full trajectory, obtained by solving Eq.~(\ref{eqgnewlambdaconst}), is determined by imposing the boundary condition $G_{k=0}(\bar\lambda)=1/m_p^2$, where $m_p$ is the Planck mass. In $d=4$, this fixes the solution for $g_k$ to
\begin{equation}\begin{split}
\label{solgadim}
&g\left(y,{\lambda }_{k};m\right)=\\
&=\frac{6 \pi  (m-1) y}{2 (5 m y+3 \pi  (m-1))-13 m^{m-1} y \left(m-2 \lambda _k\right){}^{1-m} \left((m-1) \,_2F_1\left(1,1;m+1;\frac{m}{2 \lambda _k}\right)-m\right)}
\end{split}\end{equation}
where we introduced the ratio
\begin{equation}
\label{ratioyt}
y=\frac{k^2}{m_p^2}
\end{equation}
Expanding Eq.(\ref{solgadim}) in the regime $y\gg1$ we recover the structure of Eq.(\ref{solFPgnew}) and Eq.(\ref{critgngen}).

It is convenient to introduce also the ratio $t=\frac{-\bar{\lambda}}{m_p^2}$. We focus on the regime $t\ll1$, where the running of the dimensional Newtonian coupling is dominated by the pure gravitational sector and the cosmological constant can be neglected to leading order. In this limit, since $\lambda_k=-t/y$, the dimensionless flow can be consistently evaluated at $\lambda_k=0$. In $d=4$, for general $m$, the solution reduces to
\begin{equation}
g(k;m)=\frac{6k^2\left(m-1\right)\pi}{23k^2m+6\left(m-1\right)m_p^2\pi}+O\left(\lambda_k\right)
\end{equation}
In this form, consistency of the approximation with $t\ll1$ requires $\lambda_k\ll1$, namely $|\bar\lambda| \ll k^2$, so that the cosmological constant remains subleading also with respect to the coarse-graining scale. 

In this approximation, the nontrivial fixed point regime corresponds to the limit $m_p^2\ll k^2$ at fixed $\lambda_k\ll1$,
\begin{equation}
\label{pertgk}
g(y\gg1,t\ll1;m)=\frac{6\left(m-1\right)\pi}{23m}+\frac{1}{y}\left[-\frac{36\left(m-1\right)^2\pi^2}{529m^2}+O\left(\lambda_k\right)\right]+O\left(\frac{1}{y^2}\right)
\end{equation}
The nontrivial fixed-point value remains positive and shows no dependence on $t$, this coincides with Eq.(\ref{solFPgnew}) evaluated at $\lambda_k=0$. The leading perturbation scales as $1/y$, corresponding to a critical exponent $\theta=2$ in Eq.(\ref{critgngen}).

The corresponding anomalous dimension is given by
\begin{equation}
\eta_k=-\frac{46k^2m}{23k^2m+6\left(m-1\right)m_p^2\pi}+O\left(\frac{-\bar{\lambda}}{m_p^2}\right)
\end{equation}
and reduces at the nontrivial fixed point to $\eta_\ast=-2$, independently of the cosmological constant.

These analytical results capture the main features of $g_k$ within the present approximation, and the choice of $\epsilon$ does not affect the conclusions. Accordingly, the form-factor running obtained in the mixed scheme is expected to share the same physical features as in the full $\epsilon=1$ case.

Before turning to the form factor flow, it is important to clarify the role of the condition $\lambda_k\ll1$ used in the analysis of $g_k$. This condition, or equivalently $|\bar\lambda|\ll k^2$, follows directly from the functional form of the dimensionful running Newtonian coupling in the limit $t\ll1$, upon expressing it in terms of the dimensionless coupling $g_k$. Therefore, $\lambda_k\ll1$ is specific to the dimensionless Newtonian coupling $g_k$ and should not be regarded as a general condition to be applied to the other sectors of the theory. In particular, we do not impose $\lambda_k\ll1$ on the form factor flow, as discussed below.

\subsection{Infrared regime of the fully dynamical form factor}
We now turn to the flow of the form factors. Although $\eta_k$ has a relatively simple structure, the corresponding flow equations can only be treated analytically in asymptotic regimes; nevertheless, this is sufficient to extract their main qualitative features. In contrast to the running of Newton's coupling, we do not set $\bar\lambda=0$ in this sector. The reason is that, while $\bar\lambda$ gives only subleading contributions to the flow of $G_k$ in the regime $|\bar\lambda|\ll m_p^2$, the form-factor flow depends explicitly on the ratios $\bar\lambda/k^2$, $\bar\lambda/q^2$, and $\bar\lambda/m_p^2$. The condition $|\bar\lambda|\ll m_p^2$ alone in the form factor flow does not imply that the ratios $\bar\lambda/k^2$ or $\bar\lambda/q^2$ are small. We therefore retain the assumption $|\bar\lambda|\ll m_p^2$, consistently with the approximation adopted for $G_k$, without imposing any restriction on $\bar\lambda/k^2$ or $\bar\lambda/q^2$, since the flow probes the full momentum range. Setting $\bar\lambda=0$ in the form factor flow would therefore discard contributions that can become relevant when $k^2$ or $q^2$ are comparable to or smaller than $|\bar\lambda|$.

In the presence of a running Newton constant, the relevant IR and UV regimes are controlled by the hierarchy between dynamical invariants and the Planck scale $m_p^2$. To parametrize these regimes, besides Eq.(\ref{ratioyt}), it is convenient to introduce the dimensionless variables
\begin{equation}
w=\frac{q^2}{m_p^2}, \qquad t=\frac{-\bar{\lambda}}{m_p^2}, \qquad u=\frac{q^2}{-\bar{\lambda}}.
\end{equation}
Since $t=w/u$, only two of them are independent. The IR regime corresponds to $w \ll 1$ and $t \ll 1$ at fixed $u$, while the UV regime is obtained for $(w,u)\to\infty$ at fixed $t$. The UV region also contains the fixed-point regime, which will be discussed in Sec.~\ref{secASformfactor}.

Since the flow equations cannot be solved exactly, the power series in $w\ll1$ or $w\gg1$ have to be computed before performing the integration over the RG scale $k$. This introduces a subtlety, as it cuts the regions $k^2\ll m_p^2$ and $k^2\ll q^2$ of the flow in the IR and UV limits respectively. As a result, the naïve expansion does not reproduce the full integrated flow, and a reconstruction procedure is required. This is discussed in detail in Appendix~\ref{appserie}.

The general solution in the IR regime at $k=0$ takes the form
\begin{equation}
\label{IRserie}
f_{k=0}\left(u,t=\mathrm{const},w\rightarrow0\right)=f_{k=0}^{\left(\mathrm{EFT}\right)}\left(q^2,u;\mu^2\right)+b\ln{\left(\frac{4\pi}{23}\frac{m_p^2}{\mu^2}\right)}+\sum_{n=1}^{+\infty}{h_n\left(u,t\right)w^n}
\end{equation}
where $h_n(u,t)$ denote the series coefficients. The corrections induced by the Plank mass do not modify the EFT result but add analytic terms in the ratio $q^2/m_p^2$. The logarithmic dependence on $q^2$ in the EFT term still provides the leading nonlocal contribution to the form factors. The running of the Newton constant does not introduce additional divergences, or modify the coefficient of $\ln q^2$; it instead generates an additional logarithmic dependence on the Planck scale $m_p^2$, whose origin is discussed in Appendix \ref{appserie}. The coefficients $b$ are given by
\begin{equation}
\label{bcoeff}
b^{\left(\mathrm{R}\right)}=\frac{7}{384\pi^2},\quad \quad b^{\left(\mathrm{Ricc}\right)}=\frac{7}{192\pi^2}.
\end{equation}
The resulting logarithmic terms can then be written equivalently as
\begin{equation}\begin{split}
&a_{\mathrm{EFT}}\ln{\left(\frac{q^2}{\mu^2}\right)}+b\ln{\left(\frac{4\pi}{23}\frac{m_p^2}{\mu^2}\right)}=a\ln{\left(\frac{q^2}{\mu^2}\right)}-b\ln{\left(\frac{23}{12\pi}w\right)}\\
&a_{\mathrm{EFT}}\ln{\left(\frac{q^2}{\mu^2}\right)}+b\ln{\left(\frac{4\pi}{23}\frac{m_p^2}{\mu^2}\right)}=a_{\mathrm{EFT}}\ln{\left(\frac{23}{12\pi}w\right)}+a\ln{\left(\frac{4\pi}{23}\frac{m_p^2}{\mu^2}\right)}
\end{split}\end{equation}
where $a_{\mathrm{EFT}}$ can be read from Eq.(\ref{EFTUV}) and
\begin{equation}
\label{coeffa}
a^{\left(\mathrm{R}\right)}=\frac{3}{160\pi^2},\quad \quad a^{\left(\mathrm{Ricc}\right)}=\frac{7}{120\pi^2}
\end{equation}
The two expressions differ in how the renormalization scale $\mu^2$ is distributed between the logarithmic terms. The first form isolates the dependence on $q^2/\mu^2$, while the second organizes the logarithms in terms of the dimensionless ratios $w=q^2/m_p^2$ and $m_p^2/\mu^2$.

The general structure of the solution is a local expansion in $w$, while the series coefficients $h_n^{(i)}(u,t)$ are nonlocal functions of $u$ and $t$. The explicit expressions for the coefficients are lengthy and the first terms are given in Appendix~\ref{app:techres}. To highlight their physical content, we focus here on the decoupling limit ($u\to0$) and the massless limit ($u\to\infty$), which capture the physically relevant behavior.

In the decoupling regime $u\rightarrow0$ we obtain
\begin{equation}\begin{split}
&f_{k=0}^{\left(\mathrm{R}\right)}\left(u\rightarrow0,t=\mathrm{const},w\rightarrow0\right)=\frac{3\ln{\left(\frac{q^2}{\mu^2}\right)}}{160\pi^2}+\frac{301}{5760\pi^2}-\frac{7\ln{\left(\frac{23uw}{24\pi}\right)}}{384\pi^2}+\frac{13u}{17920\pi^2}+O\left(u^2\right)+\\
&+w\left(\frac{-\frac{805}{4608\pi^3}-\frac{161\ln{\left(\frac{23t}{6\pi}\right)}}{768\pi^3}}{u}-\frac{3289}{215040\pi^3}-\frac{299\ln{\left(\frac{23t}{6\pi}\right)}}{35840\pi^3}-\frac{851u}{3870720\pi^3}+O\left(u^2\right)\right)+O(w^2)\\
&f_{k=0}^{\left(\mathrm{Ricc}\right)}\left(u\rightarrow0,t=\mathrm{const},w\rightarrow0\right)=\frac{7\ln{\left(\frac{q^2}{\mu^2}\right)}}{120\pi^2}+\frac{53}{1440\pi^2}-\frac{7\ln{\left(\frac{23uw}{24\pi}\right)}}{192\pi^2}+\frac{131u}{26880\pi^2}+O\left(u^2\right)+\\
&+w\left(\frac{-\frac{805}{2304\pi^3}-\frac{161\ln\left(\frac{23t}{6\pi}\right)}{384\pi^3}}{u}-\frac{33143}{322560\pi^3}-\frac{3013\ln{\left(\frac{23t}{6\pi}\right)}}{53760\pi^3}-\frac{1265u}{387072\pi^3}+O\left(u^2\right)\right)+O(w^2)
\end{split}\end{equation}
The corrections to the EFT contribution are analytic in the momentum. The only nonanalytic momentum dependence is contained in the leading EFT contribution through $\ln(q^2/\mu^2)$, so that the full IR series diverges in the strict $q^2\to0$ limit. The logarithmic terms appearing in the IR corrections depend only on the combination $t=-\bar{\lambda}/m_p^2$. Since $w/u=t$, the leading correction proportional to $w/u$ reduces to a momentum-independent constant. Thus, in this limit, beyond the EFT logarithm, the IR corrections do not introduce additional nonlocal momentum dependence.

In the massless limit $u\rightarrow+\infty$ of the IR regime we get 
\begin{equation}\begin{split}
&f_{k=0}^{\left(\mathrm{R}\right)}\left(u\rightarrow\infty,t=\mathrm{const},w\rightarrow0\right)=\\
&=\frac{3\ln{\left(\frac{q^2}{\mu^2}\right)}}{160\pi^2}-\frac{7\ln{\left(\frac{23w}{24\pi}\right)}}{384\pi^2}+\frac{\frac{25\ln{\left(\frac{u}{2}\right)}}{96\pi^2}-\frac{43}{144\pi^2}}{u}+O\left(\frac{1}{u^2}\right)+\\
&+w\left(\frac{31257}{2508800\pi^3}-\frac{299\ln{\left(\frac{23w}{12\pi}\right)}}{35840\pi^3}+\frac{\frac{4991}{23040\pi^3}-\frac{161\ln{\left(\frac{23w}{12\pi}\right)}}{768\pi^3}}{u}+O\left(\frac{1}{u^2}\right)\right)+O\left(w^2\right)\\
&f_{k=0}^{\left(\mathrm{Ricc}\right)}\left(u\rightarrow\infty,t=\mathrm{const},w\rightarrow0\right)=\\
&=\frac{7\ln{\left(\frac{q^2}{\mu^2}\right)}}{120\pi^2}-\frac{7\ln{\left(\frac{23w}{24\pi}\right)}}{192\pi^2}+\frac{-\frac{13\ln{\left(\frac{u}{2}\right)}}{48\pi^2}+\frac{137}{144\pi^2}}{u}+O\left(\frac{1}{u^2}\right)+\\
&+w\left(\frac{381547}{11289600\pi^3}-\frac{3013\ln{\left(\frac{23w}{12\pi}\right)}}{53760\pi^3}-\frac{\frac{161\ln{\left(\frac{23w}{12\pi}\right)}}{384\pi^3}+\frac{3979}{11520\pi^3}}{u}+O\left(\frac{1}{u^2}\right)\right)+O\left(w^2\right)
\end{split}\end{equation}
which is a nonlocal theory of gravitation. In this limit, the EFT contribution is supplemented both from local and nonlocal contributions. In particular, the IR corrections contain terms of the form $w^n\ln(w)$. In contrast to the $u\to0$ regime, in this limit the nonanalytic dependence of the form factors is therefore not restricted to the leading EFT contribution. This difference originates from the distinct role of the cosmological constant in the two limits. For $u\to0$, $\bar\lambda$ dominates over $q^2$ so that the logarithmic corrections involving the Plank mass can depend only on $t=-\bar\lambda/m_p^2$. In contrast, the $u\to\infty$ limit corresponds to $\bar\lambda\to0$ so that logarithmic terms can involve only $w$. As a consequence, as in the EFT case, in this limit there is no isolated $\ln(\bar\lambda)$ term and the contribution of $\bar\lambda$ to the form factors appears only as a power series in $\bar\lambda/q^2$.

\subsection{Ultraviolet regime of the fully dynamical form factor}
In the UV regime ($w\gg1$), the general solution at $k=0$ takes the form:
\begin{equation}
\label{UVseries}
f_{k=0}\left(u,t,w\rightarrow\infty\right)=f_{k=0}^{\left(\mathrm{EFT}\right)}\left(q^2,u;\mu^2\right)+b\ln{\left(\frac{q^2}{\mu^2}\right)}+\sum_{n=1}^{+\infty}\frac{H_n\left(t\right)+L_n(t)\ln(w)}{w^n}
\end{equation}
This corresponds to a nonlocal expansion in $w$, with coefficients depending only on $t$. The coefficients $b$ are given in Eq. (\ref{bcoeff}). As in the IR regime, the running of the Newton coupling introduces no additional divergences. In particular, the logarithmic contributions combine as $a_{\mathrm{EFT}}\ln{\left(\frac{q^2}{\mu^2}\right)}+b\ln{\left(\frac{q^2}{\mu^2}\right)}=a\ln{\left(\frac{q^2}{\mu^2}\right)}$, where $a$ is given in Eq.~(\ref{coeffa}), showing that the coefficient of the divergent logarithmic term remains universal across scales.

In this regime, only the region $t\ll1$ is physically relevant, where the solution reduces to
\begin{equation}\begin{split}
\label{serieUVloggen}
&f_{k=0}\left(u\to\infty,t\to0,w\rightarrow\infty\right)=\\
&=a\ln{\left(\frac{q^2}{\mu^2}\right)}+\sum_{n=1}^{+\infty}\frac{a_n+b_n\ln\left(\frac{u}{2}\right)}{u^n}+\sum_{n=1}^{+\infty}\frac{\sum_{m=0}^{+\infty}\left(c_{nm}+d_{nm}\ln\left(\frac{23w}{12\pi}\right)\right)t^m}{w^n}
\end{split}\end{equation}
where the series coefficients are local functions of $t$. This expression makes the structure of the solution transparent. The leading contribution is given by the universal logarithm $a\ln(q^2/\mu^2)$, while the remaining terms are suppressed by inverse powers of $w$ and $u$, with additional logarithmic momentum dependence appearing in the corresponding subleading contributions. The first terms are given by
\begin{equation}\begin{split}
&f_{k=0}^{\left(\mathrm{R}\right)}\left(u\rightarrow\infty,t\rightarrow0,w\rightarrow\infty\right)=\frac{3\ln{\left(\frac{q^2}{\mu^2}\right)}}{160\pi^2}+\frac{-\frac{43}{144\pi^2}+\frac{25\ln{\left(\frac{u}{2}\right)}}{96\pi^2}}{u}+\frac{\frac{29}{32\pi^2}+\frac{9\ln{\left(\frac{u}{2}\right)}}{16\pi^2}}{u^2}+O\left(\frac{1}{u^3}\right)+\\
&+\frac{-\frac{43}{1104\pi}+\frac{25}{736\pi}\ln{\left(\frac{23w}{12\pi}\right)}+t\left(-\frac{97}{576\pi^2}+\frac{25}{96\pi^2}\ln{\left(\frac{23w}{12\pi}\right)}\right)+O\left(t^2\right)}{w}+O\left(\frac{1}{w^2}\right)
\end{split}\end{equation}
and
\begin{equation}\begin{split}
&f_{k=0}^{\left(\mathrm{Ricc}\right)}\left(u\rightarrow\infty,t\rightarrow0,w\rightarrow\infty\right)=\frac{7\ln{\left(\frac{q^2}{\mu^2}\right)}}{120\pi^2}+\frac{\frac{137}{144\pi^2}-\frac{13\ln{\left(\frac{u}{2}\right)}}{48\pi^2}}{u}+\frac{-\frac{35}{16\pi^2}-\frac{13\ln{\left(\frac{u}{2}\right)}}{8\pi^2}}{u^2}+O\left(\frac{1}{u^3}\right)+\\
&+\frac{\frac{137}{1104\pi}-\frac{13}{368\pi}\ln{\left(\frac{23w}{12\pi}\right)}+t\left(\frac{235}{288\pi^2}-\frac{13}{48\pi^2}\ln{\left(\frac{23w}{12\pi}\right)}\right)+O\left(t^2\right)}{w}+O\left(\frac{1}{w^2}\right)
\end{split}\end{equation}
The strict $\bar\lambda\to0$ limit is regular. In particular, the
cosmological constant does not generate an isolated logarithmic contribution in this regime, and its dependence is instead encoded in the expansion in powers of $\bar\lambda/q^2$.

\section{The asymptotically safe form factors}\label{secASformfactor}
\noindent The previous computation requires renormalization to remove the divergence arising in the limit $\Lambda\to\infty$, introducing a dependence on the renormalization scale $\mu$. In the presence of a running Newton coupling, however, an alternative scenario emerges: if the form factor approaches an asymptotically safe fixed point, the theory becomes UV complete and the cutoff dependence can be removed dynamically. In this Section, we construct the asymptotically safe form factors.

\subsection{The dimensionless flow equation}
The first step toward an asymptotically safe description is to introduce dimensionless variables. We define the dimensionless form factor $F_k$ and the dimensionless momentum variable $x=q^2/k^2$. In $d=4$, the form factors are dimensionless, and we write
\begin{equation}
F_k\left(x,\lambda_k\right)=f_k\left(q^2=k^2x,\bar{\lambda}=\lambda_kk^2\right)
\end{equation}
the flow equation then reads 
\begin{equation}
\label{floweqadim}
k\partial_kF_k\left(x,\lambda_k\right)=2x\partial_xF_k\left(x,\lambda_k\right)+2\lambda_k\partial_{\lambda_k}F_k\left(x,\lambda_k\right)+H\left(\eta_k,x,\lambda_k\right)
\end{equation}
This is a linear partial differential equation. Here $\lambda_k=\bar\lambda/k^2$ is not an independent running coupling, but follows from the canonical scaling of $\bar\lambda$, yielding the term $2\lambda_k\partial_{\lambda_k}F_k$. The last term is given by
\begin{equation}
\label{eqH}
H\left(\eta_k,x,\lambda_k\right)=k\partial_kf_k\left(q^2=xk^2,\bar\lambda=\lambda_kk^2\right)
\end{equation}
This term encodes the beta function of the form factors in dimensionless variables, cf. Eq.~(\ref{eqFFm3}).

\subsection{The fixed point solution}
To determine the asymptotically safe form factor, one first needs the corresponding fixed-point solution. This amounts to imposing  $k\partial_k F_k\left(x,\lambda_k\right)=0$ where the Newtonian coupling, and consequently the anomalous dimension, is evaluated at the fixed point,
\begin{equation}
\label{FPequation}
2x\partial_xF_\ast\left(x,\lambda_k\right)+2\lambda_k\partial_{\lambda_k}F_\ast\left(x,\lambda_k\right)+H\left(\eta_\ast,x,\lambda_k\right)=0.
\end{equation}
This first-order PDE can be solved by the method of characteristics. Its structure is independent of $\epsilon$, as it depends only on the universal value $\eta_\ast$.

Imposing the boundary condition $F_\ast(x_0,\lambda_k)=h(x_0,\lambda_k)$, with $x_0$ an arbitrary reference scale and $h$ and arbitrary function, the general solution of Eq.(\ref{FPequation}) reads
\begin{equation}\begin{split}
\label{FPsol}
&F_\ast\left(x,\lambda_k;x_0\right)=h\left(x_0,x_0\frac{\lambda_k}{x}\right)-\int_{x_0}^{x}{\frac{H\left(\eta_\ast,v,\frac{\lambda_k}{x}v\right)}{2v}dv}
\end{split}\end{equation}
the explicit result for the particular solution is given by
\begin{equation}\begin{split}
\label{solFPpart}
&F_*^{(\mathrm{part},i)}(x,\lambda_k;m)\equiv -\int^{x}{\frac{H\left(\eta_\ast,v,\frac{\lambda_k}{x}v\right)}{2v}dv}=\alpha_0^{(i)}(\eta_*)\ln(x)+\beta_0^{(i)}(\eta_*)\ln{\left(m-2\lambda_k\right)}+\\
&\resizebox{1.0\hsize}{!}{$+A^{(i)}\left(x,\lambda_k;m\right)+B^{(i)}\left(x,\lambda_k;m\right)\mathrm{tanh}^{-1}{\left(\sqrt{\frac{x}{x+4m}}\right)}+C^{(i)}\left(x,\lambda_k;m\right)\mathrm{tanh}^{-1}{\left(\sqrt{\frac{x}{x-8\lambda_k+4m}}\right)}$}
\end{split}\end{equation}
where $i=\mathrm{R}$, $\mathrm{Ricc}$ and
\begin{equation}\begin{split}
&\alpha_0^{(\mathrm{R})}(\eta_*)=\frac{7\eta_\ast}{768\pi^2}-\frac{1}{1920\pi^2},\quad\quad \alpha_0^{(\mathrm{Ricc})}(\eta_*)=\frac{7\eta_\ast}{384\pi^2}-\frac{7}{320\pi^2}\\
&\beta_0^{(\mathrm{R})}(\eta_*)=\frac{7}{384\pi^2}-\frac{7\eta_\ast}{768\pi^2},\quad\quad \beta_0^{(\mathrm{Ricc})}(\eta_*)=\frac{7}{192\pi^2}-\frac{7\eta_\ast}{384\pi^2}
\end{split}\end{equation}
The structure of the coefficients $A$, $B$, $C$ is the same for all values of $m$, with only the numerical coefficients changing. The explicit expressions for $m=3$ can be read in Appendix \ref{apptech:fixedpoint}. The most relevant feature of Eq.(\ref{solFPpart}) is the logarithmic behavior, fully determined by the RG structure of the equation. In particular, using $\eta_*=0$ or $\eta_*=-2$ shows that the coefficient of the logarithmic term is identical to that obtained in the previous analytical results, confirming its universality. 

The boundary function $h$ can be expressed in terms of solutions of the homogeneous fixed-point equation. Using the ansatz $F_*(x,\lambda_k)=p(x)\,v(\lambda_k)$, one obtains power-law solutions
\begin{equation}
F_*^{(\mathrm{hom})}(x,\lambda_k)=c_E\left(\frac{x}{\lambda_k}\right)^{E/2}.
\end{equation}
The separation constant $E$ is fixed by the boundary conditions, which in general select a discrete spectrum $\{E_i\}$. The general homogeneous solution is then a superposition of these modes, with coefficients $c_E$ determined by matching to the boundary function $h$.  

The particular fixed point solutions in Eq.~(\ref{solFPpart}) can be studied in the limiting regimes $x\ll1$ and $x\gg1$. In the IR regime $x\ll1$, the solutions take the form
\begin{equation}\begin{split}
\label{eqFPx0}
&F_*^{(\mathrm{part,i})}(x\to0,\lambda_k)=\alpha_0^{(i)}(\eta_\ast )\ln(x)+\beta_0^{(i)}(\eta_\ast)\ln\left(m-2\lambda_k\right)+\frac{1}{x^{2}}\sum_{n=0}^{+\infty} a_{n}^{(i)}(\lambda_k)x^{n}
\end{split}\end{equation}
Apart from the logarithmic contribution, the solution contains only integer powers of $x$, with a leading $x^{-2}$ behavior. The explicit expression for the first coefficients can be found in Appendix \ref{apptech:fixedpoint}. The logarithmic term proportional to $\ln(x)$ represents the only nonanalytic momentum dependence in this regime. The negative-power terms are proportional to $\lambda_k$ and therefore vanish for $\lambda_k\to0$. In particular, they appear through combinations such as $\lambda_k/x$ (see Appendix~\ref{apptech:fixedpoint}), which correspond, upon restoring dimensionful variables, to powers of $\bar{\lambda}/q^2$. Although such terms would induce nonlocal contributions to the IR form factor, they are absent from the physical fixed-point solution in Eq.~(\ref{FPsol}), since they cancel in the definite integral defining the solution. Consequently, the physical fixed-point form factor has the IR behavior
\begin{equation}
\label{beFPsmallx}
F_\ast(x\to0,\lambda_k;x_0)\sim h\left(x_0,x_0\frac{\lambda_k}{x}\right)+\alpha_0^{(i)}(\eta_\ast)\ln(x)+c_0(x_0,\lambda_k)+c_1(x_0,\lambda_k)x+O(x^2)
\end{equation}
and in dimensional term no term proportional to $\bar{\lambda}/q^2$ survives.

For $x\gg1$ the solutions are
\begin{equation}\begin{split}
\label{eqFPxinf}
&F_*^{(\mathrm{part},i)}(x\to\infty,\lambda_k)=b_{0}^{\left(i\right)}+\sum_{n=1}^{+\infty } \frac{b_{n}^{\left(i\right)}(\lambda_k)+c_{n}^{(i)}(\lambda_k)\ln{\left(\frac{x}{m-2{\lambda }_{k}}\right)}}{x^{n}}
\end{split}\end{equation}
In this regime, the solutions approach a constant, with corrections suppressed by inverse powers of $x$ and modulated by logarithmic terms. The leading coefficient $b_0^{(i)}$ is independent of the cosmological constant, while $\lambda_k$ enters only through the subleading terms. The explicit expressions for the first coefficients are given in Appendix~\ref{apptech:fixedpoint}. The same asymptotic structure is preserved by the physical solution in Eq.~(\ref{FPsol}). In both regimes, the nonanalytic dependence is entirely logarithmic, although its organization is different. In contrast to the $x\ll1$ regime, where the logarithmic term appears at leading order, for $x\gg1$ the logarithmic dependence is suppressed by inverse powers of $x$. Furthermore, here the cosmological constant affects the coefficients of the expansion and the arguments of the logarithmic terms, but does not affects the power-law structure.

These asymptotic forms make explicit the logarithmic structure of the nonanalytic contributions and the corresponding power-law hierarchy in the two asymptotic regimes.

\begin{figure}[t]
     \centering
     \subfigure[]{
         \centering
         \includegraphics[width=0.45\textwidth]{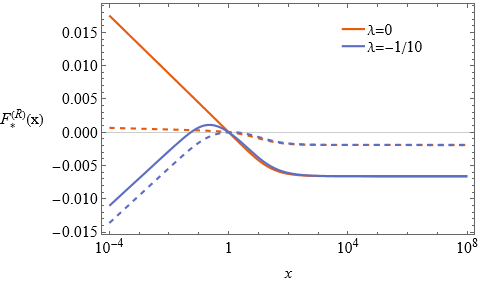}
     }
    \hspace{0.8em}
     \subfigure[]{
         \centering
         \includegraphics[width=0.45\textwidth]{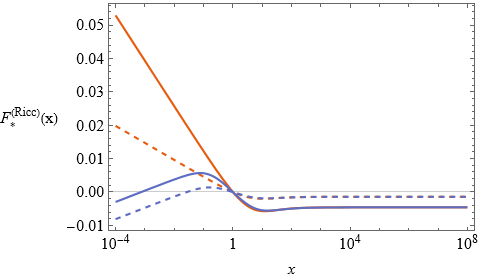}
     }
     \caption{Fixed-point form factors for $\bar\lambda \ne 0$ (blue) and $\bar\lambda = 0$ (beige), obtained with $m=3$ and boundary condition $F_{*}(x_0)=0$ at $x_0=1$. Panels (a) and (b) show $F_{*}^{(R)}(x)$ and $F_{*}^{(\mathrm{Ricc})}(x)$, respectively. Solid and dashed lines correspond to the Reuter and Gaussian fixed points of the Newton coupling.}
     \label{plotFFFP}
\end{figure}

Figure~\ref{plotFFFP} shows the solution Eq.(\ref{FPsol}) of the fixed-point equation for $\lambda_k \neq 0$ and $\lambda_k = 0$, with $x_0 = 1$ and $h=0$. The continuous and dashed lines correspond to $\eta_\ast = -2$ and $\eta_\ast = 0$, respectively. In both cases, the fixed-point solutions vanish at $x = x_0$ becoming negative for $x > x_0$ and approaching a constant asymptotic value at large $x$. The inclusion of a cosmological constant mainly affects the small-$x$ regime, where a maximum develops for $x < x_0$. The boundary function $h$ allows one to shift the solutions, and can be used to render the large-$x$ behavior positive if desired.

For $\eta_\ast=-2$ and $\lambda_k=0$, the results in Fig.~\ref{plotFFFP} can be compared with the background form factors obtained in \cite{Knorr:2021niv} using the Weyl basis and the Wetterich equation. In order to make the comparison explicit, the quadratic part of the effective action used in \cite{Knorr:2021niv} is given by
\begin{equation}
{\Gamma }_{k}^{(\mathrm{quad})}=\frac{1}{16\pi G_{k}}\int d^{d}x\sqrt{g}\left[-\frac{1}{4}\frac{d-2}{d-1}Rf_{k}^{\left(\mathrm{RBis}\right)}\left(-\Box\right)R+\frac{1}{4}\frac{d-2}{d-3}C^{\mu \nu \rho \sigma }f_{k}^{\left(\mathrm{W}\right)}\left(-\Box\right)C_{\mu \nu \rho \sigma }\right]
\end{equation}
The form factors appearing in this parametrization are related to those in the Ricci basis used in Eq.~(\ref{TE}) by
\begin{equation}\begin{split}
&f_{k}^{\left(\mathrm{RBis}\right)}\left(-\Box\right)=-16\pi G_{k}\frac{d-1}{d-2}\left[\frac{d}{4\left(d-1\right)}f_{k}^{\left(\mathrm{Ricc}\right)}\left(-\Box\right)+f_{k}^{\left(\mathrm{R}\right)}\left(-\Box\right)\right]\\
&f_{k}^{\left(\mathrm{W}\right)}\left(-\Box\right)=16\pi G_{k}f_{k}^{\left(\mathrm{Ricc}\right)}\left(-\Box\right)
\end{split}\end{equation}
The Weyl form factor is proportional to the Ricci form factor, whereas the Ricci-scalar form factor involves a linear combination of the Ricci and Weyl-basis form factors with an overall minus sign. Using these relations, we find that the corresponding $F_{*}^{\left(\mathrm{RBis}\right)}\left(x\right)$ and $F_{*}^{\left(\mathrm{W}\right)}\left(x\right)$ obtained from the results in Fig.~\ref{plotFFFP} show qualitative agreement with the background form factors of \cite{Knorr:2021niv}, both in their sign and in their overall behavior, for $h=0$ as well as for $h\neq0$: in the infrared, both exhibit a logarithmic behavior followed by an expansion in non-negative powers of $x$, $F_*(x)\sim\alpha\ln x+c_{IR}+O(x)$, while in the UV they show the trend $F_*(x)\sim c_{UV}+c_1/x+O(1/x^2)$. In particular, choosing the boundary condition such that $F_*(x\to\infty;x_0)=0$ removes the constant UV contribution and yields a purely power-law decay, making the resulting form factor more closely resemble the corresponding fluctuation form factor. 

The relation between the particular solution of the fixed-point equation and the corresponding dimensional flow, $f_k(q^2,\bar\lambda)=-\int_{k}^{\Lambda}\frac{\beta_{k_1}\left(\eta(k_1),q^2,\bar\lambda\right)}{k_1}\,dk_1$, can be made explicit by the change of variables $v=q^2/k_1^2$ (with $x_0=q^2/k_0^2$), which yields
\begin{equation}
\label{equivFP}
\int_{x_0}^{x}{\frac{H\left(\eta\left(v\right),v,\frac{\lambda_k}{x}v\right)}{2v}dv}=-\int_{k_0}^{k}{\frac{H\left(\eta\left(\frac{q^2}{k_1^2}\right),\frac{q^2}{k_1^2},\frac{\bar{\lambda}}{k_1^2}\right)}{k_1}dk_1}=\int_{k}^{k_0}{\frac{\beta_k\left(q^2,\bar{\lambda},\eta_{k_1}\right)}{k_1}dk_1}
\end{equation}
Setting $k_0=\Lambda$ and $\eta(k)=\eta_*$ shows that the fixed-point solution  corresponds to the dimensional flow evaluated along a trajectory where the anomalous dimension is taken at its fixed-point value, 
\begin{equation}
\label{eqsolpartFPdim}
F_\ast^{(\mathrm{part})}\left(x=\frac{q^2}{k^2},\lambda_k=\frac{\bar\lambda}{k^2};x_0=\frac{q^2}{\Lambda^2}\right)\equiv f_{*}\left(q^2,\bar{\lambda},\eta_\ast;k,\Lambda\right)=-\int_{k}^{\Lambda}{\frac{\beta_{k^\prime}\left(q^2,\bar{\lambda},\eta_\ast\right)}{k^\prime}dk^\prime}
\end{equation}
This establishes the equivalence between the fixed-point regime and the dimensional RG flow in the appropriate scaling limit, corresponding to the Gaussian ($m_p\to\infty$) or interacting (Reuter) fixed point ($m_p\to0$).

The limit $x\to\infty$ and $x_0\to0$ correspond to the infrared regime $k\to0$ with $\Lambda\to\infty$. In this limit, the two fixed-point solutions take the form
\begin{equation}\begin{split}
\label{FPk0}
&f_{\ast,k=0}^{\left(\mathrm{R}\right)}\left(u\right)\equiv\lim_{\substack{x\to+\infty\\ x_0\to0}}{F_\ast^{\left(\mathrm{R}\right)}\left(u,x,x_0\right)}=\left(\frac{1}{1920\pi^2}-\frac{7\eta_\ast}{768\pi^2}\right)\ln{\left(\frac{q^2}{\mu^2}\right)}+\frac{7}{768\pi^2}\left(\eta_\ast-2\right)\ln{\left(\frac{u}{2}\right)}+\\
&\resizebox{1.0\hsize}{!}{$+\frac{139}{28800\pi^2}+\frac{\left(\eta_\ast-2\right)\left(287u^2+4280u-960\right)}{23040\pi^2u^2}-\frac{\sqrt{u+8}\left(7u^2+72u-16\right)}{768\pi^2u^\frac{5}{2}}\left(\eta_\ast-2\right)\ln{\left(\frac{1+\sqrt{\frac{u}{8+u}}}{1-\sqrt{\frac{u}{8+u}}}\right)}$}\\
&f_{\ast,k=0}^{\left(\mathrm{\mathrm{Ricc}}\right)}\left(u\right)\equiv\lim_{\substack{x\to+\infty\\ x_0\to0}}{F_\ast^{\left(\mathrm{Ricc}\right)}\left(u,x,x_0\right)}=\left(\frac{7}{320\pi^2}-\frac{7\eta_\ast}{384\pi^2}\right)\ln{\left(\frac{q^2}{\mu^2}\right)}+\frac{7}{384\pi^2}\left(\eta_\ast-2\right)\ln{\left(\frac{u}{2}\right)}-\\
&\resizebox{1.0\hsize}{!}{$-\frac{31}{14400\pi^2}-\frac{\left(\eta_\ast-2\right)\left(103u^2+4640u-3840\right)}{11520\pi^2u^2}-\frac{\sqrt{u+8}\left(7u^2-80u+64\right)}{384\pi^2u^\frac{5}{2}}\left(\eta_\ast-2\right)\ln{\left(\frac{1+\sqrt{\frac{u}{8+u}}}{1-\sqrt{\frac{u}{8+u}}}\right)}$}
\end{split}\end{equation}
The fixed-point solutions at $k=0$ are nonlocal functions of $u$. Setting $\eta_\ast=0$ reproduces the EFT result in Eq.~(\ref{EFTresult}), so that the EFT flow can be interpreted as the fixed-point solution associated with the Gaussian fixed point of the Newton coupling. Conversely, $\eta_\ast = -2$ leads to a different EFT-like solution which yields the same coefficient of the logarithmic terms in both the IR and UV expansions in Eqs.~(\ref{IRserie}) and (\ref{UVseries}). This result will play an important role in the construction of the asymptotically safe solution. 

The structure of the decoupling ($u\to0$) limit and massless limit ($u\to\infty$) follow the same features of Eq.(\ref{EFTIR}) and Eq.(\ref{EFTUV}). For the illustration we present the massless limit. In the massless limit $u\to\infty$ the previous expressions yield
\begin{equation}\begin{split}
&f_{\ast,k=0}^{\left(R\right)}\left(u\to\infty\right)=\left(\frac{1}{1920\pi^2}-\frac{7\eta_\ast}{768\pi^2}\right)\ln{\left(\frac{q^2}{\mu^2}\right)}+\frac{-2314+1435\eta_\ast}{115200\pi^2}+\\
&+\left(\eta_\ast-2\right)\left[\frac{\frac{43}{288\pi^2}-\frac{25\ln{\left(\frac{u}{2}\right)}}{192\pi^2}}{u}+\frac{-\frac{29}{64\pi^2}-\frac{9\ln{\left(\frac{u}{2}\right)}}{32\pi^2}}{u^2}+O\left(\frac{1}{u^3}\right)\right]\\
&f_{\ast,k=0}^{\left(\mathrm{Ricc}\right)}\left(u\to\infty\right)=\left(\frac{7}{320\pi^2}-\frac{7\eta_\ast}{384\pi^2}\right)\ln{\left(\frac{q^2}{\mu^2}\right)}+\frac{906-515\eta_\ast}{57600\pi^2}+\\
&+\left(\eta_\ast-2\right)\left[\frac{-\frac{137}{288\pi^2}+\frac{13\ln{\left(\frac{u}{2}\right)}}{96\pi^2}}{u}+\frac{\frac{35}{32\pi^2}+\frac{13\ln{\left(\frac{u}{2}\right)}}{16\pi^2}}{u^2}+O\left(\frac{1}{u^3}\right)\right]
\end{split}\end{equation}
Apart from the leading logarithmic term, this is a power series in $\bar\lambda/q^2$ with logarithmic modulations. The limit $\bar\lambda\rightarrow0$ is regular, showing that a nonvanishing cosmological constant modifies the subleading momentum dependence without altering the leading nonlocal logarithmic term.

\subsection{The perturbations around the fixed-point solution}
Linearizing the flow equation around the fixed points allows one to study the critical properties of the theory. The perturbations of the Newton coupling around the nontrivial fixed point are obtained by expanding the solution for $g_k$ in the regime $m_p \ll k$, see Eq.~(\ref{pertgk}). This induces a corresponding expansion of the anomalous dimension and, consequently, of the dimensionless function $H$ in Eq.~(\ref{eqH}),
\begin{equation}\begin{split}
\label{expaH}
&H\left(\eta_k,x,\lambda_{k}\right)=H\left(\eta_\ast=-2,x,\lambda_{k}\right)+\sum_{j=1}^{+\infty}{\left.\partial_{m_p^2}^j H\left(\eta_k,x,\lambda_k\right)\right|_{m_p^2=0}\left(\frac{m_p}{k}\right)^{2j}}
\end{split}\end{equation}
Replacing this in the flow equation Eq.(\ref{floweqadim}) shows that the solution can be written as 
\begin{equation}\begin{split}
\label{solpert}
&F_k\left(x,\lambda_k\right)=F_\mathrm{hom}\left(x,\lambda_k,k\right)+F_\mathrm{part}\left(x,\lambda_k,k\right)\\
&F_\mathrm{part}\left(x,\lambda_k,k\right)=F_\ast\left(x,\lambda_k\right)+\sum_{j=1}^{+\infty}{F_j\left(x,\lambda_k\right)\left(\frac{m_p}{k}\right)^{2j}}
\end{split}\end{equation}
where $F_\mathrm{hom}\left(x,\lambda_k,k\right)$ is the general solution of the homogeneous equation 
\begin{equation}
\label{homeq}
kF_{\hom}^{\left(0,0,1\right)}\left(x,\lambda_k,k\right)=2xF_{\hom}^{\left(1,0,0\right)}\left(x,\lambda_k,k\right)+2\lambda_kF_{\hom}^{\left(0,1,0\right)}\left(x,\lambda_k,k\right)
\end{equation}
and $F_j\left(x,\lambda_k\right)$ are determined by
\begin{equation}
-2jF_j\left(x,\lambda_k\right)=2x\partial_xF_j\left(x,\lambda_k\right)+2\lambda_k\partial_{\lambda_k}F_j\left(x,\lambda_k\right)+\left.\partial_{m_p^2}^j H\left(\eta_k,x,\lambda_k\right)\right|_{m_p^2=0}
\end{equation}
Both equations are linear partial differential equations. The particular solution in Eq.~(\ref{solpert}) describes a deformation of the fixed-point solution and does not affect the critical manifold. Consequently, the critical exponents are fully determined by the homogeneous solution.

The general homogeneous solution can be written as
\begin{equation}
\label{homsolgen}
F_{\mathrm{hom}}\left(x,\lambda_k,k\right)=c\left(k\sqrt{x},\frac{\lambda_k}{x}\right)\,,
\end{equation}
where $c(a,b)$ is fixed by the boundary conditions. Regularity at $(x,\lambda_k)=(0,0)$ constrains $c$ and determines the spectrum of critical exponents. The local expansion near the origin reads
\begin{equation}
F_{\mathrm{hom}}\left(x,\lambda_k,k\right)=\sum_{l,m=0}^{+\infty} c_{l,m}\,k^{-\theta_{l,m}} x^l \lambda_k^m\,,
\end{equation}
with the spectrum of critical exponents determined by
\begin{equation}
\theta_{l,m} = -2(l+m)\,.
\end{equation}
This contains one marginal direction ($l+m=0$) and an infinite tower of irrelevant directions ($l+m\ge 1$). The large-$k$ asymptotics of the solutions, however, is not obtained term by term from $k^{-\theta_{l,m}}$, but follows from the global solution of the flow equation, which fixes the function $c\left(k\sqrt{x},\frac{\lambda_k}{x}\right)$. For a smooth solution in $(x,\lambda_k)\in [0,+\infty)\times[0,+\infty)$, 
for $k\sqrt{x}\gg1$, one can write 
\begin{equation}
c\left(k\sqrt{x}\to\infty,\frac{\lambda_k}{x}\right)= \sum_{l,m=0}^{\infty} d_{l,m}\, k^{\theta_{l,m}} x^{-l} \lambda_k^{-m}\,,
\end{equation}
which shows that the UV scaling is controlled by the large-argument behavior of $c$. The critical exponents $\theta_{l,m}$ encode the local stability properties of the fixed point and the structure of the UV critical manifold, while the detailed ultraviolet scaling depends on the full RG flow.

The ultraviolet scaling of the particular solution is determined by Eq.(\ref{solpert}), which shows that irrelevant directions are suppressed as powers of $1/k^2$, $1/k^4$, and higher, implying that the fixed point is ultraviolet attractive and infrared repulsive along these directions.

\begin{figure}[t]
     \centering
     \subfigure[]{
         \centering
         \includegraphics[width=0.45\textwidth]{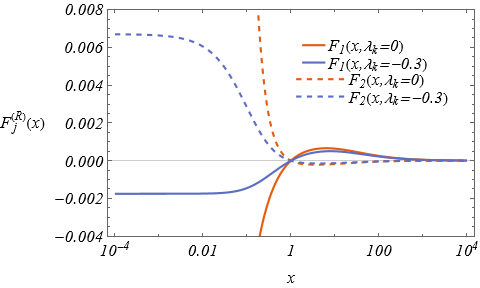}
     }
    \hspace{0.8em}
     \subfigure[]{
         \centering
         \includegraphics[width=0.45\textwidth]{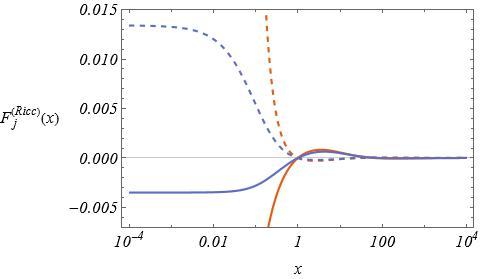}
     }
     \caption{Plots of the perturbations $F_{j}(x,\lambda_k)$ with $j=1$ (continued lines) and $j=2$ (dashed lines) around the UV fixed-point $F_*^{(\eta_*=2)}(x,\lambda_k)$. Panels (a) and (b) shows $F_j^{(\mathrm{R})}(x,\lambda_k)$ and $F_j^{(\mathrm{Ricc})}(x,\lambda_k)$ respectively. Beige and blue lines correspond to the $\lambda_k=0$ and $\lambda_k\ne0$, respectively.}
     \label{plotperFPnong}

\vspace{0.5cm}

     \centering
     \subfigure[]{
         \centering
         \includegraphics[width=0.45\textwidth]{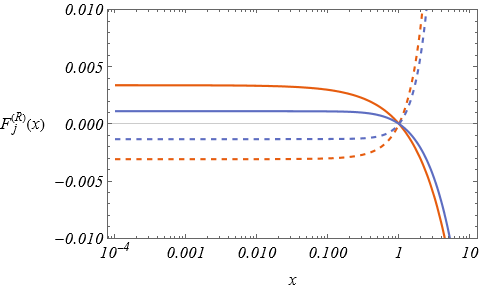}
     }
    \hspace{0.8em}
     \subfigure[]{
         \centering
         \includegraphics[width=0.45\textwidth]{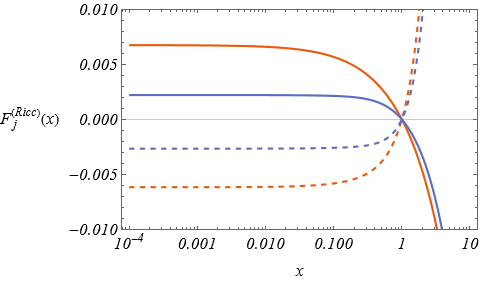}
     }
     \caption{Plots of the perturbations $F_{j}(x,\lambda_k)$ with $j=1$ (continued lines) and $j=2$ (dashed lines) around the IR fixed-point $F_*^{(\eta_*=0)}(x,\lambda_k)$. Panels (a) and (b) shows $F_j^{(\mathrm{R})}(x,\lambda_k)$ and $F_j^{(\mathrm{Ricc})}(x,\lambda_k)$ respectively. Beige and blue lines correspond to the $\lambda_k=0$ and $\lambda_k\ne0$, respectively.}
    \label{plotperFPg}
\end{figure}

The particular solutions $F_j(x,\lambda_k)$ in Eq.(\ref{solpert}) are given by
\begin{equation}
F_j(x,\lambda_k)=-x^{-j}\int_{x_0}^{x}\frac{\left.\partial_{m_p^2}^jH\!\left(\eta_k,v,\frac{\lambda_k}{x}v\right)\right|_{m_p^2=0}}{2 v^{1-j}}\, dv .
\end{equation}
The derivatives of $H$ can be read from Eq.~(\ref{expaH}). The functions $F_j$ share the same structural form as the fixed-point solution in Eq.~(\ref{solFPpart}), up to the absence of leading logarithmic contributions. Figure~\ref{plotperFPnong} shows $F_j(x,\lambda_k)$ for $j=1$ (solid lines) and $j=2$ (dashed lines), with $x_0=1$. The sign of the perturbations alternates with $j$. For $\lambda_k=0$ (beige lines), $F_j$ diverges as $x^{-j}$ for $x\to0$, while for $\lambda_k\neq0$ (blue lines) it remains finite. In both cases, the large-$x$ decay is $\sim 1/x$.

In the limits $\left(x_0,x\right)\rightarrow(0,+\infty)$, the solution Eq.(\ref{solpert}) is given by \footnote{A nonvanishing cosmological constant is required for the existence of the $x\to\infty$ limit; for $\bar\lambda=0$ the corresponding large-$x$ behavior is divergent.}
\begin{equation}
\label{sermp0k0}
f_{k=0}^{(i)}\left(u,w\right)=f_{\ast,k=0}^{\left(i\right)}\left(u\right)+\sum_{n=1}^{+\infty}\frac{L_n\left(u\right)}{w^n}
\end{equation}
where $i=\mathrm{R}$, $\mathrm{Ricc}$. The series coefficients $L_n\left(u\right)$ are nonlocal functions of $u$ as in the fixed-point case. The first terms of the series are given by
\begin{equation}\begin{split}
&L_1^{\left(R\right)}\left(u\right)\frac{\left(25u^2+208u-80\right)\ln{\left(\frac{1+\sqrt{\frac{u}{u+8}}}{1-\sqrt{\frac{u}{u+8}}}\right)}}{736\pi\sqrt{u^3\left(u+8\right)}}+\frac{\frac{60}{u}-161}{2208\pi}\\
&L_2^{\left(R\right)}=\frac{9\left(9u^2+56u-80\right)\ln{\left(\frac{1+\sqrt{\frac{u}{u+8}}}{1-\sqrt{\frac{u}{u+8}}}\right)}}{2116\sqrt u\left(u+8\right)^\frac{3}{2}}-\frac{3\left(25u^2+188u-240\right)}{8464\left(u+8\right)}
\end{split}\end{equation}
and
\begin{equation}\begin{split}
&L_1^{\left(\mathrm{Ricc}\right)}\left(u\right)=\frac{\left(-13u^2-208u+320\right)\ln{\left(\frac{1+\sqrt{\frac{u}{u+8}}}{1-\sqrt{\frac{u}{u+8}}}\right)}}{368\pi\sqrt{u^3\left(u+8\right)}}+\frac{11-\frac{15}{u}}{69\pi}\\
&L_2^{\left(\mathrm{Ricc}\right)}=\frac{3\left(13u^2+128u-960\right)}{4232\left(u+8\right)}-\frac{9\left(13u^2+16u-320\right)\ln{\left(\frac{1+\sqrt{\frac{u}{u+8}}}{1-\sqrt{\frac{u}{u+8}}}\right)}}{1058\sqrt u\left(u+8\right)^\frac{3}{2}}
\end{split}\end{equation}
Higher-order coefficients follow the same pattern. The decoupling and massless limits reproduce the same behavior of the previous cases. In the strict $u\to\infty$ limit, all $L_n^{(i)}$ approach constants.

A similar analysis applies to the Gaussian fixed point $g_\ast=0$, obtained by expanding the Newton coupling in the limit $m_p\gg k$ ($m_p\to\infty$). The beta function then admits the expansion
\begin{equation}\begin{split}
&H\left(\eta_k,x,\lambda_k\right)=H\left(\eta_\ast=0,x,\lambda_k\right)+\sum_{j=1}^{+\infty}{\left.\partial_{k^2}^jH\left(\eta_k,x,\lambda_k\right)\right|_{k^2=0}\left(\frac{k}{m_p}\right)^{2j}}
\end{split}\end{equation}
The corresponding particular solution reads
\begin{equation}
\label{pertGFPsol}
F_\mathrm{part}^{(\eta_*=0)}\left(x,\lambda_k,k\right)=F_\ast^{(\eta_*=0)}\left(x,\lambda_k\right)+\sum_{j=1}^{+\infty}{F_j^{(\eta_*=0)}\left(x,\lambda_k\right)\left(\frac{k}{m_p}\right)^{2j}}
\end{equation}
where $F_{j}^{(\eta_*=0)}\left(x,\lambda_k\right)$ satisfy
\begin{equation}
\label{eqpergauss}
2jF_j\left(x,\lambda_k\right)=2x\partial_x F_j\left(x,\lambda_k\right)+2\lambda_k\partial_{\lambda_k}F_j\left(x,\lambda_k\right)+\left.\partial_{k^2}^jH\left(\eta_k,x,\lambda_k\right)\right|_{k^2=0}
\end{equation}
The perturbations are governed by the critical exponents $\theta=-2j$, $j=1,2,\ldots$, implying that the Gaussian fixed point is IR attractive and UV repulsive in the space of form-factor perturbations. 

The solutions of Eq.~(\ref{eqpergauss}) are given by
\begin{equation}
\label{pertgaussF}
F_j(x,\lambda_k;x_0)=-x^{j}\int_{x_0}^{x}\frac{\left.\partial_{k^2}^j H\!\left(\eta_k,v,\frac{\lambda_k}{x}v\right)\right|_{k^2=0}}{2 v^{1+j}}\, dv .
\end{equation}
The functions $F_j$ exhibit the same functional form as the fixed-point solution in Eq.~(\ref{solFPpart}), with $\eta_*=0$. Figure~\ref{plotperFPg} shows $F_j(x,\lambda_k)$ for $j=1$ (solid lines) and $j=2$ (dashed lines), with $x_0=1$. As in the case of perturbations around the UV form factor fixed-point, the sign of the solutions alternates with increasing $j$. For $x>x_0$, the scaling behavior is governed by $x^j$, whereas for $x<x_0$ the solutions approach a constant value. This behavior holds both for $\lambda_k=0$ and $\lambda_k\neq 0$. 

In the limits $(x_0,x)\to(0,\infty)$, Eq.(\ref{pertgaussF}) reproduce the IR behavior of Eq.~(\ref{IRserie}), provided standard renormalization conditions at $k=\Lambda$ are imposed. Further details on the relation between flow equations, fixed points, and renormalization are given in Appendix~\ref{appserie}.

Equation~(\ref{solpert}) shows that the flow of the form factor approaches the non-Gaussian fixed point in the ultraviolet,
\begin{equation}
\label{AScondition}
\lim_{k\to\infty} F_k(x,\lambda_k)=F_\ast^{(\eta_*=-2)}(x,\lambda_k)\,,
\end{equation}
ensuring a well-defined continuum limit at the level of dimensionless variables. However, reverting to dimensional variables, $F_k(x)\to f_k(q^2)$ via Eq.~(\ref{equivFP}), a logarithmic divergence reappears in the limit $k\to\infty$ of the dimensionful form factor Eq.~(\ref{FPk0}). This indicates that asymptotic safety of the dimensionless flow does not automatically guarantee regular ultraviolet behavior of the dimensional form factor. The mismatch originates from the additional physical scale $q^2$, which prevents a direct identification of the limits $\lim_{k\to\infty} F_k(x)$ and $\lim_{k\to\infty} f_k(q^2)$. In the following, we discuss the properties of the integrated flow and show how to construct a dimensionful asymptotically safe form factor free of ultraviolet divergences.

\subsection{The properties of the adimensional flow}\label{subsec:propflow}
\begin{figure}[p]
     \centering

 \subfigure[]{
         \centering
         \includegraphics[width=0.45\textwidth]{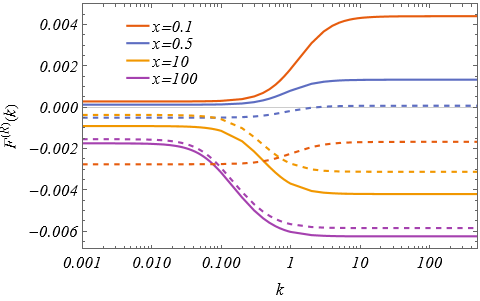}
     }
    \hspace{0.8em}
     \subfigure[]{
         \centering
         \includegraphics[width=0.45\textwidth]{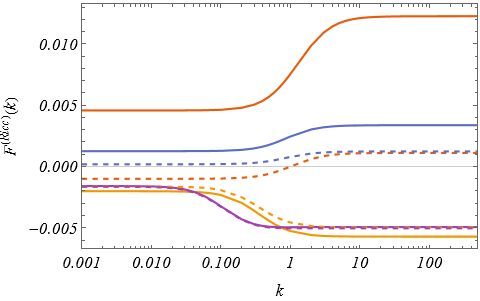}
     }
     \caption{Bulk flow of the asymptotically safe solution in Eq.~(\ref{bulkflow}) for different values of $x$. Panels (a) and (b) show $F_k^{(\mathrm{R})}(x,\lambda_k)$ and $F_k^{(\mathrm{Ricc})}(x,\lambda_k)$, respectively. The full and dashed line correspond to $\lambda_k=0$ and $\lambda_k\neq 0$, respectively. The plots are obtained fixing $m_p=1$, $x_{\mathrm{B}}=1$.}
     \label{plotflowadbulk}

\vspace{0.5cm}

     \subfigure[]{
         \centering
         \includegraphics[width=0.45\textwidth]{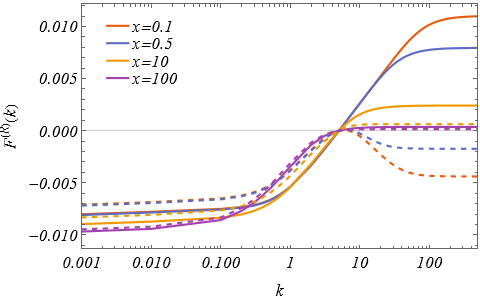}
     }
    \hspace{0.8em}
     \subfigure[]{
         \centering
         \includegraphics[width=0.45\textwidth]{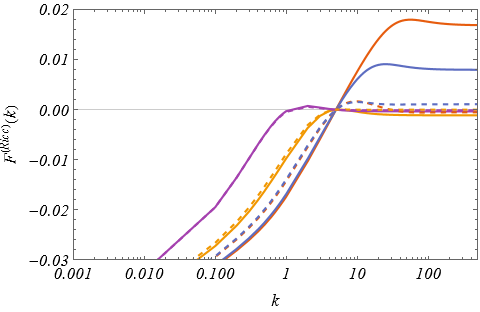}
     }
     \caption{Flow of the integrated asymptotically safe solution in Eq.~(\ref{ASflow}) for different values of $x$. Panels (a) and (b) show $F_k^{(\mathrm{R})}(x,\lambda_k)$ and $F_k^{(\mathrm{Ricc})}(x,\lambda_k)$, respectively. The full and dashed line correspond to $\lambda_k=0$ and $\lambda_k\neq 0$, respectively. The plots are obtained fixing $m_p=1$, $\Lambda=5m_p$, and $U_\Lambda=0$.}
     \label{plotflowadintegr}

   \vspace{0.5cm}
     \subfigure[]{
         \centering
         \includegraphics[width=0.45\textwidth]{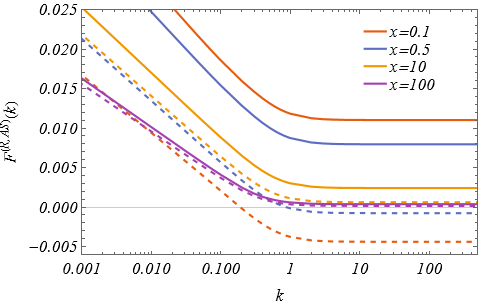}
     }
    \hspace{0.8em}
     \subfigure[]{
         \centering
         \includegraphics[width=0.45\textwidth]{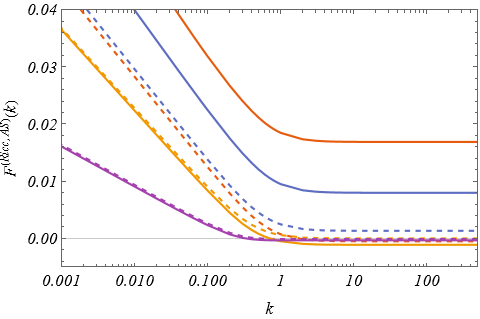}
     }
     \caption{Flow of the asymptotically safe renormalized solution in Eq.~(\ref{ASsoladim}) for different values of $x$. Panels (a) and (b) show $F_k^{(\mathrm{R})}(x)$ and $F_k^{(\mathrm{Ricc})}(x)$, respectively. The full and dashed line correspond to $\lambda_k=0$ and $\lambda_k\neq 0$, respectively. The plots are obtained fixing $m_p=1$ and $h=0$.}
     \label{plotflowad}
\end{figure}

The general solution of Eq.(\ref{floweqadim}) reads
\begin{equation}
\label{bulkflow}
F_k\left(a^2,\lambda_k\right)=c\left(k\sqrt x,\frac{\lambda_k}{x}\right)-\int_{x_{\mathrm{B}}}^{x}{\frac{H\left(\eta\left(k\sqrt\frac{x}{v}\right),v,\frac{\lambda_k}{x}v\right)}{2v}dv}
\end{equation}
where $c\left(k\sqrt x,\frac{\lambda_k}{x}\right)$ is fixed by Eq.(\ref{homsolgen}) and $x_{\mathrm{B}}$ is an arbitrary normalization scale. The integral defines the bulk solution. Figure \ref{plotflowadbulk} displays, for different values of $x$ and $\lambda_k$, the bulk asymptotically safe flow connecting the two fixed points $F_*^{(\eta_*=0)}(x,\lambda_k)$ and $F_*^{(\eta_*=-2)}(x,\lambda_k)$. Deviations from the EFT behavior set in at $k \sim m_p$, above which the flow enters the asymptotically safe regime.

The choice of boundary condition fixes $c$ and selects a specific RG trajectory within the space of solutions. Imposing that at the scale $k=\Lambda$ the solution satisfies $F_\Lambda(x,\lambda_\Lambda)=U_\Lambda(x,\lambda_\Lambda)$, with $U_\Lambda$ an arbitrary function, the resulting trajectory takes the form,
\begin{equation}\begin{split}
\label{ASflow}
&F_k\left(x,\lambda_k;\Lambda\right)=U_\Lambda\left(\frac{k^2 x}{\Lambda^2},\frac{k^2 \lambda_k}{\Lambda^2}\right)-\left[\int_{x_\mathrm{B}}^{x}{\frac{H\left(\eta\left(k\sqrt{\frac{x}{v}}\right),v,\frac{\lambda_k}{x}v\right)}{2v}dv}+\int_{\frac{k^2x}{\Lambda^2}}^{x_\mathrm{B}}{\frac{H\left(\eta\left(k\sqrt{\frac{x}{v}}\right),v,\frac{\lambda_k}{x}v\right)}{2v}dv}\right]=\\
&= U_\Lambda\left(\frac{k^2 x}{\Lambda^2},\frac{k^2 \lambda_k}{\Lambda^2}\right)
- \int_{\frac{k^2 x}{\Lambda^2}}^{x}
\frac{H\left(\eta\left(\frac{k\sqrt x}{\sqrt{v}}\right),v,\frac{\lambda_k}{x}v\right)}{2v}\, dv \,.
\end{split}\end{equation}
The combination $k^2 x/\Lambda^2$ follows from the characteristic structure of the PDE. The boundary choice enters as an additional contribution to the bulk term. This does not represent an independent dynamical sector of the RG flow, but rather encodes how the UV data at $k=\Lambda$ are propagated along the flow to lower scales. From Eq.(\ref{equivFP}) the total integral coincides the dimensional flow expressed in dimensionless variables. Indeed, one can rewrite it as
\begin{equation}
\label{equivsol}
\int_{\frac{k^2 x}{\Lambda^2}}^{x}
\frac{H\left(\eta\left(\frac{k\sqrt x}{\sqrt{v}}\right),v,\frac{\lambda_k}{x}v\right)}{2v}\, dv
= \int_k^\Lambda \frac{\beta_{k_1}\left(\eta_{k_1}, q^2, \bar{\lambda}\right)}{k_1}\, dk_1 \,,
\end{equation}
which makes explicit the equivalence with the integrated form of the flow derived in Sec.~\ref{secintegFF}. 

Figure~\ref{plotflowadintegr} illustrates, for $U_\Lambda=0$, the impact of the boundary contribution on the bulk flow. For $k \ll m_p$, the solution continues to evolve and no stationary regime is reached in the IR limit $k \to 0$, whereas for $k \gg m_p$ the flow approaches the UV fixed-point regime. 
In the Gaussian regime, the solution is equivalent to Eq.~(\ref{pertGFPsol}) with $x_0 \equiv k^2 x/\Lambda^2$,
\begin{equation}\begin{split}
\label{flowIRintegr}
&F_k\left(x,\lambda_k,\frac{k}{m_p}\ll1;\Lambda\right)=F_\ast^{\left(\eta_\ast=0\right)}\left(x,\lambda_k;\frac{k^2x}{\Lambda^2}\right)+\sum_{j=1}^{+\infty}{F_j^{\left(\eta_\ast=0\right)}\left(x,\lambda_k;\frac{k^2x}{\Lambda^2}\right)\left(\frac{k^2}{m_p^2}\right)^j}=\\
&=\sum_{j=0}^{+\infty}{\left[F_j^{\left(\eta_\ast=0\right)}\left(x,\lambda_k;x_\mathrm{B}\right)+F_j^{\left(\eta_\ast=0\right)}\left(x_\mathrm{B},\lambda_k;\frac{k^2x}{\Lambda^2}\right)\right]\left(\frac{k^2}{m_p^2}\right)^j}
\end{split}\end{equation}
In the limit $k \to 0$, the boundary term $F_{j=0}^{(\eta_\ast=0)}\!\left(x_\mathrm{B},\lambda_k;k^2 x/\Lambda^2\right)$ generates a residual logarithmic piece $a_{\mathrm{EFT}} \ln(k^2 x/\Lambda^2)$, with $a_{\mathrm{EFT}}$ given in Eq.~(\ref{EFTIR}). The small-$k$ behavior then reads
\begin{equation}
F_k\left(x,\lambda_k;\Lambda\right)=F_\ast^{\left(\eta_\ast=0\right)}\left(x,\lambda_k\right)+a_{\mathrm{EFT}}\ln\left(\frac{k^2x}{\Lambda^2}\right)+O\left(k^2\right)
\end{equation}
indicating that, although the flow is controlled by the Gaussian fixed point, the integrated RG trajectory does not become stationary in the infrared limit. This is not in contradiction with the previous analysis of the IR attractiveness of the fixed point. The residual logarithmic term does not arise from a perturbative expansion around the fixed point, but instead from the propagation of the UV data at $k=\Lambda$ to infrared scales. Moreover, it does not originate from the expansion of Eq.~(\ref{ASflow}) around $x=0$, which remains analytic as reflected in the structure of Eq.~(\ref{homsolgen}). 

The residual logarithmic dependence is necessary to reproduce the correct infrared behavior of the dimensionful form factor. Indeed, in the limits $x\to\infty$ and $k\to0$, Eq.~(\ref{flowIRintegr}) reproduces the full IR expansion of Eq.~(\ref{IRserie}) after identifying $k^2x=q^2$ and removing the cutoff $\Lambda$ via a renormalization condition. Similarly, expanding Eq.~(\ref{ASflow}) in the regime $m_p^2\ll k^2$ and taking $(x,\Lambda)\to\infty$ using $k^2x=q^2$ reproduces the UV expansion of Eq.~(\ref{UVseries}). This establishes a direct relation between the asymptotic regimes of the integrated dimensionless flow and the corresponding IR and UV expansions at $k=0$ of the dimensional flow.

\subsection{The asymptotically safe renormalized solution}\label{subsecASsol}

Although the flow is controlled by the non-Gaussian fixed point in the UV, the integrated form factor still exhibits the logarithmic behavior in the limit $\Lambda\to\infty$. As already anticipated, asymptotic safety of the dimensionless flow is not sufficient to guarantee a finite integrated form factor. An additional renormalization condition is required. The result of Sec.~\ref{secintegFF} is recovered by choosing
\begin{equation}
\label{EFTbouncon}
U_\Lambda(x,\lambda_\Lambda)=\alpha \ln\left(\frac{\mu^2}{\Lambda^2}\right)\,,
\end{equation}
where $\alpha$ is the universal coefficient in Eq.~(\ref{IRserie}). In RG terms, this choice selects the perturbatively renormalizable trajectory, yielding a well-defined $\Lambda\to\infty$ limit in the EFT sense. 

Within the asymptotic safety framework, however, physical consistency requires the absence of external renormalization scales in the ultraviolet scaling behavior. The perturbatively renormalizable trajectory is therefore not compatible with a consistent asymptotically safe flow. Consistency between ultraviolet fixed-point scaling and finiteness of the dimensional form factor instead selects the nontrivial fixed-point solution as a boundary condition,
\begin{equation}
\label{ASbouncon}
U_\Lambda(x,\lambda_\Lambda)=F_*^{(\eta_*=-2)}(x,\lambda_\Lambda;x_0)\,.
\end{equation}
We refer to this choice as an \emph{asymptotically safe renormalization condition}. With this prescription, the solution takes the form
\begin{equation}\begin{split}
\label{ASsoladim}
&F_k\left(x,\lambda_k;\Lambda,x_0\right)=\\
&=h\left(x_0,\frac{x_0\lambda_k}{x}\right)+F_\ast^{(\eta_*=-2)}\left(x,\lambda_k;x_0\right)-\int_{\frac{k^2x}{\Lambda^2}}^{x}\left[\frac{H\left(\eta\left(k\sqrt{\frac{x}{v}}\right),v,\frac{\lambda_k}{x}v\right)}{2v}-\frac{H\left(\eta_\ast=-2,v,\frac{\lambda_k}{x}v\right)}{2v}\right]dv
\end{split}\end{equation}
The key feature is the difference between the full and fixed-point beta-function contributions. This subtraction cancels the logarithmic terms in the limit $\Lambda\to\infty$ without introducing an external scale $\mu$, leading to a qualitatively different infrared behavior from the perturbative case. In contrast, if the Gaussian fixed point with $\eta_*=0$ is chosen, the subtraction reproduces the standard logarithmic behavior $\sim\ln(q^2/\mu^2)$.

The solution (\ref{ASsoladim}) still preserves the asymptotically safe UV behavior. Indeed, as $k\to\infty$ one has $\eta\!\left(\frac{k\sqrt x}{\sqrt v}\right)\to \eta_\ast$, so that the integrand vanishes and Eq.~(\ref{ASsoladim}) reduces to the non-Gaussian fixed-point solution, independently of the choice of the UV scale $\Lambda$ (see Fig. \ref{plotflowad}). 

This prescription differs from the standard perturbative one, where divergences are absorbed into couplings via counterterms and the resulting trajectories remain close to the Gaussian fixed point, displaying logarithmic running $\sim \ln(\mu^2/k^2)$. Instead, imposing Eq. (\ref{ASbouncon}) selects a trajectory on the critical surface of the non-Gaussian fixed point. The RG flow then remains finite as $\Lambda\to\infty$, with the logarithmic dependence dynamically suppressed rather than subtracted. Consequently, the dependence on the renormalization scale $\mu$ is absent in the asymptotically safe solution. The remaining parameter $x_0=q^2/k_0^2$ arises as an integration constant of the fixed-point equation and does not originate from the subtraction of ultraviolet divergences. Predictivity is controlled by the relevant directions of the fixed point. In the present one-loop approximation, no such directions arise, so that no additional free parameters appear.

This construction becomes particularly relevant when expressed in terms of dimensional variables, where the solution takes the form
\begin{equation}\begin{split}
\label{ASsol}
&F_k\left(\frac{q^2}{k^2},\frac{\bar{\lambda}}{k^2};\Lambda,\frac{q^2}{k_0^2}\right)=h\left(\frac{q^2}{k_0^2},\frac{\bar{\lambda}}{k_0^2}\right)+\int_{k_{0}}^{\Lambda}{\frac{\beta_{k^\prime}\left(q^2,\bar{\lambda},\eta_\ast=-2\right)}{k^\prime}dk^\prime}-\int_{k}^{\Lambda}{\frac{\beta_{k^\prime}\left(q^2,\bar{\lambda},\eta_{k^\prime}\right)}{k^\prime}dk^\prime}
\end{split}\end{equation}
Equation~(\ref{ASsol}) shows that the asymptotically safe dimensional form factor is obtained by subtracting the fixed-point contribution from the full RG evolution, up to the boundary term encoded in $h$. The reference scale $k_{\mathrm{0}}$ can be chosen arbitrarily, including $k_{\mathrm{0}}=0$, reflecting its role as an integration constant rather than a renormalization scale. 

Equation~(\ref{ASsol}) admits two complementary interpretations when related to the dimensional flow equation. The first arises by implementing the nonperturbative renormalization condition directly at the level of the dimensional flow,
\begin{equation}
k\partial_k f_k(q^2,\bar{\lambda})=\beta_k(q^2,\bar{\lambda}), \qquad
f_{k=\Lambda}(q^2,\bar{\lambda}) = f_{*}\left(q^2,\bar{\lambda},\eta_\ast;k_0,\Lambda\right)+h\left(\frac{q^2}{k_0^2},\frac{\bar{\lambda}}{k_0^2}\right),
\end{equation}
The AS form factor is obtained from the dimensional flow by fixing the boundary condition to the UV fixed-point solution $f_{*}\left(q^2,\bar{\lambda},\eta_\ast;k_0,\Lambda\right)$, defined in Eq.~(\ref{eqsolpartFPdim}). This ensures consistency between the dimensionless and dimensionful formulations,
$\lim_{k\to\infty} F_k(x)=\lim_{k\to\infty} f_k(q^2)$.

The second interpretation follows directly from Eq.~(\ref{ASsol}) and can be formulated entirely at the level of the beta functions,
\begin{equation}
\label{eqASbeta}
k\partial_k f_k(q^2,\bar{\lambda})
=\beta_k(q^2,\bar{\lambda}; \eta_k\neq0)
-\beta_k(q^2,\bar{\lambda}; \eta_*=-2)(1-\theta(k-k_0)), \qquad
f_{k=\Lambda}(q^2,\bar\lambda)=h\left(\frac{q^2}{k_0^2},\frac{\bar{\lambda}}{k_0^2}\right)
\end{equation}
where $\theta(t)$ is the Heaviside function. In this form, the flow is driven by the difference between the full beta function and its fixed-point counterpart, while the boundary condition at $k=\Lambda$ is encoded in $h$. This representation is particularly convenient for numerical implementations, as it avoids the explicit computation of the fixed-point solution. 

The asymptotically safe solution in Eq.~(\ref{ASsol}) is defined up to an arbitrary function $h\!\left(x_0,x_0\frac{\lambda_k}{x}\right)$, encoding the choice of boundary conditions for the fixed-point equation. This freedom allows one to select different asymptotic behaviors in the infrared and ultraviolet regimes. For $\bar\lambda=0$, $h$ reduces to a constant, $h(x_0)\equiv c$. The minimal choice $h=0$ defines a reference solution in which the form factors vanish in the ultraviolet, $\lim_{q^2\to\infty} f_k(q^2)=0$, thereby isolating the intrinsic RG running.

Setting $k=0$ with $k_0=0$ in Eq.(\ref{ASsol}) we obtain the physical AS renormalized form factors,
\begin{equation}
f_{k=0}^{(i,\mathrm{AS})}\left(q^2,\bar\lambda\right)=f_{k=0}^{(i)}(q^2,\bar\lambda)-f_{*,k=0}^{(i)}(q^2,\bar\lambda)
\end{equation}
where $i=\mathrm{R}$, $\mathrm{Ricc}$ and $f_{*,k=0}^{(i)}(q^2,\bar\lambda)$ is given in Eq.(\ref{FPk0}). In the IR regime the general structure of the solution is given by
\begin{equation}\begin{split}
\label{formASIR}
&f_{k=0}^{(i,\mathrm{AS})}\left(u,t,w\rightarrow0\right)=\tilde A_0^{(i)}(u,w)+\sum_{n=1}^{+\infty}{A_n^{(i)}\left(u,t\right)w^n}\\
\end{split}\end{equation}
where $\tilde A_0^{(i)}(u,w)$ is given for the two form factors, respectively, by 
\begin{equation}\begin{split}
&\tilde A_{0}^{\left(\mathrm{R}\right)}\left(u,w\right)=\frac{287u^2+4280u-960}{11520\pi^2u^2}-\frac{7\ln{\left(\frac{23}{6\pi}\frac{w}{u}\right)}}{384\pi^2}-\frac{\sqrt{u+8}\left(7u^2+72u-16\right)\ln{\left(\frac{1+\sqrt{\frac{u}{u+8}}}{1-\sqrt{\frac{u}{u+8}}}\right)}}{384\pi^2u^\frac{5}{2}}\\
&\tilde A_{0}^{\left(\mathrm{Ricc}\right)}\left(u,w\right)=\frac{-103u^2-4640u+3840}{5760\pi^2u^2}-\frac{7\ln{\left(\frac{23}{6\pi}\frac{w}{u}\right)}}{192\pi^2}-\frac{\sqrt{u+8}\left(7u^2-80u+64\right)\ln{\left(\frac{1+\sqrt{\frac{u}{u+8}}}{1-\sqrt{\frac{u}{u+8}}}\right)}}{192\pi^2u^\frac{5}{2}}
\end{split}\end{equation}
and $A_n$ with $n>1$ are the IR-series coefficients defined in Eq.~(\ref{IRserie}). In this regime, the logarithmic contribution persists but depends only on the combination $w/u=t=\bar\lambda/m_p^2$, effectively replacing the unphysical scale $\Lambda$ with $m_p$. The remaining properties of this regime then follows closely the results presented after Eq.~(\ref{IRserie}). The massless limit ($u\to\infty$) of $\tilde A_0^{(i)}$ admits a regular expansion in $1/u$, implying a smooth $\bar\lambda\to0$ limit. However, the limit $w\to0$ is singular due to the residual logarithmic term $\ln(w)$. Conversely, in the decoupling regime $u\to0$ ($\bar\lambda\to\infty$), the singularity at $w=0$ disappears, but the limit $\bar\lambda\to0$ becomes nonsmooth.

In the UV regime we get
\begin{equation}\begin{split}
\label{ASUV}
&f_{k=0}^{\left(\mathrm{R,AS}\right)}\left(u\rightarrow\infty,t\rightarrow0,w\rightarrow\infty\right)=\frac{86-75\ln{\left(\frac{u}{2}\right)}}{288\pi^2u}-\frac{29+18\ln{\left(\frac{u}{2}\right)}}{32\pi^2u^2}+O\left(\frac{1}{u^3}\right)\\
&+\frac{\frac{-86+75\ln{\left(\frac{23w}{12\pi}\right)}}{2208\pi}+\frac{t\left(-97+150\ln{\left(\frac{23w}{12\pi}\right)}\right)}{576\pi^2}+O\left(t^2\right)}{w}+O\left(\frac{1}{w^2}\right)\\
&f_{k=0}^{\left(\mathrm{Ricc,AS}\right)}\left(u\rightarrow\infty,t\rightarrow0,w\rightarrow\infty\right)=\frac{-137+39\ln{\left(\frac{u}{2}\right)}}{144\pi^2u}+\frac{35+26\ln{\left(\frac{u}{2}\right)}}{16\pi^2u^2}+O\left(\frac{1}{u^3}\right)\\
&+\frac{\frac{137-39\ln{\left(\frac{23w}{12\pi}\right)}}{1104\pi}+\frac{t\left(235-78\ln{\left(\frac{23w}{12\pi}\right)}\right)}{288\pi^2}+O\left(t^2\right)}{w}+O\left(\frac{1}{w^2}\right)
\end{split}\end{equation}
In comparison to the non-AS solution in Eq.(\ref{UVseries}), the asymptotically safe solution contains no isolated leading logarithmic contribution and admits an expansion in inverse powers of $q^2$ modulated by logarithmic terms. The general structure can be written as
\begin{equation}
\label{formASUV}
f_{k=0}\left(u\to\infty,t\to0,w\rightarrow\infty\right)=\sum_{n=1}^{+\infty}\frac{\tilde a_n+\tilde b_n\ln\left(\frac{u}{2}\right)}{u^n}+\sum_{n=1}^{+\infty}\frac{\sum_{m=0}^{+\infty}\left(c_{nm}+d_{nm}\ln\left(\frac{23w}{12\pi}\right)\right)t^m}{w^n}
\end{equation}
where $\tilde a_n$ and $\tilde b_n$ are determined by the difference between the EFT and fixed-point coefficients in the large-$u$ expansion.

\section{Numerical Results and Physical Properties of the Form Factors}\label{secnuman}
\noindent In this Section, we present the numerical analysis for the flow of form factors to arrive at $k=0$ for generic momenta. We first consider the case of the mixed scheme. This provides both a consistency check of the analytical results derived in the previous sections and a benchmark for understanding the numerical behavior of the full $\epsilon=1$ (scheme B) case.

\subsection{The numerical analysis of the mixed scheme case}
The general properties of the form-factor flow at finite $k$ have been analyzed in Sec.~\ref{subsec:propflow}, with Figs.~\ref{plotflowadbulk}, \ref{plotflowadintegr}, and \ref{plotflowad} showing the running of the dimensionless form factors as functions of $k$. The relation between the dimensionless and dimensional flows is given in Eq.~(\ref{equivsol}). Here, we analyze the numerical solutions of the form factor flow at $k=0$ and compare the standard with the renormalized asymptotically safe form factors of Sec. \ref{subsecASsol}. The former are obtained by numerically integrating Eq.~(\ref{eqFFm3}), whose solution is given in Eq.~(\ref{solFFdim}), while the latter are obtained by numerically integrating Eq.~(\ref{eqASbeta}), whose solution is given in Eq.~(\ref{ASsol}). In both cases, the numerical solutions considered below are defined by the UV boundary condition $f_{k=\Lambda}(q^2)=0$; we refer to them as \textit{reference form factors} and \textit{reference AS form factor}, respectively, to emphasize this choice of boundary condition. At the analytical level, the main difference between the standard and the renormalized asymptotically safe form factors at $k=0$ is encoded in their respective IR and UV asymptotic solutions, given by Eqs.~(\ref{IRserie}) and (\ref{UVseries}) and by Eqs.~(\ref{formASIR}) and (\ref{ASUV}), respectively. The numerical analysis below also allows us to compare the two solutions in the intermediate-momentum regime.

A brief comment on the choice of the boundary condition at $k=\Lambda$ is useful at this point. The solutions in Eqs.~(\ref{solFFdim}) and (\ref{ASsol}) can be written in the form
\begin{equation}
f_k(q^2,\bar\lambda)=f_\Lambda(q^2,\bar\lambda)-\int_k^\Lambda T_{k'}(q^2,\bar\lambda)dk'
\end{equation}
where $T_{k'}(q^2,\bar\lambda)$ can be read from Eqs.~(\ref{solFFdim}) and (\ref{ASsol}) in the two cases, respectively. In both cases, the UV boundary condition $f_\Lambda(q^2,\bar\lambda)$ affects the form factor at $k=0$ but its role is different, particularly in the large-$\Lambda$ limit. For the standard form factor, the minimal choice of $f_\Lambda(q^2)$ is usually fixed by the requirement of removing the divergence in the limit $\Lambda\to\infty$, thereby fixing the perturbative renormalization prescription. For the renormalized AS form factor, the limit $\Lambda\to\infty$ is smooth and $f_\Lambda(q^2)$ corresponds instead to the choice of boundary condition for the fixed-point equation, as discussed after Eq.~(\ref{ASsol}). The choice $f_\Lambda(q^2)=0$ isolates the contribution generated by the RG flow and is made here to illustrate in the simplest way the difference between the standard and renormalized AS form factors.

To perform the numerical integration, we fix $m_p=1$, $\bar{\lambda}=-1/10$, $\Lambda=150m_p$, and $m=3$. For each value of $q^2$ below the numerical cutoff, the flow is integrated from $k=\Lambda$ down to $k=10^{-8}$ with boundary condition $f_{k=\Lambda}(q^2)=0$.

\begin{figure}[p]
     \centering
     \subfigure[]{
         \centering
         \includegraphics[width=0.45\textwidth]{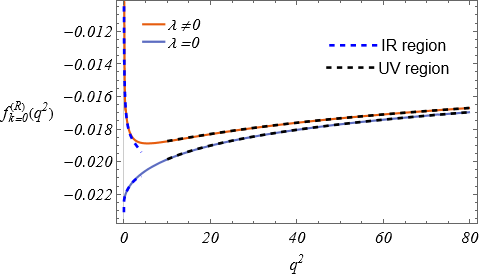}
     }
    \hspace{0.8em}
     \subfigure[]{
         \centering
         \includegraphics[width=0.45\textwidth]{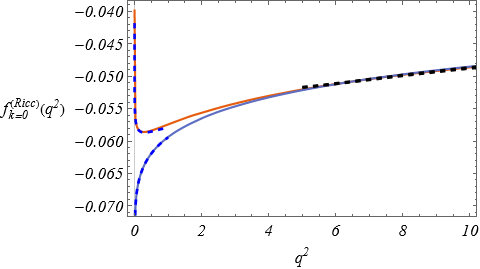}
     }
     \caption{Reference numerical form factors at $k=0$ and comparison with the asymptotic analytical results. Panels (a) and (b) show $f_{k=0}^{(R)}(q^2)$ and $f_{k=0}^{(\mathrm{Ricci})}(q^2)$, respectively. The reference form factors develop a minimum in the crossover region between the IR and UV regimes. At large $q^2$, the logarithmic behavior dominates and the dependence on $\bar\lambda$ becomes negligible.}
     \label{plotFFstan}

\vspace{0.5cm}

     \centering
     \subfigure[]{
         \centering
         \includegraphics[width=0.45\textwidth]{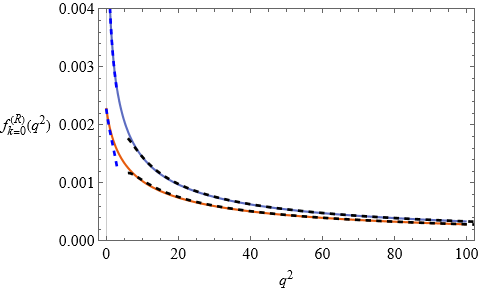}
     }
    \hspace{0.8em}
     \subfigure[]{
         \centering
         \includegraphics[width=0.45\textwidth]{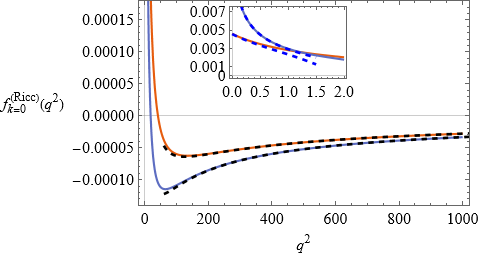}
     }
     \caption{Reference asymptotically safe form factors at $k=0$ and comparison with the asymptotic analytical results. Panels (a) and (b) show $f_{k=0}^{(\mathrm{R})}(q^2)$ and $f_{k=0}^{(\mathrm{Ricc})}(q^2)$, respectively. Both reference form factors decay as $1/q^2$ at large $q^2$. The inset in (b) zooms into the IR region of $f_{k=0}^{(Ricc)}(q^2)$.}
    \label{plotFFAS}

\vspace{0.5cm}

     \centering
     \subfigure[]{
         \centering
         \includegraphics[width=0.45\textwidth]{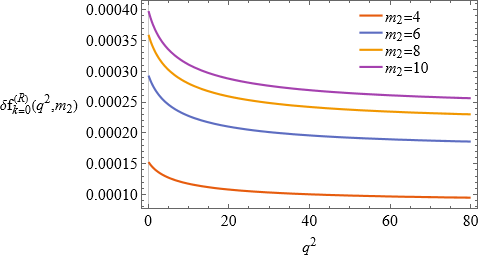}
         \label{plotw}
     }
    \hspace{0.8em}
     \subfigure[]{
         \centering
         \includegraphics[width=0.45\textwidth]{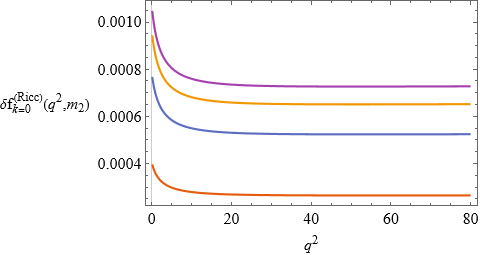}
     }
     \caption{Absolute difference $\delta f(q^2,m_1=3,m_2)$ at $k=0$ for different values of $m_2$. The residual regulator dependence is suppressed at large $q^2$.}
     \label{plotFFm}
     \end{figure}

Figure~\ref{plotFFstan} compares the numerical solutions for the reference form factors with the asymptotic expansions in Eqs.~(\ref{IRserie}) and (\ref{UVseries}) (blue and black dashed lines, respectively). The IR expansion reproduces the flow only for $q^2\ll1$, whereas the UV expansion agrees well for $q^2\gtrsim5$. Both form factors develop a minimum, located at approximately $q^2\sim4.5$ and $q^2\sim0.25$ for the Ricci-scalar and Ricci-tensor sectors, respectively.

For comparison with the B scheme, we also solve the flow for $\bar\lambda=0$ (blue curves in Fig.~\ref{plotFFstan}). In this case the minima disappear and the logarithmic growth sets in already at small momenta. At large $q^2$, where the logarithmic term dominates, the $\bar\lambda\neq0$ and $\bar\lambda=0$ solutions progressively approach each other.

Since the coefficient of the logarithmic term is positive, the reference form factors eventually cross zero and become positive at sufficiently large $q^2$. The zero can be estimated from the UV asymptotic expansion by solving $f_{k=0}(q^2)=0$ in Eq.~(\ref{UVseries}), yielding for $\Lambda=150m_p$
\begin{equation}
q_R^2\approx7.4\cdot10^{6}m_p^2,\quad\quad q^2_{Ricc}\approx3.8\cdot10^{5}m_p^2.
\end{equation}
These values lie deep in the UV region ($q^2\gg m_p^2$), as shown in Fig.~\ref{plotFFstaneps1}. However, they should not be interpreted as physical scales, since their position depends on the UV boundary condition and can be shifted arbitrarily.

To obtain the numerical reference asymptotically safe form factors, we use Eq.~(\ref{eqASbeta}) with $f_{k=\Lambda}(q^2,\bar\lambda)=0$. The red and blue curves in Fig.~\ref{plotFFAS} correspond to $\bar{\lambda}\neq0$ and $\bar{\lambda}=0$, respectively, while the dashed lines show the analytical asymptotics in Eqs.~(\ref{formASIR}) and (\ref{ASUV}), derived from Eq.~(\ref{ASsol}) and providing a nontrivial consistency check. The numerical reference AS form factor $f_{k=0}^{(\mathrm{R,AS})}(q^2)$ remains positive for all $q^2$, while $f_{k=0}^{(\mathrm{Ricc,AS})}(q^2)$ crosses zero and approaches the UV limit $f_{k=0}^{(\mathrm{Ricc,AS})}(q^2\to\infty)\to 0$ from below. As discussed in Subsec.~\ref{subsecASsol}, this behavior depends on the boundary conditions and can be shifted, e.g. by choosing $c_{\mathrm{Ricci}}$ such that $f_{\mathrm{AS}}^{(\mathrm{Ricc})}(q^2)>0$. The $\bar{\lambda}=0$ case shows the same qualitative behavior, unlike the non-AS solution.

In the EFT result of Eq.~(\ref{EFTresult}), the dependence on the cutoff parameter $m$ at $k=0$ reduces to a constant term $c(m)$, which can be removed by an appropriate choice of boundary conditions. In the full computation, this can be tested numerically by comparing solutions at different $m$ through
\begin{equation}
\delta f(q^2,m_1,m_2)=\left|f_{k=0}(q^2,m_1)-f_{k=0}(q^2,m_2)\right|.
\end{equation}
Choosing $m_1=3$ as a reference, Fig.~\ref{plotFFm} shows $\delta f(q^2,3,m_2)$ for various $m_2$. Differently from the EFT case, the $m$-dependence propagates through the full RG flow and does not reduce to a simple additive contribution. The resulting variation therefore provides a quantitative measure of the residual regulator dependence associated with the truncation. $\delta f(q^2,m_1,m_2)$ approaches a constant at large $q^2$, reflecting the fact that the nonuniversal corrections become subleading and the form factor approaches the universal logarithmic regime $\sim c(m)+\alpha\ln(q^2/\Lambda^2)$.

\subsection{The numerical analysis in the B scheme case}
In the $\epsilon=1$ scheme, the running Newton coupling is obtained numerically by solving the flow equation at $\bar{\lambda}=0$ with boundary condition $g(k_0)=k_0^2/m_p^2$, choosing $k_0=10^{-8}$ and $m_p=1$. The flow is integrated up to $k=5000\,m_p$ for different cutoff values $m=3,6,8,10$, and the form-factor flow is computed as in the previous Subsection. The deep ultraviolet regime is then extracted by fitting the numerical solution to the asymptotic expansion (\ref{UVseries}).

\begin{figure}[p]
     \centering
     \subfigure[]{
         \centering
         \includegraphics[width=0.45\textwidth]{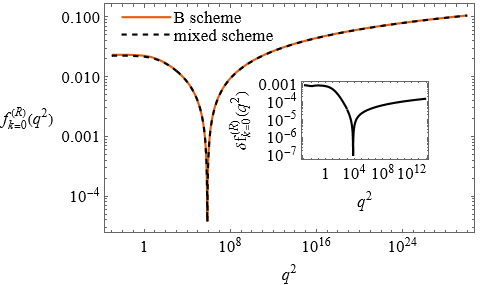}
     }
    \hspace{0.8em}
     \subfigure[]{
         \centering
         \includegraphics[width=0.45\textwidth]{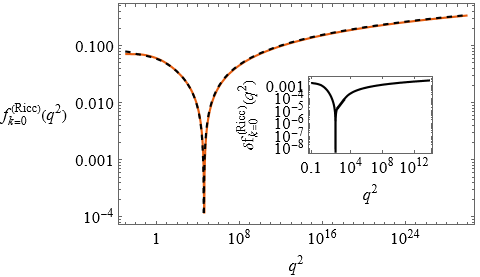}
     }
     \caption{Reference numerical form factors at $k=0$ with $\bar\lambda=0$ in the B scheme, compared with the mixed-scheme results of Fig.~\ref{plotFFstan}. Panels (a) and (b) show $f_{k=0}^{(\mathrm{R})}(q^2)$ and $f_{k=0}^{(\mathrm{Ricc})}(q^2)$, respectively, on a log-log scale. The insets display $\delta f(q^2)=|f_{\mathrm{B}}(q^2)-f_{\mathrm{MS}}(q^2)|$. The two schemes yield nearly identical reference numerical form factors, confirming the universality of the logarithmic behavior.}
     \label{plotFFstaneps1}

\vspace{0.5cm}

     \centering
     \subfigure[]{
         \centering
         \includegraphics[width=0.45\textwidth]{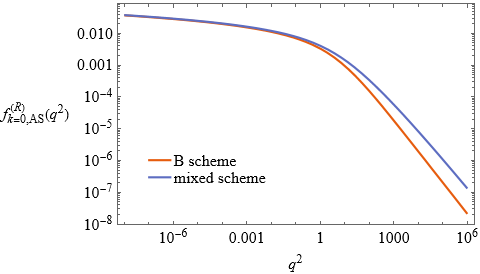}
     }
    \hspace{0.8em}
     \subfigure[]{
         \centering
         \includegraphics[width=0.45\textwidth]{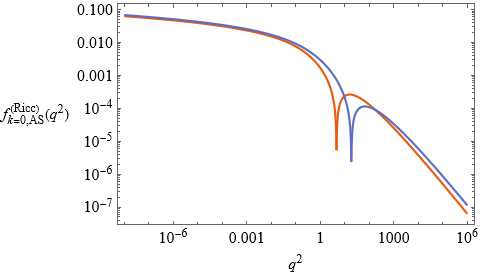}
     }
     \caption{Reference asymptotically safe form factors at $k=0$ with $\bar\lambda=0$ in the B scheme, compared with the mixed-scheme results of Fig.~\ref{plotFFAS}. Panels (a) and (b) show $f_{k=0,\mathrm{AS}}^{(\mathrm{R})}(q^2)$ and $f_{k=0,\mathrm{AS}}^{(\mathrm{Ricc})}(q^2)$, respectively, on a log-log scale. The functional structure of the reference AS form factors is independent of $\epsilon$.}
    \label{plotFFASeps1}

\vspace{0.5cm}

     \centering
     \subfigure[]{
         \centering
         \includegraphics[width=0.45\textwidth]{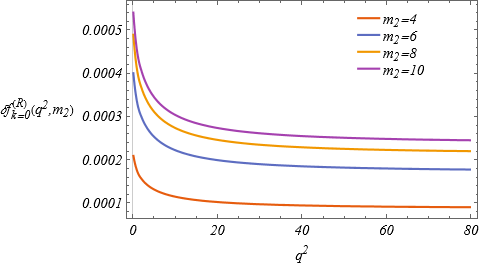}
     }
    \hspace{0.8em}
     \subfigure[]{
         \centering
         \includegraphics[width=0.45\textwidth]{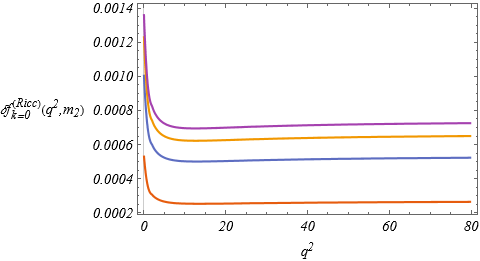}
     }
     \caption{Absolute difference $\delta f(q^2,m_1=3,m_2)$ in the B scheme at $k=0$ for different values of $m_2$. The qualitative behavior is the same as in the mixed-scheme case of Fig.~\ref{plotFFm}.}
     \label{plotFFmeps1}
     \end{figure}

Figure~\ref{plotFFstaneps1} shows the results for $m=3$, together with a comparison with the mixed-scheme (MS) case. The qualitative behavior is essentially identical in the two schemes across all momentum regimes. The difference $\delta f(q^2)=|f_{\mathrm{B}}(q^2)-f_{\mathrm{MS}}(q^2)|$ (insets of Fig.~\ref{plotFFstaneps1}) remains below $\mathcal{O}(10^{-2})$ and decreases with increasing $q^2$, indicating a weak dependence on the choice of $\epsilon$. In particular, the ultraviolet logarithmic terms exhibit universal coefficients, consistent with Eq.~(\ref{IRserie}).

The cutoff dependence is illustrated in Fig.~\ref{plotFFmeps1} through the difference $\delta f(q^2,3,m_2)$. As in the mixed scheme case, the $m$-dependence propagates through the full RG flow and does not reduce to a simple additive contribution. No qualitative differences with respect to the mixed scheme are observed. 

The reference asymptotically safe solutions are shown in
Fig.~\ref{plotFFASeps1}, together with a comparison with the mixed-scheme case. Together with Fig.~\ref{plotFFAS}, this figure presents the main result of this work for $f_{k=0}(q^2)$. The log-log plot shows that the results obtained in the B and mixed schemes display the same qualitative features. In particular, the momentum dependence of the IR and UV asymptotic behaviors in Eqs.~(\ref{formASIR}) and (\ref{ASUV}), $f_{k=0,\mathrm{AS}}(q^2\to0)\sim a_0+b_0\ln(q^2)+q^2(a_1+b_1\ln(q^2))+O(q^4)$
and $f_{k=0,\mathrm{AS}}(q^2\to\infty)\sim [a+b\ln(q^2)]/q^2$,
respectively, is insensitive to the $\epsilon$ prescription used for the running of the Newton coupling. Although the numerical values of the coefficients depend weakly on $\epsilon$, in the IR form factors are almost indistinguishable, highlighting that the logarithmic coefficient $b_0$ is independent of the $\epsilon$ prescription. In the UV, the absence of a pure logarithmic term provides an additional consistency check of the universality of the logarithmic coefficients.

The numerical analysis of this section supports the arguments presented in Sec.~\ref{subsectionscheme}. In particular, the structure of the form factors at $k=0$ is robust with respect to the choice of $\epsilon$ used in the running of the Newton constant. The scheme dependence associated with $\epsilon$ is small, with deviations below $\mathcal{O}(10^{-2})$, while the coefficients of the logarithmic terms remain scheme independent.

Asymptotically safe form factors have recently been studied in different frameworks in \cite{Pawlowski:2023dda,Assant:2026dca}. These works consider fluctuation form factors, whereas the results shown in Fig.~\ref{plotFFASeps1} concern background form factors. Although the two quantities are not directly identical, a qualitative comparison of their momentum dependence is nevertheless informative. The form factors obtained in \cite{Pawlowski:2023dda} exhibit a different asymptotic behavior from those shown in Fig.~\ref{plotFFASeps1}. They approach a constant with quadratic momentum corrections in the IR, $f(q^2\to0)\sim c_{\rm IR}+q^2$, and a constant with corrections suppressed as $1/q^2$ in the UV, $f(q^2\to\infty)\sim c_{\rm UV}+1/q^2$. In contrast, the form factors obtained here exhibit also logarithmic terms. Regarding their signs, the form factor associated with the Ricci scalar sector has the same sign as the corresponding result shown in Fig.~\ref{plotFFASeps1}, remaining positive over the entire momentum range. Conversely, the form factor associated with the Ricci tensor sector remains negative in \cite{Pawlowski:2023dda}, whereas the corresponding background form factor obtained here changes sign between the IR and UV regimes. These differences may reflect both the different nature of background and fluctuation correlation functions and the different approximations employed in the respective calculations.

The form factors obtained in \cite{Assant:2026dca} are expressed in the Weyl basis and are computed directly in Lorentzian signature. They are therefore not directly quantitatively comparable with the background form factors considered here. Nevertheless, their IR behavior exhibits the same logarithmic momentum dependence, $f(q^2\to0)\sim a_0+b_0\ln(q^2)$, although with different numerical values of $a_0$ and $b_0$. In the UV, both results decay as $1/q^2$. In the present case, however, this decay is modulated by a logarithmic correction, $f_{k=0,\mathrm{AS}}(q^2\to\infty)\sim[a+b\ln(q^2)]/q^2$, which is absent in the result of \cite{Assant:2026dca}.

\section{Conclusions}\label{secconc}
\noindent In this work, we have shown how the running of the gravitational form factors in Eq.~(\ref{TE}) can be computed within the proper-time formalism using the one-loop-improved flow equation and the nonlocal heat-kernel expansion of Barvinsky and Vilkovisky. We considered the case of a negative, nonrunning cosmological constant and derived the flow equation for the form factors in the presence of a running
Newtonian coupling,
\[
k\partial_k f_k(q^2,\bar\lambda)=A_k(q^2,\bar\lambda)
+(2-\epsilon\eta_k)B_k(q^2,\bar\lambda).
\]
Integrating this flow equation down to $k=0$ gives the physical momentum-dependent form factors, $f_{k=0}(q^2,\bar\lambda)$, which constitute the first main result of this work. The second main result follows from the corresponding dimensionless flow equation, Eq.~(\ref{floweqadim}), which allows us to identify the fixed-point structure and construct the associated asymptotically safe form factors. Their RG flow, and in particular their physical limit at $k=0$, provides the second central result of the analysis.

In the absence of a running Newton coupling ($\eta_k=0$), the flow equation for the form factors reproduces at $k=0$ the standard effective field theory results of \cite{tHooft:1974toh} (for $\bar\lambda=0$) and \cite{Codello:2015mba} (for $\bar\lambda\neq0$).

In the presence of a running Newton coupling, the physical form factors at $k=0$, shown in Figs.~\ref{plotFFstan} and \ref{plotFFstaneps1}, exhibit a nontrivial momentum dependence that interpolates between the EFT regime and the UV regime, with the Planck scale setting the crossover between the two. In the IR regime, $q^2\ll m_p^2$, the corrections to the EFT behavior are organized as an expansion in powers of $q^2/m_p^2$, as given in Eq.~(\ref{IRserie}). In the UV regime, $q^2\gg m_p^2$, the form factors instead admit an expansion in powers of $m_p^2/q^2$, as given in Eq.~(\ref{UVseries}). This nontrivial momentum dependence leads to an intrinsically nonlocal structure of the UV effective action.

The qualitative momentum dependence is insensitive to the $\epsilon$ prescription used for the running of Newton's coupling. While the numerical coefficients exhibit a weak dependence on $\epsilon$, the coefficients of the logarithmic terms are universal and independent of the RG scheme.

The dimensionless formulation reveals that the bulk form-factor flow connects two fixed-point solutions associated with the Gaussian and nontrivial fixed points of the Newtonian coupling. The Plank scale sets the crossover between the two form factor fixed point regimes. The form factor fixed point solution associated with the Gaussian fixed point ($g_*=0$) controls the IR regime and reproduces the effective field theory behavior, while the form factor fixed point solution associated with the nontrivial fixed point ($g_*\ne0$) controls the UV regime. The resulting interpolation between the two regimes is illustrated in Fig.~\ref{plotflowadbulk}.

The integrated asymptotically safe flow, obtained by imposing a boundary condition at $k=\Lambda$, exhibits a logarithmic dependence on the boundary scale $\Lambda$. In the limit $\Lambda\to\infty$, this dependence can be removed by defining a renormalized asymptotically safe flow, for which the boundary condition coincides with the UV fixed-point solution. This provides a nonperturbative renormalization prescription based on the UV fixed point. The resulting asymptotically safe form factors are independent of both the auxiliary boundary scale $\Lambda$ and the renormalization scale $\mu$. This flow is illustrated in Fig.~\ref{plotflowad}.

The renormalized asymptotically safe form factors remain dependent on the choice of boundary conditions used to solve the fixed-point equation. Different choices can modify the asymptotic behavior without affecting the universal features of the flow. Within the background approximation employed here, these boundary conditions are not fixed dynamically. However, one may expect that in a more complete framework including fluctuation form factors, consistency conditions between background and fluctuation sectors could determine them.

The renormalized asymptotically safe solutions at $k=0$, shown in Figs.~\ref{plotFFAS} and \ref{plotFFASeps1}, retain the same infrared momentum dependence as the nonasymptotically safe form factors, with $\Lambda$ replaced by $m_p$, as given in Eq.~(\ref{formASIR}). In the UV, the pure logarithmic term is absent and the form factors are described by a power series in $m_p^2/q^2$, as given in Eq.~(\ref{ASUV}). Comparison with previous results \cite{Knorr:2021niv,Pawlowski:2023dda,Assant:2026dca} shows that several qualitative features of the momentum dependence are shared across different approaches, despite the differences between background and fluctuation form factors and between Euclidean and Lorentzian formulations.

An important extension of the present analysis would be to include the running of the cosmological constant. In this case, the phase diagram in the $(\lambda_k,g_k)$ plane exhibits different RG trajectories flowing toward the nontrivial Reuter fixed point for $g_k>0$ \cite{Reuter:2001ag}. This would allow the form factors to be studied along more general asymptotically safe RG trajectories and would lead to different classes of asymptotically safe solutions distinguished by their infrared behavior at $k=0$. In this broader setting, it would be interesting to determine how the form factors depend on the choice of RG trajectory, in particular along the critical trajectory connecting the Gaussian and Reuter fixed points.

The results presented in this work rely on a one-loop improved flow equation and on the nonlocal heat-kernel expansion evaluated on the background metric, as discussed in Sec.~\ref{subsecappro}. Within this approximation, higher-derivative operators are not fed back into the Hessian and the analysis is restricted to the background effective action. Nevertheless, the framework reproduces the known effective results in the appropriate limits and captures the expected momentum dependence and asymptotic structure of the form factors.

Extending the analysis beyond the present approximation is necessary to assess which features of the form-factor flow are robust. A particularly important extension is the inclusion of fluctuation form factors, which requires going beyond the background-field approximation and solving a coupled system of integrodifferential equations. This would allow one to test whether the momentum dependence and fixed-point structure found here persist in the fluctuation sector and to clarify the relation between background and fluctuation form factors. Developing suitable techniques for this problem remains an important open challenge.

A further test would be provided by an exact FRG treatment. Since the proper-time flow equation does not define an exact renormalization-group flow for the effective average action, repeating the analysis within an exact FRG framework would allow one to disentangle genuinely scheme-independent features from truncation-dependent effects. In particular, it would test the persistence of the ultraviolet regularization mechanism and of the reconstructed dimensionful form factors beyond the present approximation. An independent cross-check would be provided by comparing the form factors obtained here with those reconstructed from the properties of the graviton propagator \cite{Bonanno:2021squ,Fehre:2021eob,Pawlowski:2025etp,Pawlowski:2023gym,Kher:2025rve}.

The physical implications of the form factors derived in this work, including the structure of the Minkowskian propagator and possible phenomenological consequences, will be presented in a forthcoming companion paper.

\section*{Acknowledgements}
\noindent E.G. would like to thank Diego Buccio and Frank Saueressig for interesting discussions on AS form factors and the formalism required to derive them. E.G. also thanks Alfio Bonanno and Oleg Melichev for valuable discussions and for the physical insights they provided. 

\section*{Data availability}
\noindent There are no publicly available research data  or software supporting this manuscript. Requests for further information or data should be sent to the author.

\newpage

\begin{appendices}

\section{Derivation of the flow equations}\label{appderbeta}
\noindent In this Appendix we derive the proper-time flow equations for the action in Eq.~(\ref{TE}). We first outline the approximation scheme employed and then present the derivation of the flow equation.

\subsection{The Hessian of the theory}
The flow equations are derived within the background field formalism using the linear split
\begin{equation}
g_{\mu\nu}=\bar g_{\mu\nu}+h_{\mu\nu}\,,
\end{equation}
where barred quantities denote background fields. 

For later convenience, we introduce the parametrization
\begin{equation}
\frac{1}{16\pi G_k}=\frac{Z_{Nk}}{16\pi G_0}=2\kappa^2 Z_{Nk}\,,
\end{equation}
where $Z_{Nk}$ denotes the wave-function renormalization associated with Newton's constant.

The gravitational action is supplemented by the standard gauge-fixing term,
\begin{equation}
S_{\mathrm{gf}}\left[h;\bar{g}\right]=\frac{1}{\alpha}\int{d^dx\sqrt{\bar{g}}{\bar{g}}^{\mu\nu}F_\mu\left(h;\bar{g}\right)F_\nu\left(h;\bar{g}\right)}
\end{equation}
where
\begin{equation}\begin{split}
&F_\mu\left(h;\bar{g}\right)=\kappa\mathcal{F}_\mu^{\alpha\beta}\left[\bar{g}\right]h_{\alpha\beta},\\ &\mathcal{F}_\mu^{\alpha\beta}\left[\bar{g}\right]h_{\alpha\beta}=\left(\delta_\mu^\beta{\bar{g}}^{\alpha\gamma}{\bar{\nabla}}_\gamma-\bar{\omega}{\bar{g}}^{\alpha\beta}{\bar{\nabla}}_\mu\right)h_{\alpha\beta}={\bar{\nabla}}^\alpha h_{\alpha\mu}-\bar{\omega}{\bar{\nabla}}_\mu h
\end{split}\end{equation}
here $\alpha$ and $\bar\omega$ are the gauge parameters.

The gauge fixing yields the standard ghost action term
\begin{equation}
S_\mathrm{gh}\left[h;\bar{g}\right]=-\frac{1}{\kappa}\int{d^dx\sqrt{\bar{g}}{\bar{C}}_\mu{\bar{g}}^{\mu\nu}\frac{\partial F_\nu}{\partial h_{\alpha\beta}}\mathcal{L}_C\left({\bar{g}}_{\alpha\beta}+h_{\alpha\beta}\right)},
\end{equation}
the Lie derivative is given by $\mathcal{L}_vT_{\alpha\beta}=v^\rho\partial_\rho T_{\alpha\beta}+T_{\rho\beta}\partial_\alpha v^\rho+T_{\rho\alpha}\partial_\beta v^\rho$. In what follows we assume that the ghost action does not run.

In order to obtain the Hessian, we compute
\begin{equation}
\delta^2\mathcal{L}=\delta^2\left\{\sqrt g\left[\frac{1}{16\pi G_k}\left(-R+2\bar\lambda_k\right)+R f_k^{(\mathrm{R})}(-\Box)R+R_{\mu\nu}f_k^{(\mathrm{Ricc})}(-\Box)R^{\mu\nu}\right]\right\}+\delta^2\mathcal{L}_\mathrm{gf}+\delta^2\mathcal{L}_\mathrm{gh}\,.
\end{equation}
Denoting collectively the curvature by $\mathcal{R}=\{R, R_{\mu\nu}\}$, the second variation of the form-factor terms reads \footnote{To obtain Eq.~(\ref{var2lagFF}) we use $\int d^dx\sqrt g\,\mathcal R\,f(\Delta)\,\delta\mathcal R=\int d^dx\sqrt g\,(\delta\mathcal R)\,f(\Delta)\,\mathcal R+\mathrm{cov.div.}$, which is readily verified using the Laplace representation $f(\Delta)=\int_{-\infty}^{\infty}ds\,f(s)e^{-s\Delta}$.}
\begin{equation}\begin{split}
\label{var2lagFF}
&\delta^2\mathcal{L}_f
=(\delta^2\sqrt g)\,\mathcal{R}f(-\Box)\mathcal{R}
+2(\delta\sqrt g)\left[\mathcal{R}(\delta f(-\Box))\mathcal{R}
+2(\delta\mathcal{R})f(-\Box)\mathcal{R}\right]+\\
&+\sqrt g\bigg\{
4(\delta\mathcal{R})(\delta f(-\Box))\mathcal{R}
+2\left[(\delta\mathcal{R})f(-\Box)(\delta\mathcal{R})
+(\delta^2\mathcal{R})f(-\Box)\mathcal{R}\right]+\mathcal{R}(\delta^2 f(-\Box))\mathcal{R}
\bigg\}\,.
\end{split}\end{equation}
Collecting all contributions, we obtain
\begin{equation}\begin{split}
\label{secvaract}
&\delta^2\mathcal{L}=-\left(\delta^2\sqrt g\right)\left(\kappa^2Z_{Nk}\left(R-2{\bar{\lambda}}_k\right)-Rf_k^{\left(R\right)}\left(-\Box\right)R-R_{ab}f_k^{\left(Ricc\right)}\left(-\Box\right)R^{ab}\right)-\\
&-\left(2\left(\delta\sqrt g\right)\delta R+\delta^2R\right)\left(\kappa^2Z_{Nk}-2f_k^{\left(R\right)}\left(-\Box\right)R\right)+\\
&+\sqrt g\left[2\left(\delta R\right)f_k^{\left(R\right)}\left(-\Box\right)\left(\delta R\right)+2\left(\delta R_{ab}\right)f_k^{\left(Ricc\right)}\left(-\Box\right)\left(\delta R^{ab}\right)\right]+\\
&+\left(\delta\sqrt g\right)4\left(\delta R_{ab}\right)f_k^{\left(\mathrm{Ricc}\right)}\left(-\Box\right)R^{ab}+\sqrt g2\left(\delta^2R_{ab}\right)f_k^{\left(\mathrm{Ricc}\right)}\left(-\Box\right)R^{ab}+\\
&+2\left(\delta\sqrt g\right)\left[R\left(\delta f_k^{\left(R\right)}\left(-\Box\right)\right)R+R_{ab}\left(\delta f_k^{\left(\mathrm{Ricc}\right)}\left(-\Box\right)\right)R^{ab}\right]+\\
&+\sqrt g\left\{4\left[\left(\delta R\right)\left(\delta f_k^{\left(R\right)}\left(-\Box\right)\right)R+\left(\delta R_{ab}\right)\left(\delta f_k^{\left(\mathrm{Ricc}\right)}\left(-\Box\right)\right)R^{ab}\right]\right\}\\
&+\sqrt g\left[R\left(\delta^2f_k^{\left(R\right)}\left(-\Box\right)\right)R+R_{ab}\left(\delta^2f_k^{\left(\mathrm{Ricc}\right)}\left(-\Box\right)\right)R^{ab}\right]+\delta^2\mathcal{L}_{gf}+\delta^2\mathcal{L}_{gh}
\end{split}\end{equation}
in the next lines, we discuss separately each piece.

The first and second lines contain the Einstein-Hilbert contribution and the corrections induced to it by the form-factors. Including the gauge-fixing term, one obtains,
\begin{equation}\begin{split}
&\Gamma_{k}^{(2),EH}\left[g\right]=\kappa^2Z_{Nk}\int{d^dx}\sqrt{\bar{g}}h_{\mu\nu}\Bigg\{-K_{\rho\sigma}^{\mu\nu}\left(\alpha;\omega\right)\Box+U_{\rho\sigma}^{\mu\nu}\left(-\Box\right)+\\
&+\left(1-\frac{2f_k^{\left(R\right)}\left(-\Box\right)R}{\kappa^2Z_{Nk}}-\frac{1}{\alpha}\right)\delta_\rho^\nu{\bar{\nabla}}^\mu{\bar{\nabla}}_\sigma-\left(1-\frac{2f_k^{\left(R\right)}\left(-\Box\right)R}{\kappa^2Z_{Nk}}-\frac{2\bar{\omega}}{\alpha}\right){\bar{g}}^{\mu\nu}\nabla_\rho{\bar{\nabla}}_\sigma\Bigg\}h^{\rho\sigma}
\end{split}\end{equation}
where
\begin{equation}\begin{split}
&K_{\rho\sigma}^{\mu\nu}\left(\alpha,\bar{\omega}\right)=\frac{1}{4}\left(1-\frac{2f_k^{\left(\mathrm{R}\right)}\left(-\Box\right)R}{\kappa^2Z_{Nk}}\right)\left(\delta_\rho^\mu\delta_\sigma^\nu+\delta_\sigma^\mu\delta_\rho^\nu\right)-\left(\frac{1}{2}\left(1-\frac{2f_k^{\left(\mathrm{R}\right)}\left(-\Box\right)R}{\kappa^2Z_{Nk}}\right)-\frac{{\bar{\omega}}^2}{\alpha}\right){\bar{g}}^{\mu\nu}{\bar{g}}_{\rho\sigma}\\
&U_{\rho\sigma}^{\mu\nu}\left(-\Box\right)=\frac{1}{4}\left(\delta_\rho^\mu\delta_\sigma^\nu+\delta_\sigma^\mu\delta_\rho^\nu-g^{\mu\nu}g_{\rho\sigma}\right)\left(R-2{\bar{\lambda}}_k-\frac{Rf_k^{\left(\mathrm{R}\right)}\left(-\Box\right)R}{\kappa^2Z_{Nk}}-\frac{R_{ab}f_k^{\left(\mathrm{Ricc}\right)}\left(-\Box\right)R^{ab}}{\kappa^2Z_{Nk}}\right)+\\
&+\left[-\frac{1}{4}\left(R_\sigma^\nu\delta_\rho^\mu+R_\sigma^\mu\delta_\rho^\nu+R_\rho^\nu\delta_\sigma^\mu+R_\rho^\mu\delta_\sigma^\nu\right)+g^{\mu\nu}R_{\rho\sigma}-\frac{1}{2}\left(R_{\rho\sigma}^{\mu\nu}+R_{\sigma\rho}^{\mu\nu}\right)\right]\left(1-\frac{2f_k^{\left(R\right)}\left(-\Box\right)R}{\kappa^2Z_{Nk}}\right)
\end{split}\end{equation}
the Hessian of ghost action is given by
\begin{equation}
S^{(2)}_\mathrm{gh}=-\int{d^dx\sqrt{\bar{g}}\left[{\bar{C}}_\mu\left(-\bar \Box+\frac{\bar{R}}{d}\right)C^\mu+{\bar{C}}_\mu\left(1-2\bar{\omega}\right){\bar{\nabla}}^\mu{\bar{\nabla}}_\nu C^\nu\right]}
\end{equation}
in these pieces, only the form factor related to Ricci scalar gives a contribution to the Einstein-Hilbert piece.

The third and fourth line of Eq.(\ref{secvaract}) can be written in the form
\begin{equation}
\int{d^dx\sqrt g\left\{h_{\mu\nu}\left[H_{\rho\sigma}^{\mu\nu}\left(-\Box\right)+\left(f_k^{\left(\mathrm{Ricc}\right)}\left(-\Box\right)R^{ab}\right)L_{\rho\sigma ab}^{\mu\nu}\left(-\Box\right)\right]h^{\rho\sigma}\right\}}
\end{equation}
where
\begin{equation}\begin{split}
&H_{\rho\sigma}^{\mu\nu}\left(-\bar\Box\right)=2\left(\bar R^{\mu\nu}-\bar\nabla^\mu\bar\nabla^\nu+\bar g^{\mu\nu}\bar\Box\right)f_k^{\left(\mathrm{R}\right)}\left(-\bar\Box\right)\left(\bar R_{\rho\sigma}-\bar\nabla_\rho\bar\nabla_\sigma+\bar g_{\rho\sigma}\bar\Box\right)+\\
&+\frac{1}{2}\left(\bar g^{\mu\nu}\bar\nabla_a\bar\nabla_b-\bar g_b^\mu\bar\nabla_a\bar\nabla^\nu-\bar g^\mu_{\ a}\bar\nabla_b\bar\nabla^\nu+\bar g_a^\nu\bar g_b^\mu\bar\Box\right)f_k^{\left(\mathrm{Ricc}\right)}\left(-\bar\Box\right)\left(\bar g_{\rho\sigma}\bar\nabla^a\bar\nabla^b-\bar g_\sigma^b\bar\nabla^a\bar\nabla_\rho-\bar g_\sigma^a\bar\nabla^b\bar\nabla_\rho+\bar g_\rho^a\bar g_\sigma^b\bar\Box\right)\\
&L_{\rho\sigma ab}^{\mu\nu}\left(-\bar\Box\right)=\bar g^{\mu\nu}
\left(-\bar g_{\rho\sigma}\bar\nabla_a\bar\nabla_b+\bar g_{\sigma b}\bar\nabla_a\bar\nabla_\rho+\bar g_{\sigma a}\bar\nabla_b\bar\nabla_\rho-\bar g_{\rho a}\bar g_{\sigma b}\bar\Box\right)+\bar g_{a\sigma}\left(\bar g^{\mu\nu}\bar g_{b\rho}+\bar g_b^\nu\bar g_\rho^\mu\right)\bar\Box\\
&+\bar g_\rho^\mu\bar g_\sigma^\nu\bar\nabla_a\bar\nabla_b-\bar g_a^\nu\bar g_{\rho\sigma}\bar\nabla_b\bar\nabla^\mu-\bar g_b^\nu\bar g_{\rho\sigma}\bar\nabla_a\bar\nabla^\mu\\
&-2\Big(\bar g_{b\rho}\bar g_\sigma^\mu\bar\nabla^\nu\bar\nabla_a+\bar g_a^\mu\big(\bar g_b^\nu\bar\nabla_\sigma\bar\nabla_\rho+\bar g_\rho^\nu\bar\nabla_b\bar\nabla_\sigma\big)+\bar g_{a\rho}\big(\bar g_\sigma^\mu\bar\nabla^\nu\bar\nabla_b-\bar g_{b\sigma}\bar\nabla^\nu\bar\nabla^\mu\big)+\bar g_b^\nu\big(\bar g_\rho^\mu\bar\nabla_a\bar\nabla_\sigma+\bar g_{a\rho}\bar\nabla_\sigma\bar\nabla^\mu\big)\Big)\,.
\end{split}\end{equation}
These terms introduce two additional derivatives acting on $h_{\mu\nu}$, together with nonlocal interactions. In flat spacetime, $f_k^{(\mathrm{Ricc})}(-\Box) R^{ab}=0$, and the derivatives in $H_{\rho\sigma}^{\mu\nu}(-\Box)$ commute with the form factors. This reproduces the results of \cite{Modesto:2011kw,Biswas:2011ar,Frolov:2015usa,Biswas:2013cha}.

The terms involving the variation of the form factors in Eq.~(\ref{secvaract}) do not vanish in curved spacetimes. They can be computed using the technology developed in \cite{Knorr:2019atm}. The two contributions are given, respectively, by
\begin{equation}\begin{split}
&\int d^dx\sqrt g\,\mathcal{R}_1\,\delta f\left(\Delta\right)\mathcal{R}_2
=\int d^dx\sqrt g\,\frac{f\left(\Delta_1\right)-f\left(\Delta_2\right)}{\Delta_1-\Delta_2}\left(\delta\Delta_2\right)\mathcal{R}_1\mathcal{R}_2,\\
&\int d^dx\sqrt g\,\mathcal{R}_1\,\delta^2 f\left(\Delta\right)\mathcal{R}_2=\\
&=\int d^dx\sqrt g\,\bigg\{
2\left[\frac{f\left(\Delta_1\right)-f\left(\Delta_2\right)}{\left(\Delta_1-\Delta_2\right)^2}
-\frac{f'\left(\Delta_2\right)}{\Delta_1-\Delta_2}\right]\left(\delta\Delta_2\right)^2+\frac{f\left(\Delta_1\right)-f\left(\Delta_2\right)}{\Delta_1-\Delta_2}\,\delta^2\Delta_2
\bigg\}\mathcal{R}_1\mathcal{R}_2.
\end{split}\end{equation}
Here a prime denotes a derivative with respect to $\Delta=-\Box$, and the subscripts $1$ and $2$ indicate on which curvature tensor the operator $\Box$ acts. The explicit expressions in terms of $h_{\mu\nu}$ are rather lengthy and will not be displayed here. 


\subsection{The propertime flow}
To bring the Hessian in the standard form, we choose to work with the Lorentz gauge $\alpha=1$ and $\bar{\omega}=1/2$. In this way the propertime flow equations are obtained from
\begin{equation}\begin{split}
\label{PTFLOW}
k&\partial_k\Gamma_k\left[g\right]=-\frac{1}{2}\int_{0}^{+\infty}\frac{ds}{s}\rho_m^{\mathrm{grav}}\left(s,2\kappa^2Z_{Nk}k^2\right)\mathrm{Tr}
\left[e^{-s2\kappa^2Z_{Nk}(-\Box\mathbf{1}_T-2\bar\lambda_k+\mathbf{U}^{EH})}\right]+\\
&+\int_{0}^{+\infty}\frac{ds}{s}\rho_m^{\mathrm{ghost}}\left(s,k^2\right)\mathrm{Tr}\left[e^{-s(-\Box\mathbf{1}_V+\mathbf{U}^{gh})}\right]
\end{split}\end{equation}
where $\mathbf{1}_T$ and $\mathbf{1}_V$ refer to the identity in the spin $2$-tensor and vectorial space. The two regulators are given by
\begin{equation}\begin{split}
&\rho_m^{\mathrm{grav}}(s,2\kappa^2Z_{Nk}k^2)= \left(2-\epsilon\eta_k\right)\frac{(s\, m\, k^2 2\kappa^2Z_{Nk})^m}{\Gamma(m)}e^{-s\, m\, k^2 2\kappa^2Z_{Nk}}\,,\\
&\rho_m^{\mathrm{ghost}}(s,k^2)= 2\frac{(s\, m\, k^2)^m}{\Gamma(m)}e^{-s\, m\, k^2}\,,
\end{split}\end{equation}
where $\eta_k=-k\partial_k Z_{Nk}/Z_{Nk}$.

It is convenient to absorb the overall factor $2\kappa^2Z_{Nk}$ of the gravitational sector in the propertime variable by performing the substitution $s^\prime=s2\kappa^2Z_{Nk}$ in the first integral. Since the propertime measure is invariant under this transformation, $ds/s=ds'/s'$ and $\rho_m^{\mathrm{grav}}(s,2\kappa^2Z_{Nk}k^2)=\rho_m^{\mathrm{grav}}(s',k^2)$, we get, after renaming $s'\to s$,
\begin{equation}\begin{split}
\label{PTflowsimp}
&k\partial_k\Gamma_k\left[g\right]=-\frac{1}{2}\int_{0}^{+\infty}\frac{ds}{s}\rho_m^{\mathrm{grav}}\left(s,k^2\right)e^{2s\bar\lambda_k}\mathrm{Tr}\left[e^{-s(-\bar \Box\mathbf{1}_T+\mathbf{U}^{EH})}\right]+\\
&+\int_{0}^{+\infty}\frac{ds}{s}\rho_m^{\mathrm{ghost}}\left(s,k^2\right)\mathrm{Tr}\left[e^{-s(-\bar \Box\mathbf{1}_V+\mathbf{U}^{gh})}\right]
\end{split}\end{equation}
As a consequence, the Newton coupling does not appear as an explicit overall multiplicative factor in the resulting flow equation. Its scale dependence is nevertheless retained through the anomalous dimension $\eta_k$ in $\rho_m^{\mathrm{grav}}\left(s,k^2\right)$. The resulting expression is now in a standard form that can be evaluated using heat kernel techniques.

The left-hand side of the flow equation is given by
\begin{equation}\begin{split}
\label{LHSFLOW}
&k\partial_k\Gamma_k=\int{d^dx\sqrt g\left[-R(2\kappa^2k\partial_k Z_{Nk})+2k\partial_k\left(Z_{Nk}{\bar{\lambda}}_k\right)\right]}+\\
&+\int d^dx \sqrt x\left[R (k\partial_k f_k^{(\mathrm{R})}(-\Box))R+R_{\mu\nu} (k\partial_k f_k^{(\mathrm{Ricc})}(-\Box))R^{\mu\nu}\right]
\end{split}\end{equation}
to get the flow equations, the right-hand side of Eq.(\ref{PTFLOW}) has to be expanded in powers of curvatures or equivalently in powers of $h_{\mu\nu}$.

The expansion in curvatures can be obtained via a vertex expansion of the effective action. On the right-hand side this induces an expansion of the heat kernel trace,
\begin{equation}\begin{split}
&\mathrm{Tr}\left[e^{-s(- \Box\mathbf{1}+\mathbf{U})}\right]=\left.\mathrm{Tr}\left[e^{-s(- \Box\mathbf{1}+\mathbf{U})}\right]\right|_{h=0}+\int d^dx \sqrt{\bar g} \left.\frac{\delta \mathrm{Tr}\left[e^{-s(- \Box\mathbf{1}+\mathbf{U})}\right]}{\delta h_{\mu\nu}(x)}\right|_{h=0}\delta h_{\mu\nu}(x)+\\
&+\frac{1}{2}\int d^dx d^dy \sqrt{\bar{g}(x)\bar{g}(y)} \left.\frac{\delta^2 \mathrm{Tr}\left[e^{-s(- \Box\mathbf{1}+\mathbf{U})}\right]}{\delta h_{\mu\nu}(x)\delta h_{\rho\sigma}(y)}\right|_{h=0}\delta h_{\mu\nu}(x)\delta h_{\rho\sigma}(y)+\cdots.
\end{split}\end{equation}
The first line contains the background contribution together with the term linear in the fluctuation field, while the second line collects the quadratic contribution. 
The flow of the Newtonian and cosmological constants is obtained from the background contribution, whereas the flow of the form factors arises from the second-order term. The computation of these derivatives is rather involved and was developed in detail in \cite{Codello:2012kq}. Keeping the full Hessian structure, the second functional derivative of the trace can be decomposed diagrammatically. In momentum space this yields schematically,
\begin{equation}
\label{integroDE}
\frac{\delta^2 \mathrm{Tr}\left[e^{-s(- \Box\mathbf{1}+\mathbf{U})}\right]}{\delta h_{\mu\nu}(-q)\delta h_{\rho\sigma}(q)}\sim \int \frac{d^dp}{(2\pi)^d} \, e^{-s \Gamma_k^{(2)}(p^2)}\,\Gamma_k^{(4)}(p,q)+ \int \frac{d^dp}{(2\pi)^d} \, e^{-s \Gamma_k^{(2)}(p^2)}\,\Gamma_k^{(3)}(p,q)\,e^{-s \Gamma_k^{(2)}((p+q)^2)}\,\Gamma_k^{(3)}(p,q) \, .
\end{equation}
Here $\Gamma_k^{(3)}$ and $\Gamma_k^{(4)}$ denote the three- and four-graviton vertices obtained from the scale-dependent effective action, and $\Gamma_k^{(2)}(p^2)$ is the inverse propagator (including the effect of the form factors). The two terms correspond to the usual one-loop topologies: a tadpole diagram and a two-vertex (self-energy) diagram. The momentum $p$ represents the loop momentum. Inserting this structure into the vertex expansion of the proper-time flow equation yields a system of coupled integrodifferential equations for the running form factors.

These integrodifferential equations are highly non-linear due to the implicit dependence of $\Gamma_k^{(n)}$ on the running form factors. To obtain a tractable system, we neglect the dependence of the form factors in the Hessian entering the second variation of the trace and evaluate the corresponding vertices within the Einstein-Hilbert truncation. Diagrammatically, this amounts to retaining only the leading contribution in which the nonlocal form factors do not feed back into the inverse propagators of Eq.~(\ref{integroDE}). Within this approximation, the vertex expansion in powers of the fluctuation field $h_{\mu\nu}$ can be reorganized into a curvature expansion \cite{Codello:2012kq}, allowing the use of the standard nonlocal heat-kernel expansion of Barvinsky and Vilkovisky \cite{Barvinsky:1990up, Barvinsky1987BeyondTS, Barvinsky:1990uq, Barvinsky:1993en, Avramidi:2000bm}. Inserting this expansion into the proper-time flow equation then yields a closed system for the running form factors.

The nonlocal heat-kernel expansion of Barvinsky and Vilkovisky then reads,
\begin{equation}\begin{split}
\label{NLHK}
&\mathrm{Tr}\left[e^{-s(-\bar\Box\mathbf{1}+\bar{\mathbf{U}})}\right]=\\
&=\frac{1}{(4\pi s)^{\frac{d}{2}}}\int d^dx\sqrt{\bar g}\,\mathrm{tr} \bigg\{
\mathbf{1}
+s\,\frac{\bar R}{6}\mathbf{1}
-s\,\bar{\mathbf{U}}+s^2\Big[
\bar R\,f_R\!\left(-s\bar\Box\right)\bar R\,\mathbf{1}
+\bar R_{\mu\nu}f_{Ric}\!\left(-s\bar\Box\right)\bar R^{\mu\nu}\mathbf{1}+\\
&\qquad +\bar R\,f_{RU}\!\left(-s\bar\Box\right)\bar{\mathbf{U}}
+\bar{\mathbf{U}}\,f_U\!\left(-s\bar\Box\right)\bar{\mathbf{U}}
+\bar\Omega_{\mu\nu}f_\Omega\!\left(-s\bar\Box\right)\bar\Omega^{\mu\nu}
\Big]
+\mathcal{O}(\bar R^3)
\bigg\}.
\end{split}\end{equation}
This is an expansion in curvature terms evaluated on the background metric. Its validity is determined by the relation $\nabla\nabla \mathcal{R}\gg \mathcal{R}$, where $\mathcal{R}$ is any element of the set $\{\mathrm{U},R, R_{\mu\nu},\Omega_{\mu\nu}\}$ \cite{Barvinsky:1990up}. The tensor $\Omega_{\mu\nu}$ is defined as follows: $\left[\nabla_\mu,\nabla_\nu\right]\psi=\Omega_{\mu\nu}\psi$. The heat kernel structure functions are found to be,
\begin{equation}\begin{split}
&f_R\left(x\right)=\frac{1}{32}f\left(x\right)+\frac{1}{8x}f\left(x\right)-\frac{7}{48x}-\frac{1}{8x^2}\left[f\left(x\right)-1\right]\\
&f_{Ric}\left(x\right)=\frac{1}{6x}+\frac{1}{x^2}\left[f\left(x\right)-1\right]\\
&f_{RU}\left(x\right)=-\frac{1}{4}f\left(x\right)-\frac{1}{2x}\left[f\left(x\right)-1\right],\quad \quad f_U\left(x\right)=\frac{1}{2x},\quad \quad f_\Omega\left(x\right)=-\frac{1}{2x}\left[f\left(x\right)-1\right]
\end{split}\end{equation}
where the basic heat kernel structure function $f(x)$ is defined in terms of the parametric integral,
\begin{equation}
f\left(x\right)=\int_{0}^{1}{d\xi e^{-x\xi\left(1-\xi\right)}}
\end{equation}
to compute the flow equations we need to apply this expansion to both pieces of Eq.(\ref{PTflowsimp}). 

In our case
\begin{equation}\begin{split}
&\left(\Omega_{\mu\nu}^{(EH)}\right)_{\ \ \ \   \alpha\beta}^{\gamma\sigma}=-2\bar R_{\ \ \ \ \alpha\mu\nu}^{(\gamma}\delta_{\beta)}^{\sigma)}=-\left(\delta_\alpha^\gamma \bar R_{\mu\nu\ \ \ \beta}^{\ \ \ \ \ \sigma}+\delta_\alpha^\sigma \bar R_{\mu\nu\ \ \ \beta}^{\ \ \ \ \ \gamma}+\delta_\beta^\gamma \bar R_{\mu\nu\ \ \ \alpha}^{\ \ \ \ \ \sigma}+\delta_\beta^\sigma \bar R_{\mu\nu\ \ \ \alpha}^{\ \ \ \ \ \gamma}\right),\ \\
&\left(\Omega_{\mu\nu}^{(gh)}\right)_{\ \ \ \beta}^\alpha=\bar R_{\ \ \ \beta\mu\nu}^\alpha
\end{split}\end{equation}
respectively, for the gravitational and ghost parts.

To simplify the background trace, we decompose the fluctuation $h_{\mu\nu}$ into a traceless part $\tilde h_{\mu\nu}$ and its trace $\phi \equiv \bar g^{\mu\nu} h_{\mu\nu}$,
\begin{equation}
h_{\mu\nu}=\tilde h_{\mu\nu}+\frac{1}{d}\,\bar g_{\mu\nu}\,\phi, 
\qquad \bar g^{\mu\nu}\tilde h_{\mu\nu}=0 \, .
\end{equation}
To further simplify the computation, we work on a maximally symmetric background, for which
\begin{equation}
\bar R_{\mu\nu\rho\sigma}=\frac{\bar R}{d(d-1)}\left(\bar g_{\mu\rho}\bar g_{\nu\sigma}-\bar g_{\mu\sigma}\bar g_{\nu\rho}\right), 
\qquad 
\bar R_{\mu\nu}=\frac{\bar R}{d}\,\bar g_{\mu\nu} \, .
\end{equation}
In this setting, the background trace then yields,
\begin{equation}\begin{split}
&k\partial_k\bar\Gamma_k\left[\bar g\right]=-\frac{1}{2}\int_{0}^{\infty}\frac{ds}{s}\rho_m(s,k^2)\left.\mathrm{Tr}\left[e^{-s(- \Box\mathbf{1}+\mathbf{U})}\right]\right|_{h=0}=-\frac{1}{2}\int_{0}^{+\infty}\frac{ds}{s}\rho_m\left(s,k^2\right)\mathrm{Tr}_T\left[e^{-s(-\bar\Box-2\bar\lambda_k+C_T\bar R)}\right]-\\
&-\frac{1}{2}\int_{0}^{+\infty}\frac{ds}{s}\rho_m\left(s,k^2\right)\mathrm{Tr}_S\left[e^{-s(-\bar\Box-2\bar\lambda_k+C_S\bar R)}\right]+\int_{0}^{+\infty}\frac{ds}{s}\rho_m\left(s,k^2\right)\mathrm{Tr}_V\left[e^{-s(-\bar\Box+C_V \bar R)}\right] \, .
\end{split}\end{equation}
where
\begin{equation}
C_T=\frac{d(d-3)+4}{d(d-1)},\qquad 
C_S=\frac{d-4}{d},\qquad 
C_V=\frac{1}{d} \, .
\end{equation}
The subscripts on the traces indicate the space of fields on which $-\bar\Box$ acts. The corresponding multiplicities are
\begin{equation}
\mathrm{Tr}_T[\mathbf{1}]=\frac{(d-1)(d+2)}{2},\qquad 
\mathrm{Tr}_V[\mathbf{1}]=d, \qquad 
\mathrm{Tr}_S[\mathbf{1}]=1 \, .
\end{equation}
Using Eq.~(\ref{NLHK}) and expanding the result to first order in $\bar R$, comparison with Eq.~(\ref{LHSFLOW}) yields the flow of Newton's constant and the cosmological constant.

The use of a maximally symmetric background serves only to unambiguously project onto the invariants $\int d^dx \sqrt{\bar g}$ and $\int d^dx \sqrt{\bar g}\,\bar R$. The resulting flow equations are independent of this choice of background. For the nonlocal terms of the form $\mathcal{R}\, f(-\Box)\, \mathcal{R}$, such a restriction is not necessary, and in the following we return to a general background metric.

Computing the second-order terms in the curvature expansion of Eq.~(\ref{NLHK}) we obtain \cite{book}
\begin{equation}
\begin{aligned}
{\mathrm{tr}}_T\!\left[\bar R\,f_R\!\left(-s\bar\Box\right)\bar R\right]
&= \frac{d(1+d)}{2}\,\bar R\,f_R\!\left(-s\bar\Box\right)\bar R,
\quad &
{\mathrm{tr}}_T\!\left[\bar R_{\mu\nu}f_{Ric}\!\left(-s\bar\Box\right)\bar R^{\mu\nu}\right]
&= \frac{d(1+d)}{2}\,\bar R_{\mu\nu}f_{Ric}\!\left(-s\bar\Box\right)\bar R^{\mu\nu}
\\[6pt]
{\mathrm{tr}}_V\!\left[\bar R\,f_R\!\left(-s\bar\Box\right)\bar R\right]
&= d\,\bar R\,f_R\!\left(-s\bar\Box\right)\bar R,
\quad &
{\mathrm{tr}}_V\!\left[\bar R_{\mu\nu}f_{Ric}\!\left(-s\bar\Box\right)\bar R^{\mu\nu}\right]
&= d\,\bar R_{\mu\nu}f_{Ric}\!\left(-s\bar\Box\right)\bar R^{\mu\nu}
\end{aligned}
\end{equation}
and
\begin{equation}\begin{split}
&\mathrm{tr}\left[\bar R\, f_{RU}\!\left(-s\bar\Box\right)\bar{\mathbf{U}}^{(EH)}\right]
=\frac{d(d-1)}{2}\,\bar R\,f_{RU}\!\left(-s\bar\Box\right)\bar R,\\
&\mathrm{tr}\left[\bar R\, f_{RU}\!\left(-s\bar\Box\right)\bar{\mathbf{U}}^{(gh)}\right]
=-\,\bar R\,f_{RU}\!\left(-s\bar\Box\right)\bar R,
\end{split}\end{equation}
and
\begin{equation}\begin{split}
&\mathrm{tr}\left[\bar{\mathbf{U}}^{(EH)} f_U\!\left(-s\bar\Box\right)\bar{\mathbf{U}}^{(EH)}\right]=\\
&\quad=\frac{d^2-8d+4}{d-2}\,\bar R_{cd}f_U\!\left(-s\bar\Box\right)\bar R^{cd}
+\frac{4+8d-5d^2+d^3}{2(d-2)}\,\bar R\,f_U\!\left(-s\bar\Box\right)\bar R
+3\,\bar R_{cdef}f_U\!\left(-s\bar\Box\right)\bar R^{cdef},\\
&\mathrm{tr}\left[\bar{\mathbf{U}}^{(gh)} f_U\!\left(-s\bar\Box\right)\bar{\mathbf{U}}^{(gh)}\right]
=\bar R_{ab}f_U\!\left(-s\bar\Box\right)\bar R^{ab},
\end{split}\end{equation}
and
\begin{equation}\begin{split}
&\mathrm{tr}\left[\Omega_{\mu\nu}^{(EH)} f_{\Omega}\!\left(-s\bar\Box\right)\Omega^{(EH)\,\mu\nu}\right]
=-(d-2)\,\bar R_{abcd}f_{\Omega}\!\left(-s\bar\Box\right)\bar R^{abcd},\\
&\mathrm{tr}\left[\Omega_{\mu\nu}^{(gh)} f_{\Omega}\!\left(-s\bar\Box\right)\Omega^{(gh)\,\mu\nu}\right]
=\bar R_{abcd}f_{\Omega}\!\left(-s\bar\Box\right)\bar R^{abcd}.
\end{split}\end{equation}

To eliminate the Riemann tensor we use the identity \cite{Codello:2012kq}
\begin{equation}
\label{idRieman}
\bar R_{abcd}f(-s\bar\Box)\bar R^{abcd}
=4\,\bar R_{ab}f(-s\bar\Box)\bar R^{ab}
-\bar R\,f(-s\bar\Box)\bar R+\mathcal{O}(\bar R^3).
\end{equation}
Collecting all contributions, we obtain
\begin{equation}\begin{split}
&\frac{\delta^2 \mathrm{Tr}\left[e^{-s(- \bar\Box\mathbf{1}+\mathbf{U}^{\mathrm{(EH)}})}\right]}{\delta h_{\mu\nu}(-q)\delta h_{\rho\sigma}(q)}=\\
&=\frac{2s^2}{(4\pi s)^{d/2}}\int d^dx\sqrt{\bar g}\,\bigg[
\bar R\left(\frac{d(1+d)}{4}f_R+\frac{d(d-1)}{4}f_{RU}
+\frac{16+d(2+(d-5)d)}{4(d-2)}f_{U}+\frac{d+2}{2}f_{\Omega}\right)\bar R\\
&+\bar R_{\mu\nu}\left(\frac{d(1+d)}{4}f_{Ric}
+\left(3-\frac{4}{d-2}+\frac{d}{2}\right)f_{U}
-2(d+2) f_{\Omega}\right)\bar R^{\mu\nu}\bigg]\\
&\frac{\delta^2 \mathrm{Tr}\left[e^{-s(- \bar\Box\mathbf{1}+\mathbf{U}^{(\mathrm{gh})})}\right]}{\delta h_{\mu\nu}(-q)\delta h_{\rho\sigma}(q)}
=\frac{2s^2}{(4\pi s)^{d/2}}\int d^dx\sqrt{\bar g}\,\bigg[
\bar R\left(-d\,f_R+f_{RU}-f_{\Omega}\right)\bar R
+\bar R_{\mu\nu}\left(-d\,f_{Ric}-f_{U}+4f_{\Omega}\right)\bar R^{\mu\nu}
\bigg].
\end{split}\end{equation}
Inserting these expressions into the right-hand side of Eq.~(\ref{PTFLOW}) and comparing with Eq.~(\ref{LHSFLOW}), we obtain the flow equations for the form factors. 

\subsection{The flow equations}
Introducing the dimensionless variables $g_k=k^{d-2}G_k$ and $\lambda_k=k^{-2}\bar\lambda_k$, the flow equations for the Newtonian constant and cosmological constants are given by
\begin{equation}\begin{split}
&k\partial_k\lambda_k=-g_k\frac{m^\frac{d}{2}d\left(8-\left(d+1\right)\left(2-\epsilon\eta\right)\left(1-\frac{2\lambda_k}{m}\right)^{\frac{d}{2}-m}\right)}{2\left(4\pi\right)^{\frac{d}{2}-1}}\frac{\Gamma\left(m-\frac{d}{2}\right)}{\Gamma\left(m\right)}-\lambda_k\left(2-\eta_k\right)\\
&k\partial_kg_k=g_k\left(d-2+\eta_k\right)
\end{split}\end{equation}
where the gravitational anomalous dimension is given by
\begin{equation}
\label{eqeta1}
\eta_k=-\frac{m^{\frac{d}{2}-1}g_k\left(4\left(d+6\right)+d\left(5d-7\right)\left(1-\frac{2\lambda_k}{m}\right)^{\frac{d}{2}-m-1}\right)}{3\left(4\pi\right)^{\frac{d}{2}-1}\left(1-\epsilon\frac{d\left(5d-7\right)m^{\frac{d}{2}-1}g_k\left(1-\frac{2\lambda_k}{m}\right)^{\frac{d}{2}-m-1}}{6\left(4\pi\right)^{\frac{d}{2}-1}}\frac{\Gamma\left(m-\frac{d}{2}+1\right)}{\Gamma\left(m\right)}\right)}\ \frac{\Gamma\left(m-\frac{d}{2}+1\right)}{\Gamma\left(m\right)}.
\end{equation}
The flow equations for the two form factors are given by
\begin{equation}\begin{split}
\label{flowequationsFF}
&k\partial_kf_k^{\left(i\right)}\left(z\right)=\alpha^{(i)}(k;m)\left[A^{(i)}\left(k,z;m\right)+\int_{0}^{1}{d\xi B^{(i)}\left(k,z;m,\xi\right)}\right]
\\
&+\left(2-\epsilon\eta\right)\beta^{(i)}(k;m)\left[C^{(i)}\left(k,z,\bar{\lambda};m\right)+\frac{1}{d-2}\int_{0}^{1}{d\xi D^{(i)}\left(k,z,\bar{\lambda};m,\xi\right)}\right]\\
\end{split}\end{equation}
where $i=\mathrm{R}$, $\mathrm{Ricc}$, $z=-\Box$ and
\begin{equation}\begin{split}
&\alpha^{(\mathrm{R})}(k;m)=\frac{k^{d-4}m^{\frac{d}{2}-2}}{8\left(4\pi\right)^\frac{d}{2}}\frac{\Gamma\left(m-\frac{d}{2}\right)}{\Gamma\left(m\right)},\quad\quad \alpha^{(\mathrm{Ricc})}(k;m)=\frac{\ k^{d-4}m^{\frac{d}{2}-2}}{2\left(4\pi\right)^\frac{d}{2}}\frac{\Gamma\left(m-\frac{d}{2}\right)}{\Gamma\left(m\right)}\\
&\beta^{(\mathrm{R})}(k;m)=\frac{k^{d-4}m^{\frac{d}{2}-2}}{2^7\left(4\pi\right)^\frac{d}{2}}\frac{\Gamma\left(m-\frac{d}{2}\right)}{\Gamma\left(m\right)},\quad\quad \beta^{(\mathrm{Ricc})}(k;m)=\frac{k^{d-4}m^{\frac{d}{2}-2}}{16\left(4\pi\right)^\frac{d}{2}}\frac{\Gamma\left(m-\frac{d}{2}\right)}{\Gamma\left(m\right)}
\end{split}\end{equation}
and
\begin{equation}\begin{split}
&A^{(\mathrm{R})}\left(k,z;m\right)=\frac{d}{3}\left(7m\left(1-\frac{d}{2m}\right)\frac{k^2m}{z}\ -6\left(\frac{k^2m}{z}\right)^2\right)\\
&A^{(\mathrm{Ricc})}\left(k,z;m\right)=-\frac{2d}{3}\left(m\left(1-\frac{12}{d}\right)\left(1-\frac{d}{2m}\right)\frac{k^2m}{z}-6\left(\frac{k^2m}{z}\right)^2\right) 
\end{split}\end{equation}
and
\begin{equation}\begin{split}
&C^{(\mathrm{R})}\left(k,z,\bar{\lambda};m\right)=\frac{\left(\frac{k^2m}{z}\right)^2}{3\left(1-\frac{2\lambda}{k^2m}\right)^{m-\frac{d}{2}}}\left(12d\left(d+1\right)+\frac{2m\left(17d\left(d+1\right)+96\right)\left(1-\frac{d}{2m}\right)}{\left(1-\frac{2\bar{\lambda}}{k^2m}\right)\frac{k^2m}{z}}\right)\\
&C^{(\mathrm{Ricc})}\left(k,z,\bar{\lambda};m\right)=\frac{\left(\frac{k^2m}{z}\right)^2}{3\left(1-\frac{2\bar{\lambda}}{k^2m}\right)^{m-\frac{d}{2}}}\left(-12d\left(d+1\right)+2m\frac{\left(\left(d-23\right)d-48\right)\left(1-\frac{d}{2m}\right)}{\left(1-\frac{2\bar{\lambda}}{k^2m}\right)\frac{k^2m}{z}}\right)
\end{split}\end{equation}
and
\begin{equation}\begin{split}
&B^{(\mathrm{R})}\left(k,z;m,\xi\right)=\\
&=\frac{\left(\frac{k^2m}{z}\right)^2}{\left(1-\left(\xi-1\right)\xi\frac{z}{k^2m}\ \right)^{m-\frac{d}{2}}}\left(2d-\frac{2md\left(1-\frac{d}{2m}\right)}{\left(1-\left(\xi-1\right)\xi\frac{z}{k^2m}\ \right)\frac{k^2m}{z}}-\frac{m\left(m+1\right)\left(1+\frac{d}{2}\right)\left(1-\frac{d}{2m}\right)\left(1-\frac{d}{2\left(m+1\right)}\right)}{\left(1-\left(\xi-1\right)\xi\frac{z}{k^2m}\ \right)^2\left(\frac{k^2m}{z}\right)^2}\right)\\
&B^{(\mathrm{Ricc})}\left(k,z;m,\xi\right)=\\
&=-\frac{2\left(\frac{k^2m}{z}\right)^2}{\left(1-\left(\xi-1\right)\xi\frac{z}{k^2m}\right)^{m-\frac{d}{2}}}\left(2d+\frac{4m\left(1-\frac{d}{2m}\right)}{\left(1-\left(\xi-1\right)\xi\frac{z}{k^2m}\right)\frac{k^2m}{z}}+\frac{m\left(m+1\right)\left(1-\frac{d}{2m}\right)\left(1-\frac{d}{2\left(m+1\right)}\right)}{4\left(1-\left(\xi-1\right)\xi\frac{z}{k^2m}\right)^2\left(\frac{k^2m}{z}\right)^2}\right)
\end{split}\end{equation}
and
\begin{equation}\begin{split}
&D^{(\mathrm{R})}\left(k,z,\bar{\lambda};m,\xi\right)=-\frac{4\left(\frac{k^2m}{z}\right)^2}{\left(1-\frac{2\bar{\lambda}}{k^2m}-\left(\xi-1\right)\xi\frac{z}{k^2m}\right)^{m-\frac{d}{2}}}\Bigg(d\left(d+1\right)\left(d-2\right)+\\
&+\frac{m\left(3d\left(d+1\right)+16\right)\left(d-2\right)\left(1-\frac{d}{2m}\right)}{\left(1-\frac{2\bar{\lambda}}{k^2m}-\left(\xi-1\right)\xi\frac{z}{k^2m}\right)\frac{k^2m}{z}}+\frac{\left(d\left(9d^2-57d+14\right)+256\right)\left(1-\frac{d}{2m}\right)\left(1-\frac{d}{2\left(m+1\right)}\right)}{16\left(1-\frac{2\bar{\lambda}}{k^2m}-\left(\xi-1\right)\xi\frac{z}{k^2m}\right)^2\left(\frac{k^2m}{z}\right)^2}\Bigg)\\
&D^{(\mathrm{Ricc})}\left(k,z,\lambda;m,\xi\right)=\frac{\left(\frac{k^2m}{z}\right)^2}{\left(1-\frac{2\bar{\lambda}}{k^2m}-\left(\xi-1\right)\xi\frac{z}{k^2m}\right)^{m-\frac{d}{2}}}\Bigg(4d\left(d+1\right)\left(d-2\right)+\\
&+\frac{16m\left(d+2\right)\left(d-2\right)\left(1-\frac{d}{2m}\right)}{\left(1-\frac{2\bar{\lambda}}{k^2m}-\left(\xi-1\right)\xi\frac{z}{k^2m}\right)\frac{k^2m}{z}}+\frac{\left(d\left(d+4\right)-20\right)\left(1-\frac{d}{2m}\right)\left(1-\frac{d}{2\left(m+1\right)}\right)}{\left(1-\frac{2\bar{\lambda}}{k^2m}-\left(\xi-1\right)\xi\frac{z}{k^2m}\right)^2\left(\frac{k^2m}{z}\right)^2}\Bigg)
\end{split}\end{equation}
Generally, as in $d=2$ case of \cite{Glaviano:2024hie}, for generic $m$ the integrals over $\xi$ involve complicated combinations of hypergeometric functions. For integer $m$, these combinations reduce to Eq.(\ref{eqFFm3}). 

\section{Technique to derive the asymptotic series}\label{appserie}
\noindent In this Appendix, we derive the asymptotic expansions in Eqs.~(\ref{IRserie}) and (\ref{UVseries}) without explicitly evaluating the exact $k$ integral. The construction applies whenever the flow equation can be written as
\begin{equation}
\label{eqdiffeta}
k\partial_k f_k(q^2,\{m_i^2(k)\}) + A_k f_k(q^2,\{m_i^2(k)\}) = H_k(q^2,\{m_i^2(k)\})\,,
\end{equation}
where $\{m_i^2(k)\}$ denotes a set of $k$-dependent quantities. In the present case, $\{m_i^2(k)\}=\{m_p^2,\bar\lambda(k)\}$. The solution of Eq.~(\ref{eqdiffeta}) can be written as
\begin{equation}\begin{split}
&f_{k=\Lambda}\left(q^2,\{m_i^2(k)\}\right)-f_{k=k_0}\left(q^2,\{m_i^2(k)\}\right)
= \int_{k_0}^{\Lambda} \frac{1}{k} T_k\left(q^2,\{m_i^2(k)\}\right) dk\,,\\
&T_k\left(q^2,\{m_i^2(k)\}\right)
= e^{\int_{k_0}^{k}\frac{A_x}{x}dx}\, H_k\left(q^2,\{m_i^2(k)\}\right)\,.
\end{split}\end{equation}
Generally, $A_k\neq0$ accounts for a nonvanishing anomalous dimension.

Asymptotic limits are obtained through Taylor expansions. Here, since the flow equations cannot be solved exactly, the asymptotic expansion must be performed before integrating over the RG scale $k$. This truncates part of the RG flow, because the limits $q^2\to\infty$ or $m_p\to0$ alter the hierarchy between external and running scales, enforcing $k^2\ll q^2$ and $m_p^2\ll k^2$, respectively. After integration, this generates spurious contributions, such as terms $\sim\Lambda^2$, $\Lambda^4$ for $\Lambda\to\infty$ or $\sim1/k^2$, $1/k^4$ for $k\to0$. The correct asymptotic behavior is recovered by reconstructing the missing part of the flow through a matching between the truncated (“cut”) flow and the asymptotic series, effectively implementing a Wilsonian reconstruction of the fluctuation contribution.

The origin of the spurious behaviors is related to the non-commutativity between the limiting procedures and the RG integration,
\begin{equation}\begin{split}
&\int_{k}^{\Lambda}{\frac{1}{k_1}\lim_{m_p\rightarrow0}{T_{k_1}\left(q^2,m_p^2\right)dk_1}}\ne\lim_{m_p\rightarrow0}\int_{k}^{\Lambda}{\frac{1}{k_1}T_{k_1}\left(q^2,m_p^2\right)dk_1}\\
&\int_{k}^{\Lambda}{\frac{1}{k_1}\lim_{q\rightarrow\infty}{T_{k_1}\left(q^2,m_p^2\right)dk_1}}\ne\lim_{q\rightarrow\infty}\int_{k}^{\Lambda}{\frac{1}{k_1}T_{k_1}\left(q^2,m_p^2\right)dk_1}
\end{split}\end{equation}
If the two operations commute, the truncated flows would vanish and the procedure would reduce to a single integral evaluation.

In the presence of a cosmological constant, the limit $q^2\to\infty$ is not equivalent to the limit $m_p\to0$. In the following Subsections we treat separately the regimes $q^2\to\infty$ and $m_p\to0$.

For notational convenience, we denote asymptotic expansions around a variable $x$ simply as $f_k(x\to0)$ or $f_k(x\to\infty)$. More generally, throughout the following, the presence of an arrow in a given quantity will always denote an asymptotic expansion around the corresponding limit.

From the previous setup, it follows that the qualitative asymptotic behavior is independent of $\epsilon$ and controlled only by the UV-attractive fixed point of the Newton coupling and the ratios $k^2/m_p^2$, $q^2/m_p^2$, and $q^2/k^2$. Accordingly, Eqs.~(\ref{IRserie}) and (\ref{UVseries}) retain the same asymptotic structure for all $\epsilon$, although the coefficients of the series--except for the universal logarithmic term--may still depend on it.

Before proceeding, let us comment on an alternative evaluation of Eq.~(\ref{flowequationsFF}). One may interchange the integrations over $\xi$ and $k$ and perform the $k$ integral first. Although this partially simplifies the expression, the resulting integrand cannot in general be integrated analytically over $\xi$, so that only asymptotic regimes remain accessible. As in the previous approach, performing Taylor expansions before the $\xi$ integration truncates part of the flow and may generate spurious divergences at $\xi=0$ and $\xi=1$. A matching procedure is therefore still required. We do not discuss this implementation further, but it provides an independent consistency check and reproduces the same asymptotic results obtained from the matching procedure in the $k$ representation.

\subsection{The asymptotic series around \texorpdfstring{$m_p=0$}{N}}
We start the discussion with the expansion around the limit $m_p \to 0$, as this case is technically simpler and clearly illustrates the matching procedure. Moreover, it provides useful physical insight into the structure of the flow.

Taking the limit $m_p \to 0$ in the beta function effectively selects the regime $m_p \ll k$. This removes the contribution of infrared fluctuations in the region $k \lesssim m_p$, and naturally splits the flow into two domains, $(0,k_1)$ and $(k_1,\Lambda)$, with $k_1 \sim m_p$,
\begin{equation}
\begin{split}
&f_{k=\Lambda}(q^2,m_p^2\to0) - f_{k=0}(q^2,m_p^2\to0)= \int_{0}^{k_1} \frac{T_k(q^2,m_p^2\to0)}{k}dk
+\int_{k_1}^{\Lambda} \frac{T_k(q^2,m_p^2\to0)}{k}dk\equiv\\ &\equiv\int_{IR}{\frac{T_{k_0}\left(q^2,m_p^2\to0\right)}{k_0}dk_0}+\int_{UV}{\frac{T_{k_{UV}}\left(q^2,m_p^2\to0\right)}{k_{UV}}dk_{UV}}.
\end{split}
\end{equation}
This decomposition is always possible, independently of the expansion. To obtain at $k=0$ the ultraviolet regime $m_p^2 \ll q^2$, one must consistently select the UV sector of the fluctuations in both contributions. 

In the UV contribution, $m_p^2 \ll q^2 \ll k_{\mathrm{UV}}^2$, so that the flow can be expanded around $m_p \to 0$ (or $k \to \infty$), restricting the integration to $k_1 \gtrsim m_p$. In other words,
\begin{equation}
\int_{UV}{\frac{T_{k_{UV}}\left(q^2,m_p^2\to0\right)}{k_{UV}}dk_{UV}}=F_\ast^{\left(\eta_\ast=-2\right)}\left(\frac{q^2}{k^2},\frac{\bar{\lambda}}{k^2};\frac{q^2}{\Lambda^2}\right)+\sum_{j=1}^{+\infty}{F_j^{\left(\eta_\ast=-2\right)}\left(\frac{q^2}{k^2},\frac{\bar{\lambda}}{k^2};\frac{q^2}{\Lambda^2}\right)\left(\frac{m_p}{k}\right)^{2j}}
\end{equation}
the UV sector contains the flow of the UV fixed-point and its perturbations.

In the IR contribution, where $k_{\mathrm{IR}}^2 \ll m_p^2 \ll q^2$, one introduces the rescaling $k_{\mathrm{IR}} = m_p X$ and rewrites the upper integration limit as $X_1^2 = k_1^2/m_p^2$. Expanding around $m_p \to 0$, one obtains
\begin{equation}
\int_{IR}{\frac{T_{k_{\mathrm{IR}}}\left(q^2,m_p^2\to0\right)}{k_{\mathrm{IR}}}dk_{\mathrm{IR}}}
\equiv f^{(IR)}_{k=0}\left(q^2,\frac{k_1^2}{m_p^2}\right)\,.
\end{equation}
To make the procedure consistent, this expression must be reexpanded after performing the integration, i.e.
\begin{equation}
f^{(IR)}_{k=0}\left(q^2,\frac{k_1^2}{m_p^2}\right)
\;\longrightarrow\;
f^{(IR)}_{k=0}\left(q^2,\frac{k_1^2}{m_p^2\to0}\right)\,.
\end{equation}

Now let us consider the limit $k_1 \to 0$ at fixed $m_p$. This suppresses the IR contribution, since the integration domain shrinks, and enhances the relative weight of the UV sector. In this limit, both contributions develop divergent terms. However, because the exact integral has been split into two parts and a change of variables has been performed in one of them, these divergences are identical but enter with opposite sign, leading to their cancellation and yielding a finite flow,
\begin{equation}
f_{k=\Lambda}\left(q^2,m_p^2\to0\right)-f_{k=0}\left(q^2,m_p^2\to0\right)=\lim_{k_1\to0}\left[f^{(IR)}_{k=0}\left(q^2,\frac{k_1^2}{m_p^2\to0}\right)+\int_{k_1}^{\Lambda}{\frac{T_k\left(q^2,m_p^2\to0\right)}{k}dk}\right]
\end{equation}
As a result, expanding the integrand around $m_p \to 0$ and taking the limit $k_1 \to 0$, the IR region $k \in (0,k_1)$ reproduces the singular structure of the UV region $(k_1,\Lambda)$, leading to an exact cancellation of spurious divergences. In the absence of genuine divergences, this is equivalent to taking the limit $k \to 0$ without encountering singular behavior. The final result is therefore the expansion of the exact integral around $m_p = 0$, which reproduces Eq.~(\ref{sermp0k0}) as a special case.

If the limit $m_p \to 0$ commutes with the integral, the IR contribution vanishes and the computation reduces to the standard single-integral result.

Beyond their role in canceling divergences, the IR contributions also encode nonanalytic terms in the expansion. In the limits $\Lambda \to \infty$ and $k \to 0$, these terms reproduce the expected logarithmic behavior $\sim \ln(q^2/m_p^2)$.

\subsection{The asymptotic series for large momentum}\label{subappaslargep}
The limit $q^2 \to \infty$ probes the UV regime of the flow. In this regime, in the beta functions $k^2 \ll q^2$, so that the contribution to the flow for $k^2$ above $q^2$ is cut. As for the $m_p\to0$ case, the flow effectively splits into IR and UV contributions separated at a scale $k_1$,
\begin{equation}\begin{split}
&\int_{IR}{\frac{1}{k_0}T_{k_0}\left(q^2\to\infty,m_p^2\right)dk_0}=\int_{0}^{k_1}{\frac{1}{k}T_k\left(q^2\to\infty,m_p^2\right)dk}\\
&\int_{UV}{\frac{T_{k_{UV}}\left(q^2\rightarrow\infty,m_p^2\right)}{k_{UV}}dk_{UV}}=\int_{k_1}^{\Lambda}{\frac{1}{k}T_k\left(q^2\to\infty,m_p^2\right)dk}
\end{split}\end{equation}
In the UV region, the cuts implies that in the limit $\Lambda\to\infty$ spurious contributions emerge. The reconstruction of the correct flow proceeds treating the UV region via the rescaling $k=qX$, followed by an expansion in $q^2\to\infty$ before and after integration,
\begin{equation}
\int_{UV}{\frac{T_{k_{UV}}\left(q^2\rightarrow\infty,m_p^2\right)}{k_{UV}}dk_{UV}}=\int_{\frac{k_1}{q}}^{\frac{\Lambda}{q}}{\frac{T_{k=X(q\rightarrow\infty)}\left(q^2\rightarrow\infty,m_p^2\right)}{X}dX}\equiv f_{\frac{k_1}{q\to\infty}}^{\left(UV\right)}\left(q^2\rightarrow\infty,m_p^2,\frac{\Lambda^2}{q^2\rightarrow\infty}\right)
\end{equation}
The full flow is obtained by combining IR and UV contributions. In the limit $k_1 \to \infty$, the reconstruction is effectively dominated by the IR sector. Nevertheless, due to the splitting of the exact integral and the change of variables in the UV part, both sectors contain identical divergent contributions with opposite sign, which cancel in the sum,
\begin{equation}\begin{split}
&f_{k=\Lambda}\left(q^2\to\infty,m_p^2\right)-f_{k=0}\left(q^2\to\infty,m_p^2\right)=
\\
&\lim_{k_1\to\infty}\left[\int_{0}^{k_1}{\frac{1}{k}T_k\left(q^2\to\infty,m_p^2\right)dk}+f_{\frac{k_1}{q\to\infty}}^{\left(UV\right)}\left(q^2\rightarrow\infty,m_p^2,\frac{\Lambda^2}{q^2\rightarrow\infty}\right)\right]
\end{split}\end{equation}
this reproduces Eq.(\ref{UVseries}).

Apart from the regularization terms, the UV region contains nonanalytic but smooth contributions. In the limits $\Lambda\to\infty$ and $k\to0$, these reduce to $\sim q^{-2n}\ln(q^2/m_p^2)$ with $n=0,1,2,\ldots$.

\subsection{The asymptotic series for \texorpdfstring{$m_p\to\infty$}{N}}
The expansion for a large Planck mass, $m_p \to \infty$, describes the IR regime. This limit truncates the flow at large $k$ and generates spurious divergences in the limit $\Lambda \to \infty$, analogous to the $q^2\to\infty$ case. The same reconstruction procedure can therefore be applied, with $q^2$ replaced by $m_p^2$, leading to
\begin{equation}\begin{split}
\label{regIR}
&f_{k=\Lambda}\left(q^2,m_p^2\to\infty\right)-f_{k=0}\left(q^2,m_p^2\to\infty\right)=
\\
&\lim_{k_1\to\infty}\left[\int_{0}^{k_1}{\frac{1}{k}T_k\left(q^2,m_p^2\to\infty\right)dk}+f_{\frac{k_1}{m_p\to\infty}}^{\left(UV\right)}\left(q^2,m_p^2\to\infty,\frac{\Lambda^2}{m_p^2\rightarrow\infty}\right)\right]
\end{split}\end{equation}
this reproduces Eq.(\ref{IRserie}). 

The UV fluctuation region generates the nonanalytic contributions to the flow, which in the limit $\Lambda\to\infty$ and $k\to0$ reduce at leading order to $\sim \ln(\Lambda^2/m_p^2)$, with subleading terms behaving as $\ln(m_p^2/\bar\lambda)$.

In the IR region of the flow, both the momentum $q$ and the coarse-graining scale $k$ are much smaller than $m_p$. The regime can be described as an expansion at fixed $x=q^2/k^2$ in the limit $k\ll m_p$, corresponding to the IR fixed point and its perturbations,
\begin{equation}
\label{IRflow}
\int_{k}^{\Lambda}{\frac{1}{k}T_k\left(q^2,m_p^2\to\infty\right)dk}=F_\ast^{\left(\eta_\ast=0\right)}\left(\frac{q^2}{k^2},\frac{\bar{\lambda}}{k^2};\frac{q^2}{\Lambda^2}\right)+\sum_{j=1}^{+\infty}{F_j^{\left(\eta_\ast=0\right)}\left(\frac{q^2}{k^2},\frac{\bar{\lambda}}{k^2};\frac{q^2}{\Lambda^2}\right)\left(\frac{k}{m_p}\right)^{2j}}
\end{equation}
According to Sec. \ref{secASformfactor}, the IR fixed point reproduces the EFT flow, while the perturbations arise from the Gaussian running of the Newton constant. In this regime the flow generates divergences of the form $\Lambda^2$, $\Lambda^4$, etc. The finite flow in Eq.~(\ref{regIR}) shows how the full nonperturbative running cures these divergences. As the scale $k$ approaches the Planck mass, the UV-attractive non- Gaussian fixed point becomes relevant and tames the growth of the flow, replacing the divergent contributions with finite nonanalytic terms.

\section{Technical results}\label{app:techres}
\noindent In this Appendix we collect long expressions and technical analytical results of this work.

\subsection{The flow equation for the form factors}
After solving the $\xi$ integral in Eq.(\ref{flowequationsFF}), the flow equation for the form factors involve complicated combinations of hypergometric functions. For integer $m$, the flow equations for the two form factors take the form
\begin{equation}\begin{split}
&k\partial_kf_k^{\left(i\right)}\left(z\right)=A^{(i)}\left(k,z;m\right)+B^{(i)}\left(k,z;m\right)\tanh^{-1}\left(\sqrt{\frac{z}{4mk^2+z}}\right)+\\
&+\left(2-\epsilon\eta_k\right)\left(C^{(i)}\left(k,z,\bar{\lambda}_k;m\right)+D^{(i)}\left(k,z,\bar{\lambda}_k;m\right)\tanh^{-1}\left(\sqrt{\frac{z}{4mk^2-8\bar\lambda_k+z}}\right)\right)
\end{split}\end{equation}
where $i=\mathrm{R}$, $\mathrm{Ricc}$ and $z=-\Box$. With $m=3$, the coefficients read
\begin{equation}\begin{split}
&A^{(\mathrm{R})}\left(k,z\right)=-\frac{k^2\left(1296k^6-72k^4z+78k^2z^2+z^3\right)}{32\pi^2z^2\left(12k^2+z\right)^2},\\
&A^{(\mathrm{Ricc})}\left(k,z\right)=\frac{k^2\left(5184k^6+1008k^4z+6k^2z^2+z^3\right)}{16\pi^2z^2\left(12k^2+z\right)^2}
\end{split}\end{equation}
and
\begin{equation}\begin{split}
&B^{(\mathrm{R})}\left(k,z\right)=\frac{27k^6\left(72k^4-5z^2\right)}{4\pi^2\left(z\left(12k^2+z\right)\right)^\frac{5}{2}},\\
&B^{(\mathrm{Ricc})}\left(k,z\right)=\frac{27k^6\left(288k^4+72k^2z+7z^2\right)}{2\pi^2\left(z\left(12k^2+z\right)\right)^\frac{5}{2}}
\end{split}\end{equation}
and
\begin{equation}\begin{split}
&C^{(\mathrm{R})}\left(k,z,{\bar{\lambda}}_k\right)=\frac{9k^6}{128\pi^2z^2\left(3k^2-2\bar{\lambda_k}\right)^2\left(12k^2-8{\bar{\lambda}}_k+z\right)^2}\Bigg(3240k^6+36k^4\left(67z-180{\bar{\lambda}}_k\right)-\\
&-960{\bar{\lambda}}_k^3+6k^2\left(720{\bar{\lambda}}_k^2+73z^2-536{\bar{\lambda}}_kz\right)+25z^3-292{\bar{\lambda}}_kz^2+1072{\bar{\lambda}}_k^2z\Bigg)\\
&C^{(\mathrm{Ricc})}\left(k,z,{\bar{\lambda}}_k\right)=-\frac{9k^6}{64\pi^2z^2\left(3k^2-2{\bar{\lambda}}_k\right)^2\left(12k^2-8{\bar{\lambda}}_k+z\right)^2}\Bigg(12960k^6+576k^4\left(7z-45{\bar{\lambda}}_k\right)-\\
&-3840{\bar{\lambda}}_k^3+6k^2\left(2880{\bar{\lambda}}_k^2+13z^2-896{\bar{\lambda}}_kz\right)+13z^3-52{\bar{\lambda}}_kz^2+1792{\bar{\lambda}}_k^2z\Bigg)
\end{split}\end{equation}
and
\begin{equation}\begin{split}
&D^{(\mathrm{R})}\left(k,z,\bar{\lambda}_k\right)=-\frac{27k^6\left(180k^4+48k^2\left(3z-5{\bar{\lambda}}_k\right)+80{\bar{\lambda}}_k^2+13z^2-96\lambda z\right)}{16\pi^2\left(z\left(12k^2-8\lambda+z\right)\right)^\frac{5}{2}}\\
&D^{(\mathrm{Ricc})}\left(k,z,\bar{\lambda}_k\right)=\frac{27k^6\left(720k^4+24k^2\left(11z-40{\bar{\lambda}}_k\right)+320{\bar{\lambda}}_k^2+35z^2-176{\bar{\lambda}}_kz\right)}{8\pi^2\left(z\left(12k^2-8{\bar{\lambda}}_k+z\right)\right)^\frac{5}{2}}
\end{split}\end{equation}
Different values of the cutoff parameter $m$ modify the form of the coefficients but leave the structure of the flow equations unchanged.

\subsection{The IR series}\label{app:techresIRserie}
The general solution in the IR regime at $k=0$ takes the form
\begin{equation}
f_{k=0}\left(u,t=\mathrm{const},w\rightarrow0\right)=f_{k=0}^{\left(\mathrm{EFT}\right)}\left(q^2,u;\mu^2\right)+b\ln{\left(\frac{4\pi}{23}\frac{m_p^2}{\mu^2}\right)}+\sum_{n=1}^{+\infty}{h_n\left(u,t\right)w^n}
\end{equation}
where $h_n(u)$ denote the series coefficients. The coefficients $b$ are given by
\begin{equation}
b^{\left(\mathrm{R}\right)}=\frac{7}{384\pi^2},\quad \quad b^{\left(\mathrm{Ricc}\right)}=\frac{7}{192\pi^2}
\end{equation}
The first series coefficients are given by
\begin{equation}\begin{split}
&h_1^{\left(\mathrm{R}\right)} \left(u,t\right)=\frac{23\left(12231u^3+245420u^2+1675520u-268800\right)}{22579200\pi^3u^3}-\\
&-\frac{23\left(39u+980\right)\ln{\left(\frac{23t}{6\pi}\right)}}{107520\pi^3u}-\frac{23\left(39u^2+512u-80\right)\left(u+8\right)^\frac{3}{2}\ln{\left(\frac{1+\sqrt{\frac{u}{u+8}}}{1-\sqrt{\frac{u}{u+8}}}\right)}}{107520\pi^3u^\frac{7}{2}}\\
&h_2^{\left(\mathrm{R}\right)}\left(u,t\right)=\frac{529\left(-12902400+103326720u+23927400u^2+2386116u^3+75823u^4\right)}{7315660800\pi^4u^4}\\
&-\frac{529\left(8+u\right)^\frac{5}{2}\left(-80+664u+37u^2\right)\ln{\left(\frac{1+\sqrt{\frac{u}{u+8}}}{1-\sqrt{\frac{u}{u+8}}}\right)}}{5806080\pi^4u^\frac{9}{2}}-\frac{529\left(17640+1404u+37u^2\right)\ln{\left(\frac{23t}{6\pi}\right)}}{5806080\pi^4u^2}
\end{split}\end{equation}
and
\begin{equation}\begin{split}
&h_1^{\left(\mathrm{Ricc}\right)}\left(u,t\right)=\frac{23\left(12231u^3+245420u^2+1675520u-268800\right)}{22579200\pi^3u^3}-\\
&-\frac{23\left(39u^2+512u-80\right)\left(u+8\right)^\frac{3}{2}\ln{\left(\frac{1+\sqrt{\frac{u}{u+8}}}{1-\sqrt{\frac{u}{u+8}}}\right)}}{107520\pi^3u^\frac{7}{2}}-\frac{23\left(39u+980\right)\ln{\left(\frac{23t u}{6\pi}\right)}}{107520\pi^3u}\\
&h_2^{\left(\mathrm{Ricc}\right)}\left(u,t\right)=\frac{529\left(51609600-111390720u-4256280u^2+3483684u^3+305045u^4\right)}{3657830400\pi^4u^4}-\\
&-\frac{529\left(8+u\right)^\frac{5}{2}\left(320-784u+275u^2\right)\ln{\left(\frac{1+\sqrt{\frac{u}{u+8}}}{1-\sqrt{\frac{u}{u+8}}}\right)}}{2903040\pi^4u^\frac{9}{2}}+\frac{529\left(17640+4716u+275u^2\right)\ln{\left(\frac{23t}{6\pi}\right)}}{2903040\pi^4u^2}
\end{split}\end{equation}
The form factors exhibit the same nonlocal structure, which extends to higher-order coefficients $h_i(u)$. The IR regime shares the same functional structure as in the EFT result.

\subsection{The fixed point solutions}\label{apptech:fixedpoint}
The form factor fixed point solutions are given by
\begin{equation}\begin{split}
&F_*^{(\mathrm{part},i)}(x,\lambda_k)=\alpha_0^{(i)}(\eta_*)\ln(x)+\beta_0^{(i)}(\eta_*)\ln{\left(3-2\lambda_k\right)}+\\
&+A^{(i)}\left(x,\lambda_k\right)+B^{(i)}\left(x,\lambda_k\right)\mathrm{arctanh}{\left(\sqrt{\frac{x}{x+12}}\right)}+C^{(i)}\left(x,\lambda_k\right)\mathrm{arctanh}{\left(\sqrt{\frac{x}{x-8\lambda_k+12}}\right)}
\end{split}\end{equation}
where $i=\mathrm{R}$, $\mathrm{Ricc}$ and
\begin{equation}\begin{split}
&\alpha_0^{(\mathrm{R})}(\eta_*)=\frac{7\eta_\ast}{768\pi^2}-\frac{1}{1920\pi^2},\quad\quad \alpha_0^{(\mathrm{Ricc})}(\eta_*)=\frac{7\eta_\ast}{384\pi^2}-\frac{7}{320\pi^2}\\
&\beta_0^{(\mathrm{R})}(\eta_*)=\frac{7}{384\pi^2}-\frac{7\eta_\ast}{768\pi^2},\quad\quad \beta_0^{(\mathrm{Ricc})}(\eta_*)=\frac{7}{192\pi^2}-\frac{7\eta_\ast}{384\pi^2}
\end{split}\end{equation}
For $m=3$ the other coefficients are given by
\begin{equation}\begin{split}
&A^{(\mathrm{R})}\left(x,\lambda_k\right)=\frac{29x^2+60x-216}{320\pi^2x^2\left(x+12\right)}+\\
&+\frac{\left(\eta_\ast-2\right)\left(4\left(3-2\lambda_k\right)^2\left(8\lambda_k+9\right)+x^2\left(33-72\lambda_k\right)+2x\left(280\lambda_k^2-534\lambda_k+171\right)\right)}{256\pi^2x^2\left(2\lambda_k-3\right)\left(x-8\lambda_k+12\right)}\\
&A^{(\mathrm{Ricc})}\left(x,\lambda_k\right)=\frac{19x^2+120x+864}{160\pi^2x^2\left(x+12\right)}\\
&+\frac{\left(\eta_\ast-2\right)\left(-16\left(3-2\lambda_k\right)^2\left(8\lambda_k+9\right)+x^2\left(80\lambda_k-81\right)-144x\left(4\lambda_k^2-8\lambda_k+3\right)\right)}{128\pi^2x^2\left(2\lambda_k-3\right)\left(x+12-8\lambda_k\right)}
\end{split}\end{equation}
and
\begin{equation}\begin{split}
&B^{(\mathrm{R})}\left(x,\lambda_k\right)=\frac{3888-864x-918x^2-306x^3-17x^4}{480\pi^2x^\frac{5}{2}\left(x+12\right)^\frac{3}{2}}\\
&B^{(\mathrm{Ricc})}\left(x,\lambda_k\right)=-\frac{15552+3024x+378x^2+126x^3+7x^4}{240\pi^2x^\frac{5}{2}\left(x+12\right)^\frac{3}{2}}
\end{split}\end{equation}
and
\begin{equation}\begin{split}
&C^{(\mathrm{R})}\left(x,\lambda_k\right)=\frac{\eta_\ast-2}{384\pi^2x^\frac{5}{2}\left(x-8\lambda_k+12\right)^\frac{3}{2}}\Bigg[8\left(3-2\lambda_k\right)^2\left(32\lambda_k^2+24\lambda_k+27\right)+\\
+&16x\left(272\lambda_k^3-630\lambda_k^2+243\lambda_k+135\right)-18x^2\left(88\lambda_k^2-128\lambda_k+21\right)+2x^3\left(92\lambda_k-63\right)-7x^4\Bigg]\\
&C^{(\mathrm{Ricc})}\left(x,\lambda_k\right)=\frac{\eta_\ast-2}{192\pi^2x^\frac{5}{2}\left(x+12-8\lambda_k\right)^\frac{3}{2}}\Bigg[32\left(3-2\lambda_k\right)^2\left(32\lambda_k^2+24\lambda_k+27\right)+\\
&+16x\left(256\lambda_k^3-648\lambda_k^2+270\lambda_k+189\right)+x^2\left(-768\lambda_k^2+432\lambda_k+378\right)+x^3\left(126-32\lambda_k\right)+7x^4\Bigg]
\end{split}\end{equation}
These expressions form the general structure of the particular fixed point solution.

In the IR regime $x\ll1$, the solutions take the form
\begin{equation}\begin{split}
&F_*^{(\mathrm{part,R})}(x\to0,\lambda_k)=\left(\frac{7\eta_\ast}{768\pi^2}-\frac{1}{1920\pi^2}\right)\ln{\left(x\right)}+\left(\frac{7}{384\pi^2}-\frac{7\eta_\ast}{768\pi^2}\right)\ln{\left(3-2\lambda_k\right)}+\\
&+\frac{\left(\eta_\ast-2\right)\lambda_k^2}{24\pi^2x^2}+\frac{107\left(\eta_\ast-2\right)\lambda_k}{576\pi^2x}+\frac{21856\lambda_k^2-52968\lambda_k+20826-35\eta_\ast\left(344\lambda_k^2-852\lambda_k+369\right)}{115200\pi^2\left(3-2\lambda_k\right)^2}+\\
&+\frac{-480\lambda_k^3+2056\lambda_k^2-2772\lambda_k+918+13\eta_*\left(4\lambda_k^2-18\lambda_k+27\right)}{17920\pi^2\left(2\lambda_k-3\right)^3}x+\\
&+\frac{5440\lambda_k^4-32640\lambda_k^3+72996\lambda_k^2-70776\lambda_k+21546+111\eta_\ast\left(2\lambda_k^2-12\lambda_k+27\right)}{1451520\pi^2\left(3-2\lambda_k\right)^4}x^2+O\left(x^3\right)\\
&F_*^{\mathrm{(part,Ricc)}}(x\to0,\lambda_k)=\left(\frac{7\eta_\ast}{384\pi^2}-\frac{7}{320\pi^2}\right)\ln{\left(x\right)}+\left(\frac{7}{192\pi^2}-\frac{7\eta_\ast}{384\pi^2}\right)\ln{\left(3-2\lambda_k\right)}-\\
&-\frac{\left(\eta_\ast-2\right)\lambda_k^2}{3\pi^2x^2}-\frac{29\left(\eta_\ast-2\right)\lambda_k}{72\pi^2x}+\frac{8976\lambda_k^2-14328\lambda_k-8154+\eta_*\left(-4240\lambda_k^2+6420\lambda_k+4635\right)}{57600\pi^2\left(3-2\lambda_k\right)^2}+\\
&+\frac{-736\lambda_k^3+2264\lambda_k^2-252\lambda_k-4590+131\eta_\ast\left(4\lambda_k^2-18\lambda_k+27\right)}{26880\pi^2\left(2\lambda_k-3\right)^3}x+\\
&+\frac{3008\lambda_k^4-18048\lambda_k^3+37308\lambda_k^2-20808\lambda_k-29322+825\eta_\ast\left(2\lambda_k^2-12\lambda_k+27\right)}{725760\pi^2\left(3-2\lambda_k\right)^4}x^2+O\left(x^3\right)
\end{split}\end{equation}
in the $\lambda_k\to0$ limit the negative powers vanish.

In the UV $x\gg1$ the solutions take the form
\begin{equation}
F_*^{(\mathrm{part},i)}(x\to\infty,\lambda_k)=b_{0}^{\left(i\right)}+\sum_{n=1}^{+\infty } \frac{b_{n}^{\left(i\right)}+c_{n}\ln{\left(\frac{x}{3-2{\lambda }_{k}}\right)}}{x^{n}}
\end{equation}
The first terms are given by
\begin{equation}\begin{split}
&F_*^{(\mathrm{part,R)}}(x\to\infty,\lambda_k)=\frac{17\ln{3}}{960\pi^2}+\frac{\frac{-112\lambda_k^2+744\lambda_k-414+\eta_\ast\left(56\lambda_k^2-384\lambda_k+225\right)}{768\pi^2\left(2\lambda_k-3\right)}+\frac{25\left(\eta_\ast-2\right)\lambda_k\ln{\left(\frac{x}{3-2\lambda_k}\right)}}{192\pi^2}}{x}+O\left(\frac{1}{x^2}\right)\\
&F_*^{(\mathrm{part,Ricc})}(x\to\infty,\lambda_k)=\frac{7\ln{3}}{480\pi^2}+\frac{\frac{-2\left(56\lambda_k^2+60\lambda_k-99\right)+\eta_\ast\left(56\lambda_k^2+72\lambda_k-117\right)}{384\pi^2\left(2\lambda_k-3\right)}-\frac{13\left(\eta_\ast-2\right)\lambda_k\ln{\left(\frac{x}{3-2\lambda_k}\right)}}{96\pi^2}}{x}+O\left(\frac{1}{x^2}\right)
\end{split}\end{equation}

\end{appendices}

\bibliography{refformfact}
\end{document}